\documentclass[preprint]{aastex} 

\shorttitle{Pulsating B-type stars in the open cluster NGC 884}
\shortauthors{Saesen et al.}

\usepackage{graphicx}
\usepackage{txfonts}
\usepackage{lscape}
\usepackage{natbib}
\usepackage{longtable}
\usepackage[normalem]{ulem}
\bibpunct{(}{)}{;}{a}{}{,}
\usepackage{placeins}
 
\begin{document}

   \title{Pulsating B-type stars in the open cluster NGC 884: frequencies, mode
     identification and asteroseismology}

\author{S.~Saesen\altaffilmark{1,2}, M.~Briquet\altaffilmark{2,3,6}, C.~Aerts\altaffilmark{2,4}, A.~Miglio\altaffilmark{5}, F.~Carrier\altaffilmark{2}}

\altaffiltext{1}{Observatoire de Gen\`eve, Universit\'e de Gen\`eve, Chemin des Maillettes 51, 1290 Sauverny, Switzerland; \\sophie.saesen@unige.ch}
\altaffiltext{2}{Instituut voor Sterrenkunde, Katholieke Universiteit Leuven, Celestijnenlaan 200 D, 3001 Leuven, Belgium}
\altaffiltext{3}{Institut d'Astrophysique et de G\'eophysique de l'Universit\'e de
Li\`ege, All\'ee du 6 Ao\^ut 17, 4000 Li\`ege, Belgium}
\altaffiltext{4}{Department of Astrophysics, Radboud University Nijmegen, POBox 9010, 6500 GL Nijmegen, The Netherlands}
\altaffiltext{5}{School of Physics and Astronomy, University of Birmingham, Edgbaston, Birmingham, B15 2TT, UK}
 \altaffiltext{6}{Postdoctoral Researcher of the Fund for Scientific Research, Fonds de la Recherche Scientifique -- FNRS, Belgium}

\begin{abstract}
  Recent progress in the seismic interpretation of field $\beta$~Cep stars has
    resulted in improvements of the physics in the stellar structure and
    evolution models of massive stars. Further asteroseismic constraints can be
    obtained from studying ensembles of stars in a young open cluster, which all
    have similar age, distance and chemical composition.  We present an
    observational asteroseismology study based on the discovery of numerous 
    multi-periodic and mono-periodic B-stars in the open cluster NGC~884.
    We describe a thorough investigation of the pulsational properties of all
    B-type stars in the cluster.  Overall, our detailed frequency analysis
    resulted in 115~detected frequencies in 65~stars. We found 36~mono-periodic,
    16~bi-periodic, 10~tri-periodic, and 2~quadru-periodic stars and one star
    with 9~independent frequencies.  We also derived the amplitudes and phases
    of all detected frequencies in the $U$, $B$, $V$ and $I$ filter, if
    available.  We achieved unambiguous identifications of the mode degree for
    twelve of the detected frequencies in nine of the pulsators.  Imposing the
    identified degrees and measured frequencies of the radial, dipole and
    quadrupole modes of five pulsators led to a seismic cluster age estimate of
    $\log(\mathrm{age/yr}) =7.12-7.28$ from a comparison with stellar models.
    Our study is a proof-of-concept for and illustrates the current status of
    ensemble asteroseismology of a young open cluster.
\end{abstract}



\section{INTRODUCTION}
\label{sect:intro}

Main-sequence stars of spectral type {O9 to B2} are very interesting targets for
asteroseismology, i.e., the study of the internal stellar structure by
interpreting the observed oscillation characteristics. Indeed, these stars have
a convective core which strongly determines the evolution of the star, while
they do not suffer from a strong stellar wind. Moreover, the internal rotation
of these stars is not well known, but it causes mixing of the chemical elements
which may also have an important effect on their evolution.  And finally, as
they are the progenitors of core-collapse supernovae, these stars will
chemically enrich the Universe.

In this paper, we present an observational asteroseismology study of pulsating
B-stars in the cluster NGC~884. This study is a particular observational part of a global photometric
study of that cluster already introduced by \citet{saesenI} and can be seen as a
multi-colour photometric analogue of the recent spectroscopic study of B-stars in
the cluster by \citet{marsh2012}.

The simultaneous exploitation of oscillation frequencies of stars in the same
environment obviously has advantages compared to the study of single stars.
Imposing the same age and chemical composition will give much tighter
constraints when interpreting the observed oscillation spectra.  For this
reason, various observational initiatives were taken {in the past decade}
\citep[e.g.,][for recent examples of young open clusters with pulsating
B-stars]{handler_2008, majewska, michalska, handler_2011, jerzykiewicz}.  Based
on observational results in the literature, the study by \citet{balona_clusters}
led to a {derivation of a frequency -- age -- mass relation} for three young open clusters,
from averaging out detected but unidentified pulsation frequencies of cluster
$\beta\,$Cep stars and relying on evolutionary models with one fixed set of the
initial hydrogen fraction and metallicity of $(X,Z)=(0.70,0.02)$, ignoring core
overshooting (i.e., taking $\alpha_{\rm ov}=0.0$). Despite these restrictions,
the authors came to the conclusion that such age estimation is already more
precise than the one based on isochrone fitting and could be improved if the
detected pulsation modes could be identified in the future.

Putting more efforts into ground-based young open cluster campaigns is still
relevant, even with the enormous database now at hand gathered by the satellites
CoRoT and {\it Kepler\/}.  Indeed, {\it Kepler\/} did not reveal many early
B-stars with strong pulsations \citep{balona_kepler}. Amongst 48~variable {\it
  Kepler\/} B-type stars, fifteen are pulsating in the g-mode regime, of which
seven show weak frequencies in the p-mode range. These are not good candidates
for asteroseismic studies, since the observations in white light and the lack of
frequency splittings do not allow for mode identification. Moreover, their
pulsation amplitudes are too low for ground-based follow-up studies. CoRoT did
observe several pulsating B-stars, but dedicated long-term ground-based
photometry and time-resolved spectroscopy were necessary to perform stellar
modelling \citep[e.g.,][]{betacepcorot,briquetcorot,V1449aq}.

While asteroseismology of the few open clusters in the field-of-view of {\it
  Kepler\/} based on solar-like oscillations in red giants is progressing fast
\citep{Stello2011,hekker_clusters,miglio_clusters}, leading to seismic
constraints on mass loss on the red giant branch, it concerns clusters of {a few Gyr}
old. {Neither CoRoT nor {\it Kepler\/} are observing} young open clusters with
$\beta\,$Cep pulsators. Therefore, our ground-based photometric study of the
B-type stars in NGC~884 combined with their recent spectroscopic analogues
\citep{strom,huang,marsh2012} offers an interesting and different approach to
the advancement of understanding such young massive objects.

\citet{saesenI} presented a description of the global multi-site study of
NGC~884 in terms of differential time-resolved multi-colour CCD photometry of a
selected field of the cluster. This study also contained a global stellar
variability study among the 3165~cluster members through an automated frequency
analysis in the $V$ filter, leading to the identification of 36~multi-periodic
and 39~mono-periodic B-stars, 19 multi-periodic and 24~mono-periodic A- and
F-stars, and 20~multi-periodic and 20~mono-periodic variable stars of unknown
nature. Moreover, 15~irregular variable stars were found of which 6 are
  supergiant stars, 8 are Be-stars and 1 star is a B-star with a similar
  light-curve behaviour as the Be-stars. Also 6~new and 4~candidate eclipsing
binaries were detected, apart from 2~known cases.

The present paper exploits the full photometric multi-site campaign data set of
the periodic B-stars and couples them to published ground-based spectroscopy of
those stars for a global interpretation in terms of variability.  We present a
detailed frequency analysis of all discovered periodic B-stars using all
filters. Moreover, we attempt to derive the mode degree $\ell$ of all the
detected oscillation frequencies by means of the well-known method of amplitude
ratios \citep[e.g.,][Chapter~6]{bookaerts}, making use of the multi-colour
photometric time series. Finally, we select the $\beta$~Cep stars in NGC~884 and
use them to compare the observed properties of their oscillations with those
predicted for models considering their measured rotational velocities. This
allowed us to deduce a seismic age range for the cluster which is compatible
with the one deduced previously from an eclipsing binary in the cluster.


\section{DATA DESCRIPTION}

\label{sect:data}

For a detailed description of the extensive overall 12-site campaign, which
resulted in almost 77\,500 images and 92~hours of photo-electric data in $UBVI$,
spread over three observing seasons, we refer to \citet{saesenI}. That paper
also contains a comprehensive report of the calibration and reduction
process, which is thus omitted here, but we recall that the resulting light
curves of the brighter stars have a precision of 5.7~mmag in $V$, 6.9~mmag in
$B$, 5.0~mmag in $I$ and 5.3~mmag in $U$.  In the current paper, we focus
exclusively on the variable B-stars found in the cluster and we perform the
first detailed analyses of their light curves.

{First, we manually checked the automated merging of the light curves of all sites, which are a combination of Johnson ($B$, $V$, $I$), Bessell ($U$, $B$, $V$, $I$), Geneva ($U$, $B$, $V$) and Cousins ($I$) filter systems, and we applied a magnitude shift where needed.} Then, to avoid second-order extinction and seeing
effects, we calculated the residuals of a preliminary harmonic solution, for
which we did not include possibly spurious alias frequencies. For every
observing site separately, we then fitted a linear function of the airmass to
these residuals, subtracted it from these residuals and finally fitted a linear
function of the seeing, where we used the full-width half-max values of the
star's point spread function on the CCD as measure. Both linear trends were then
removed from the original data.  This {led} to better results than those from the
automated procedures adopted in \citet{saesenI} as illustrated by the case of
Oo~2444, for which the frequency close to 1~d$^{-1}$ does not occur any more
(compare Fig.~\ref{fig:freq} with Fig.~11 of Saesen et al.\ 2010).  Such manual
corrections were applied for all B-stars treated in this paper.

{For some selected stars (Oo~2246, Oo~2299, Oo~2444, Oo~2488, Oo~2566, Oo~2572 and Oo~2649), we also assembled photo-electric data.} Here, we merged
them with the other $UBVI$ data sets in \citet{saesenI}. Simultaneous
Str\"omgren $uvby$ and Geneva $UB_1BB_2V_1VG$ measurements of (suspected)
$\beta$~Cep stars were collected using the Danish photometer at Observatorio
Astron\'omico Nacional de San Pedro M\'artir (OAN-SPM) and the P7 photometer at
Observatorio del Roque de los Muchachos (ORM), respectively. The reduction
method for the photo-electric data taken at OAN-SPM is described in
\citet{redspm} {and references therein}.  For the reduction of the
photo-electric measurements of ORM with P7, we refer to \citet{redp7_1,
  redp7_2}. We combined the filters $U$ and $u$, $B$ and $v$, and $V$ and $y$,
since the effective wavelengths of these filters are similar. We compared the
photometric zero-points and adjusted them if necessary. In the end, we obtained
an ultra-violet, blue, visual and near-infrared light curve, which we refer to
as $U$, $B$, $V$ and $I$ hereafter.


\section{BASIC STELLAR PARAMETERS}
\label{sect:par}
We obtained the basic stellar parameters, the effective temperature
$T_\mathrm{eff}$ and the {luminosity $\log(L/L_{\sun})$}, from different
sources. First, we have absolute photometry of NGC~884 in the seven Geneva
filters at our disposal (see Sect.~7 of \citet{saesenI} 
for more information on the data
and their reduction). The calibrations of the Geneva system allow a physically
meaningful classification of early-type stars, by the determination of {the
  stellar parameter $T_\mathrm{eff}$}, based on the code {\sc calib}
\citep{kunzli} and the estimates of the average values of the six Geneva colours
$(U-B)$, $(B_1-B)$, $(B_2-B)$, $(V_1-B)$, $(V-B)$ and $(G-B)$.  {This
  calibration is based on the calculation of synthetic colours using Kurucz
  atmosphere models with solar metallicities.} The effective temperature values
and their interpolation errors in the grid of atmosphere models, as provided by
{\sc calib}, are presented in Table~\ref{table:hrcalib}.  We adopted 1000~K on
$T_\mathrm{eff}$ as realistic $1\sigma$-uncertainties including the systematic,
statistical and interpolation errors \citep{morel,Bmercator}, except if the
interpolation error was larger.

To determine the luminosity of the stars on the basis of our absolute
photometric measurements, we first calculated the {interstellar} reddening with the
formula
\begin{equation}
A_V = [3.3 + 0.28 (B-V)_{\mathrm{J},0}+0.04
E(B-V)_\mathrm{J}]\,E(B-V)_\mathrm{J}
\end{equation}
\citep{arenou}. For the reddening
$E(B-V)_\mathrm{J}$ in the Johnson system, we used the non-uniform reddening
values discussed in 
Sect.~7.2 of \citet{saesenI}. 
These reddening values are around $E(B-V)_\mathrm{J} = 0.53 \pm 0.05$,
which is well in agreement with the recent literature values listed in
\citet{southworth_hper} and the value deduced by \citet{currie}. The dereddened
$(B-V)_{\mathrm{J},0}$-colour was computed as 
\begin{equation}
(B-V)_{\mathrm{J},0} = 1.362
(B_2-V_1)_{\mathrm{G},0} + 0.197
\end{equation}
 \citep{meylan} with
$(B_2-V_1)_{\mathrm{G},0}$ the dereddened $(B_2-V_1)$-colour in the Geneva
system. With the interstellar reddening $A_V$ and the adopted distance modulus
$\mu = 5\log(d)-5$ of 11.71~$\pm$~0.1~mag (see \citet{saesenI}, Sect.~7.2), we
calculated the absolute visual magnitude $M_V = V - \mu - A_V$. The used
distance estimate falls well in the range of recent values listed by
\citet{southworth_hper} and the ones deduced by \citet{currie}. Finally, the
luminosity was then given by
\begin{equation}
\log(L/L_{\sun}) = -0.4 (M_\mathrm{bol} - M_{\mathrm{bol},\sun}),
\end{equation}
where we used the bolometric correction of \citet{flower} for $M_\mathrm{bol} =
M_V + \mathrm{BC}$ (see \citet{torres} for the correct coefficients) and
$M_{\mathrm{bol},\sun}=4.73$~mag \citep{torres}.

Table~\ref{table:hrcalib} contains the derived values for the luminosity
together with their errors, for which uncertainties on the absolute photometry,
the reddening, the effective temperature and the distance were taken into
account. We note that, by using the values for the reddening and the distance of
the cluster, we implicitly assumed that the stars are cluster members.

\begin{table*}
\footnotesize
\caption{The basic stellar parameters $T_\mathrm{eff}$ and
$\log(L/L_{\sun})$ determined for the {periodic} $B$-stars based on our mean
Geneva colours.\label{table:hrcalib}}
\centering\begin{tabular}{cr@{ $\pm$ }lr@{ $\pm$ }l | cr@{ $\pm$ }lr@{
$\pm$ }l
| cr@{ $\pm$ }lr@{ $\pm$ }l}
\tableline
Star ID &  \multicolumn{2}{c}{$T_\mathrm{eff}$} &
\multicolumn{2}{c|}{$\log(L/L_{\sun})$} & Star ID & 
\multicolumn{2}{c}{$T_\mathrm{eff}$} & \multicolumn{2}{c|}{$\log(L/L_{\sun})$} &
Star ID &  \multicolumn{2}{c}{$T_\mathrm{eff}$} &
\multicolumn{2}{c}{$\log(L/L_{\sun})$} \\
~          & \multicolumn{2}{c}{(K)}                         &
\multicolumn{2}{c|}{(dex)} & ~          & \multicolumn{2}{c}{(K)}               
         & \multicolumn{2}{c|}{(dex)} & ~          & \multicolumn{2}{c}{(K)}    
                    & \multicolumn{2}{c}{(dex)} \\
\tableline
\textit{Oo~1990}  &  \textit{13750}  &   \textit{160}  &  \textit{2.27}  & 
\textit{0.10} &
Oo~2267  &  14380  &    90  &  2.39  &  0.10 &
Oo~2426  &  12810  &    70  &  2.08  &  0.11 \\

Oo~2006  &  15470  &   120  &  2.67  &  0.09 &
Oo~2285  &  19670  &   220  &  2.96  &  0.08 &
Oo~2429  &  16560  &   130  &  2.66  &  0.09 \\

Oo~2037  &  14590  &   210  &  2.51  &  0.10 &
Oo~2299  &  30200  &   580  &  4.75  &  0.08 &
\textit{Oo~2444}  &  \textit{22850}  &  \textit{1180}  &  \textit{4.34}  & 
\textit{0.09} \\

Oo~2091  &  18080  &   170  &  3.30  &  0.09 &
Oo~2309  &  15000  &    90  &  2.60  &  0.10 &
Oo~2455  &  17450  &   160  &  2.85  &  0.09 \\

\textit{Oo~2094}  &  \textit{21220}  &   \textit{340}  &  \textit{3.33}  & 
\textit{0.08} &
Oo~2319  &  14440  &    80  &  2.44  &  0.10 &
Oo~2462  &  20360  &   200  &  3.42  &  0.08 \\

Oo~2110  &  15710  &   110  &  2.74  &  0.09 &
Oo~2323  &  11770  &    50  &  1.79  &  0.11 &
Oo~2482  &  11140  &    50  &  1.68  &  0.11 \\

Oo~2114  &  21810  &   210  &  3.67  &  0.08 &
Oo~2324  &  12270  &    50  &  1.96  &  0.11 &
Oo~2488  &  25450  &   660  &  4.22  &  0.08 \\

Oo~2116  &  13860  &    80  &  2.29  &  0.10 &
Oo~2345  &  15730  &   120  &  2.54  &  0.09 &
Oo~2507  &  17110  &   150  &  2.88  &  0.09 \\

Oo~2139  &  22830  &   230  &  3.59  &  0.08 &
Oo~2349  &  15580  &   120  &  2.63  &  0.09 &
Oo~2515  &  14710  &   100  &  2.31  &  0.10 \\

Oo~2185  &  17030  &   150  &  3.48  &  0.09 &
Oo~2350  &  14960  &   100  &  2.39  &  0.10 &
Oo~2524  &  12740  &    60  &  2.06  &  0.11 \\

Oo~2189  &  18290  &   180  &  3.04  &  0.09 &
\textit{Oo~2352}  &  \textit{15680}  &   \textit{130}  &  \textit{2.85}  & 
\textit{0.09} &
Oo~2531  &  11950  &    40  &  1.90  &  0.11 \\

Oo~2191  &  18970  &   170  &  3.33  &  0.09 &
Oo~2370  &  12280  &    60  &  1.92  &  0.11 &
Oo~2562  &  12810  &    70  &  1.94  &  0.11 \\

Oo~2228  &  11740  &    70  &  2.08  &  0.11 &
Oo~2371  &  24400  &   600  &  4.49  &  0.08 &
\textit{Oo~2566}  &  \textit{30060}  &   \textit{720}  &  \textit{4.25}  & 
\textit{0.08} \\

\textit{Oo~2235}  &  \textit{22570}  &   \textit{840}  &  \textit{4.37}  & 
\textit{0.08} &
Oo~2372  &  22140  &   240  &  3.47  &  0.08 &
Oo~2572  &  23180  &   550  &  4.14  &  0.08 \\

Oo~2242  &  19790  &   420  &  3.63  &  0.08 &
Oo~2377  &  23730  &   410  &  3.75  &  0.08 &
Oo~2579  &  18790  &   160  &  3.18  &  0.09 \\

Oo~2246  &  24440  &   270  &  4.22  &  0.08 &
Oo~2406  &  13020  &    60  &  2.07  &  0.11 &
Oo~2601  &  21600  &   220  &  3.85  &  0.08 \\

Oo~2253  &  15330  &   120  &  2.70  &  0.09 &
Oo~2410  &  12140  &    80  &  1.88  &  0.11 &
\multicolumn{5}{c}{~} \\

Oo~2262  &  24230  &   600  &  3.98  &  0.08 &
Oo~2414  &  10680  &    60  &  1.87  &  0.11 &
\multicolumn{5}{c}{~} \\
\tableline
\end{tabular}
\tablecomments{The first column denotes 
the star number, the second column
the effective temperature in Kelvin and the third column the luminosity in dex.
The noted uncertainties on the effective temperature are interpolation errors 
in the grid of atmosphere models as provided by {\sc calib}. The adopted and
more realistic errors are taken as $\Delta T_\mathrm{eff} = 1000$~K, except if
the
interpolation error would exceed this value. Values denoted in italics are found
by extrapolation outside the calibration tables and should be used with
caution.}
\end{table*}

For the B-stars that were not observed in absolute photometry, we only have
the differential photometry to derive their $T_\mathrm{eff}$ and
$\log(L/L_{\sun})$ estimates. For this purpose, we searched for the star
with most similar mean $V$, $B-V$ and $V-I$ measurements and for which we
obtained the basic stellar parameters, excluding binary stars. We then
made use of the fact that we are working in a cluster and adopted these
calibrations for the star of interest. Their effective temperature and
luminosity values, together with the star on which we based these
estimates, are noted in Table~\ref{table:hrsimil}.

\begin{table*}
\footnotesize
\caption{Same as Table~\ref{table:hrcalib}, but for the B-stars for which we did
not obtain absolute photometry.\label{table:hrsimil}} 
\centering\begin{tabular}{cr@{ $\pm$ }lr@{ $\pm$ }l | cr@{ $\pm$ }lr@{ $\pm$ }l}
\tableline
Star ID &  \multicolumn{2}{c}{$T_\mathrm{eff}$} &
\multicolumn{2}{c|}{$\log(L/L_{\sun})$} & Star ID & 
\multicolumn{2}{c}{$T_\mathrm{eff}$} & \multicolumn{2}{c}{$\log(L/L_{\sun})$} \\
~          & \multicolumn{2}{c}{(K)}                         &
\multicolumn{2}{c|}{(dex)} & ~          & \multicolumn{2}{c}{(K)}               
         & \multicolumn{2}{c}{(dex)}  \\
\tableline
Oo~1898 (Oo~2224)  &  14830  &   110  &  2.23 &  0.10 &
Oo~2520 (Oo~2601)  &  21600  &   220  &  3.84  &  0.08 \\

Oo~1973 (Oo~2126)  &   9740  &   120  &  1.44  &  0.11 &
\textit{Oo~2563 (Oo~2566)}  &  \textit{30060}  &   \textit{720}  &  \textit{4.21}  &  \textit{0.08} \\

Oo~1980 (Oo~2337)  &  12540  &    60  &  2.26  &  0.11 &
Oo~2611 (Oo~2507)  &  17110  &   150  &  2.86  &  0.09 \\

Oo~2019 (Oo~2189)  &  18290  &   180  &  3.04  &  0.09 &
Oo~2616 (Oo~2406)  &  13020  &    60  &  2.06  &  0.11 \\

Oo~2086 (Oo~2091)  &  18080  &   170  &  3.30  &  0.09 &
Oo~2622 (Oo~2053)  &  19640  &   250  &  3.26  &  0.08 \\

\textit{Oo~2089 (Oo~2200)}  &  \textit{16450}  &   \textit{160}  &  \textit{2.77}  &  \textit{0.09} &
Oo~2633 (Oo~2053)  &  19640  &   250  &  3.23  &  0.08 \\

Oo~2141 (Oo~2211)  &  14630  &    90  &  2.53  &  0.10 &
Oo~2649 (Oo~2262)  &  24230  &   600  &  3.96  &  0.08 \\

Oo~2146 (Oo~2442)  &  12920  &    60  &  2.10  &  0.11 &
\textit{Oo~2694 (Oo~2444)}  &  \textit{22850}  &  \textit{1180}  &  \textit{4.34}  &  \textit{0.09} \\

\textit{Oo~2151 (Oo~2412)}  &   \textit{9060}  &    \textit{90}  &  \textit{2.02}  &  \textit{0.10} &
\textit{Oo~2725 (Oo~1990)}  &  \textit{13750}  &   \textit{160}  &  \textit{2.27}  &  \textit{0.10} \\

Oo~2245 (Oo~2455)  &  17450  &   160  &  2.88  &  0.09 &
Oo~2752 (Oo~2110)  &  15710  &   110  &  2.75  &  0.09 \\

\textit{Oo~2342 (Oo~2037)}  &  \textit{14590}  &   \textit{210}  &  \textit{2.47}  &  \textit{0.10} &
Oo~2753 (Oo~2501)  &  12590  &    50  &  2.07  &  0.11 \\

Oo~2448 (Oo~2211)  &  14630  &    90  &  2.52  &  0.10 &
\multicolumn{5}{c}{~} \\
\tableline
\end{tabular}
\tablecomments{The stars with most similar values for $V$,
$B-V$ and $V-I$, on which the calibrations are based, are denoted in brackets
after the star identification.}
\end{table*}

In the literature, we also find absolute Geneva colours and $T_\mathrm{eff}$
values for some B-stars in NGC~884. \citet{waelkens_ngc884} observed the
brighter cluster stars with the Geneva photometer attached to the 76-cm
telescope at Jungfraujoch in Switzerland.  We used their absolute photometry,
their dereddened visual magnitudes, and our distance estimate to determine
$T_\mathrm{eff}$ and $\log(L/L_{\sun})$ values using the same calibrations as
described above. The results are noted in Table~\ref{table:hrlit}. The study of
\citet{slesnick} derived the effective temperature for many stars, for some
based on photometry, for others based on spectroscopy. However, because they do
not provide the uncertainties, we will not take these values into account. The
spectroscopic determinations of the effective temperature of \citet{huang} and
\citet{marsh2012} are also noted in Table~\ref{table:hrlit}, the latter study
being a continuation and extension of the former. In Table~\ref{table:hrlit} we
also list our photometric derivations with realistic error bars along with the
values from the literature, for comparison. In the last column, we give the
final estimates for $T_\mathrm{eff}$ and $\log(L/L_{\sun})$. These values are
calculated as the centre of an error box encompassing all the individual
$2\sigma$-error boxes. The noted uncertainty gives the borders of the large
overall error box. These final {and conservative} estimates will be used for the
mode identification performed in Sect.~\ref{sect:mode}.

It is noteworthy that the projected rotational velocities of cluster members
derived by \citet{strom}, \citet{huang}, and \citet{marsh2012}, which are shown
in our Fig.\,\ref{fig:amprot} below, are typically less than half of the
critical velocity. In that case, the above calibrations and those used in the
literature, which are all based on the assumption of non-deformed stars, are
justified \citep[see, e.g., Fig.\,1 of][which shows the value of the local
radius, gravity, effective temperature and luminosity as a function of
co-latitude for various values of $v \sin i/v_{\rm crit}$]{aertslamers}.

\begin{table*}
\footnotesize
\tabcolsep=4pt
\caption{The basic stellar parameters $T_\mathrm{eff}$ and
$\log(L/L_{\sun})$ for stars that have, besides our estimates, other 
parameter estimates from photometric
calibrations or spectroscopic literature values.\label{table:hrlit}}
\centering\begin{tabular}{c|r@{ $\pm$ }lr@{ $\pm$ }l | r@{ $\pm$ }lr@{ $\pm$ }l
| r@{ $\pm$ }l | r@{ $\pm$ }l | r@{ $\pm$ }lr@{ $\pm$ }l}
\tableline
~ & \multicolumn{4}{c|}{Our data}  &
\multicolumn{4}{c|}{Waelkens et al.}& \multicolumn{2}{c|}{Huang \& Gies}
& \multicolumn{2}{c|}{Marsh Boyer et al.} & \multicolumn{4}{c}{Final} \\
Star ID &  \multicolumn{2}{c}{$T_\mathrm{eff}$} &
\multicolumn{2}{c|}{$\log(L/L_{\sun})$} &\multicolumn{2}{c}{$T_\mathrm{eff}$} &
\multicolumn{2}{c|}{$\log(L/L_{\sun})$} & \multicolumn{2}{c|}{$T_\mathrm{eff}$}
& \multicolumn{2}{c|}{$T_\mathrm{eff}$} &\multicolumn{2}{c}{$T_\mathrm{eff}$} &
\multicolumn{2}{c}{$\log(L/L_{\sun})$} \\
~          & \multicolumn{2}{c}{(K)}                         &
\multicolumn{2}{c|}{(dex)}      & \multicolumn{2}{c}{(K)}                       
 & \multicolumn{2}{c|}{(dex)}      & \multicolumn{2}{c|}{(K)}                  
     & \multicolumn{2}{c|}{(K)}      & \multicolumn{2}{c}{(K)}                 
      & \multicolumn{2}{c}{(dex)}  \\
\tableline
Oo~2091 & 18080 & 1000 & 3.30 & 0.09 & \multicolumn{4}{c|}{\nodata} &
\multicolumn{2}{c|}{\nodata} & 18800 &  1790 & 18800 & 3600 & 3.30 & 0.18 \\

Oo~2094 & 21220 &1000 & 3.33 & 0.08 & \multicolumn{4}{c|}{\nodata} & 
\multicolumn{2}{c|}{\nodata} & 20350 & 50 & 21200 & 2000 & 3.33 & 0.16 \\

Oo~2185 & 17030 & 1000 & 3.48 & 0.09 & 16860 & 1000 & 3.45 & 0.07 & 16520 & 
180 & 17000 & 200 & 17000 & 2100 & 3.48 & 0.18 \\

Oo~2262 & 24230 & 1000 & 3.98 & 0.08 & 27950 & 1000 & 4.33 & 0.06 & 
\multicolumn{2}{c|}{\nodata} & 24900 & 1400 & 26100 & 3900 & 4.14 & 0.32 \\

Oo~2520 & 21600 & 1000 & 3.84 & 0.08 & 22940 & 1000 & 4.12 & 0.06 & 
\multicolumn{2}{c|}{\nodata} & 23200 & 350 & 22300 & 2700 & 3.96 & 0.28 \\

Oo~2563 & 30060 & 1000 & 4.21 & 0.08 & \multicolumn{4}{c|}{\nodata} & 
\multicolumn{2}{c|}{\nodata} & 26100 & 1790 & 27300 & 4800 & 4.21 & 0.16 \\

Oo~2622 & 19640 & 1000 & 3.26 & 0.08 & \multicolumn{4}{c|}{\nodata} & 19250 & 400 & 
17200 & 200 & 19200 & 2400 & 3.26 & 0.16 \\

Oo~2299	& 30200 & 1000 & 4.75 & 0.08 & 24790 & 1000 & 4.62 & 0.06 & 25930 & 420
& \multicolumn{2}{c|}{\nodata} & 27500 & 4700 & 4.71 & 0.21 \\

Oo~2371 & 24400 & 1000 & 4.49 & 0.08 & 26530 & 1000 & 4.68 & 0.06 &
\multicolumn{2}{c|}{\nodata} & \multicolumn{2}{c|}{\nodata} & 25500 & 3100 & 4.57 & 0.24 \\

Oo~2377 & 23730 & 1000 & 3.75 & 0.08 & 22160 & 1000 & 3.76 & 0.06 & 
\multicolumn{2}{c|}{\nodata} & \multicolumn{2}{c|}{\nodata} & 22900 & 2800 & 3.75 & 0.16 \\

Oo~2444	& 22850 & 1180 & 4.34 & 0.09 & \multicolumn{4}{c|}{\nodata} & 24900 & 360 &
\multicolumn{2}{c|}{\nodata} & 23100 & 2600 & 4.34 & 0.18 \\

Oo~2488	& 25450 & 1000 & 4.22 & 0.08 & 24920 & 1000 & 4.24 & 0.06 & 24480 &
260 & \multicolumn{2}{c|}{\nodata} & 25200 & 2300 & 4.22 & 0.16 \\

Oo~2520 & 21600 & 1000 & 3.84 & 0.08 & 22940 & 1000 & 4.12 & 0.06 &
\multicolumn{2}{c|}{\nodata} & 23160 & 350 & 22300 & 2700 & 3.96 & 0.28 \\

Oo~2563 & 30060 & 1000 & 4.21 & 0.08 & \multicolumn{4}{c|}{\nodata} & \multicolumn{2}{c|}{\nodata} & 25820 & 1130 & 27810 & 4250 & 4.21 & 0.16 \\

Oo~2572 & 23180 &  1000	 & 4.14  & 0.08 & 22830 & 1000 & 4.18 & 0.06 &
\multicolumn{2}{c|}{\nodata} & \multicolumn{2}{c|}{\nodata} & 23000 & 2200 & 4.14 & 0.16 \\

Oo~2601 & 21600 & 1000 & 3.85 & 0.08 & \multicolumn{4}{c|}{\nodata} & 23840 & 240 &
\multicolumn{2}{c|}{\nodata} & 22000 & 2400 & 3.85 & 0.16 \\

Oo~2622 & 19640 & 1000 & 3.26 & 0.08 & \multicolumn{4}{c|}{\nodata} & 19250 & 400 & 17230 & 200 & 19200 & 2400 & 3.26 & 0.16 \\
\tableline
\end{tabular}
 \tablecomments{The first column denotes the
star number, the following columns the effective temperature in Kelvin and the
luminosity in dex for our photometric calibrations, for the photometric
calibrations of \citet{waelkens_ngc884}, for the
spectroscopy of \citet{huang}
and \citet{marsh2012}, and the final adopted values, respectively. For all
estimates, the uncertainties noted are $1\sigma$-error bars, for the final
estimates, these are the overall $2\sigma$-error boxes.}
\end{table*}

\section{FREQUENCY ANALYSIS}
\label{sect:freq}

\begin{figure*}
\centering
\includegraphics[width=\columnwidth]{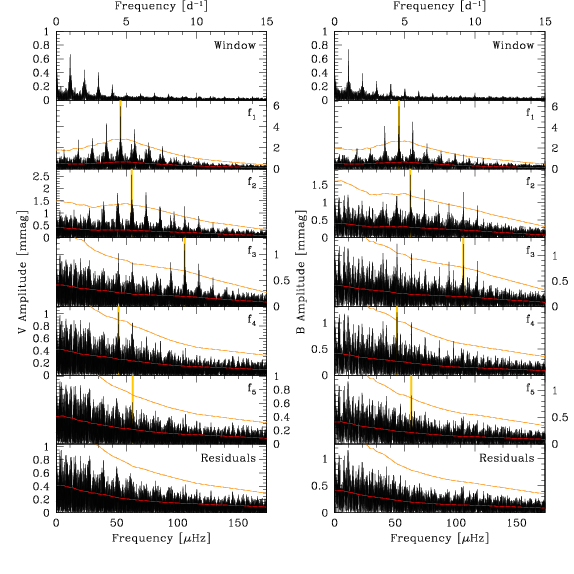}
\caption{Weighted frequency analysis of Oo~2444. We show the spectral window
(top) and amplitude spectra in the different steps of subsequent prewhitening in
the $V$ (left) and $B$ (right) filter. The accepted frequencies are marked by a
yellow band, the red line corresponds to the noise level and the orange line to
the 4~S/N-level.\label{fig:freq}}
\end{figure*}

We performed the frequency analysis with {\sc period04} \citep{period04}. This
code was developed to extract individual frequencies of self-driven oscillation
modes in the approximation of an infinite mode lifetime, by means of a discrete
Fourier transform algorithm. Moreover, it allows to apply simultaneous
multi-frequency sine-wave fitting in the least-squares sense and iterative
prewhitening.  The advantage is that the code is especially dedicated to the
analysis of large astronomical time series containing gaps and that it is able
to calculate optimal light-curve fits for multi-periodic signals with the
inclusion of harmonic and combination frequencies.

We searched for frequencies in the $V$-, $B$-, $I$- and $U$-filter data. For the
$V$-data we used a weighted and non-weighted analysis, for the other filters
only a weighted frequency analysis was carried out. For the point weights we
used the inverse of the square of the observational uncertainties on each data
point, taken from \cite{saesenI}. For most stars we restricted the computation
of the amplitude spectra to the frequency interval [0-15]~d$^{-1}$, since the
typical B-star pulsations range from 0.3 to 1~d$^{-1}$ (for slowly pulsating B
stars, SPBs hereafter) and 3 to 12~d$^{-1}$ (for $\beta$~Cep stars) and since it
was found in \citet{saesenI} that no power occurs beyond 15~d$^{-1}$.  We worked
in frequency steps of 0.000\,05~d$^{-1}$. The choice for this step relied on the
time base and noise properties of the data, but its precise value is not of
importance since the frequency values are optimised in the frequency analysis
procedure.

We considered a frequency peak significant when its amplitude exceeds four times
the noise level \citep{breger_sn}. We computed this noise level as a running
average of the amplitude spectrum.  As already {argued} in \citet{saesenI}, we
considered different intervals over which we evaluated the noise level,
according to the frequency value: we used an interval of 1\,d$^{-1}$ for $f \in
[0-3]$\,d$^{-1}$, of 1.9\,d$^{-1}$ for $f \in [3-6]$\,d$^{-1}$, of 3.9\,d$^{-1}$
for $f \in [6-11]$\,d$^{-1}$ and of 5\,d$^{-1}$ for $f \in [11-15]$\,d$^{-1}$.
We allow lower significance levels for harmonic and combination frequencies, or
if the frequency value occurs in different filters.  We then relax the criterion
to a signal-to-noise (S/N) level of 3.6 instead of 4.0, as is justified for
B-stars \citep{nueri_phot, 12lac1, handler2012}.  With these different intervals
for the noise computation, we are stricter than taking the conventional approach
of computing the noise over a 5\,d$^{-1}$-range \citep[e.g.,][]{handler2012},
while still being computationally efficient.

We describe here the different steps of our frequency analysis by means of one
example star, Oo~2444, which was suspected to be variable with a period of
0.17~days by \citet{krzesinski1} and confirmed as $\beta$~Cep star by
\citet{ngc6910_2}, but no frequencies were reported so far. We started our
analysis by calculating the spectral window and the amplitude spectrum of the
data (see first two panels of Fig.~\ref{fig:freq} for the weighted analysis of
the $V$ and $B$ data). In the amplitude spectrum, one clear peak $f_1$
dominates. The peak structure, which is not explained by the spectral window, is
a signature of additional frequencies.  This first frequency also appears as a
significant peak in all other filters.

We prewhitened the data by subtracting a sine-wave from the original data. The
frequency, amplitude and phase of this sine-wave were optimised in the $V$
filter, since the time span of the light curves is largest in this filter and
yields most data points. For the $B$, $I$ and $U$ light curve, we fixed the
frequency value and optimised only the amplitude and phase of the variation. 

We then proceeded by searching for frequencies in the residual light curve by
again calculating its amplitude periodogram (see third panel of
Fig.~\ref{fig:freq}). We detect a second signal at frequency $f_2$ in $V$, $B$
and $I$, but not in $U$. This is not surprising given the limited number of
measurements in the $U$ filter. We prewhitened the original data, again by
optimising the frequency values of $f_1$ and $f_2$ simultaneously in the $V$
filter and imposing these frequencies in the multi-frequency fit of $B$, $I$ and
$U$, leaving only the amplitude and phase as free parameters.

The subsequent frequency in the residual light curve $f_3$ is equal to 2$f_1$
within its uncertainty (see fourth panel of Fig.~\ref{fig:freq}). This signal
was again significantly present in both the $V$, $I$ and $B$ data set. We
thus removed a three-frequency fit from the data, using the same
optimisation method as before and fixing the third frequency to the exact
harmonic of the first frequency. Hereafter we retain $f_4$ (see fifth panel of
Fig.~\ref{fig:freq}), which is significant in $V$ and $B$, but not in $I$ and
$U$.

In the residuals of the four-frequency fit, no significant frequencies are left,
except in the $B$-filter, where we retrieve frequency $f_5$ (see {sixth} panel
of Fig.~\ref{fig:freq}). We verified that including $f_5$ in the frequency
solution for $V$ and $I$ did not change the fitted amplitude for the frequencies
$f_1$ to $f_4$. Optimising all frequencies in the $B$ filter did not converge to
the same frequency solution as in the $V$ filter within the uncertainty,
therefore we optimise $f_1$ to $f_4$ in $V$ and $f_5$ in $B$. The residuals
after prewhitening with the five-frequency fit do not show any significant
frequencies anymore (see {seventh} panel of Fig.~\ref{fig:freq}).

An overview of the properties of the final five-frequency fit is given in
Table~\ref{table:indiv}. The uncertainties on the frequency, amplitude and phase
values are computed from the error matrix of the least-squares harmonic fitting,
as provided by {\sc period04}, where we {decoupled the} frequencies and phases. 

Another error estimate for the frequency $f_i$ is given by $\Delta f_i =
\sqrt{6} \sigma / \sqrt{N} \pi T A_i$, where $\sigma$ is the standard deviation
of the final residuals, $N$ is the number of data points, $T$ is the total time
span and $A_i$ is the amplitude of frequency $f_i$ \citep{montgomery}. This
analytical formula gives an overestimation of the accuracy. A more realistic
value is obtained when taking the correlations of the data into account, by
multiplying the uncertainties by $\sqrt{\mathrm{D}}$, where D is the number of
consecutive data points which are correlated \citep{schwarzenberg}. For our data
set of Oo~2444, this gives errors of {0.6, 1, 3 and 5 x$10^{-5}$~d$^{-1}$} for
$f_1$, $f_2$, $f_4$ and $f_5$, respectively, so that these error
estimates deviate less than a factor 6 from the formal least-squares
uncertainties listed in Table~\ref{table:indiv}.

\tabcolsep=4pt
\begin{table*}
\footnotesize
\caption{Results of the multi-frequency fit to the $U$, $B$, $V$ and $I$ light
curves of all B-stars.\label{table:indiv}} 
\centering\begin{tabular}{l || lll | lll | lll| lll | c}
\tableline
~ & \multicolumn{3}{c|}{$U$} & \multicolumn{3}{c|}{$B$} & \multicolumn{3}{c|}{$V$} & \multicolumn{3}{c|}{$I$} & \\
$f_i$ & $A_i$ & $\phi_i$ & S/N & $A_i$ & $\phi_i$ & S/N & $A_i$ & $\phi_i$ & S/N & $A_i$ & $\phi_i$ & S/N & $\ell$ \\
(d$^{-1}$) & (mmag) & (rad) & ~ & (mmag) & (rad) & ~ & (mmag) & (rad) & ~ & (mmag) & (rad) & ~ & \\
\tableline
\multicolumn{14}{l}{}\\

\multicolumn{14}{l}{\textbf{Oo~2094}}\\
$f_1 = 0.37596(4)$ & 7(1) & 0.49(3) & 3.7 & 4.9(5) & 0.46(2) & 5.0 & 4.5(2) &
0.439(9) & 4.1 & 4.3(5) & 0.46(2) & 4.8 &1,2,{4}\\
$f_2 = 2f_1$       & 2(1) & 0.4(1)  & \nodata    & 2.2(5) & 0.47(4) & 3.9 & 1.3(2) &
0.24(3)  & 3.2 & 1.6(5) & 0.37(5) &   \nodata  &\nodata \\
\multicolumn{14}{l}{}\\

\multicolumn{14}{l}{\textbf{Oo~2246}}\\
$f_1 = 5.429245(7)$ & 6(1) & 0.17(3) & 5.5 & 6.0(3) & 0.169(9) & 8.6 & 5.7(1) &
0.168(4) & 8.9 & 5.1(2) & 0.166(7) & 7.3 & 2\\
$f_2 = 5.85616(1)$   & 3(1) & 0.01(8) & 3.6 & 3.3(3) & 0.07(2)   & 8.7 & 3.2(1)
& 0.078(7) & 8.9 & 3.4(2) & 0.07(1)   & 8.9 & 4\\
$f_3 = 5.02348(4)$   & 2(1) & 0.5(1)   & 3.2 & 1.0(3) & 0.51(5)   & 4.2 & 0.6(1)
& 0.51(4)   & 3.7 & 0.7(2) & 0.53(5)   & 3.4 & 0\\
\multicolumn{14}{l}{}\\

\multicolumn{14}{l}{\textbf{Oo~2444}}\\
$f_1 = 4.58161(2)$  & 7(2) & 0.63(5) & 7.2 & 6.1(3) & 0.622(7) & 9.5 & 5.9(1) &
0.624(4) & 8.9 & 4.8(2) & 0.624(8) & 7.8 & 1\\
$f_2 = 5.39333(5)$  & 2(2) & 0.3(2)   &   \nodata   & 1.9(3) & 0.30(2)   & 6.0 & 2.3(1)
& 0.297(9) & 7.9 & 1.8(2) & 0.29(2)   & 5.5 & /\\
$f_3 = 2f_1$            & 1(2) & 0.5(2)    & 3.3 & 1.2(3) & 0.59(4)  & 6.4 & 1.2(1) & 0.57(2)   & 7.1 & 0.9(2) & 0.57(4)   & 5.0 & \nodata\\
$f_4 = 4.4494(1)$    & 1(2) & 0.2(4)   &    \nodata  & 1.2(3) & 0.13(4)   & 4.7 &
1.0(1) & 0.12(2)   & 4.6 & 0.8(2) & 0.15(4)  & 3.3 &
0,\textbf{1},\textbf{2},3,4\\
$f_5 = 5.4643(3)$    & 1(2) & 0.4(3)   &  \nodata    & 1.0(3) & 0.44(5)   & 4.2 &
0.6(1) & 0.49(3)   & 3.4 & 0.8(2) & 0.44(5)  & 3.7 & /\\
\multicolumn{14}{l}{}\\
\tableline
\end{tabular}
\tablecomments{The signal is written as $C+\sum_{i=1}^N A_i \sin(2\pi f_i
(t-t_0) +2 \pi \phi_i)$ with $t_0$=HJD 2453000. $A_i$ stands for amplitude and
is expressed in millimag, $2 \pi \phi_i$ is the phase expressed in radians.
Uncertainties in units of the last given digit on the amplitudes and phases are
denoted in brackets. We also indicate the signal-to-noise {level} of the peak,
if it is larger than 3. In the last column we note the mode identification possibilities of the degree $\ell$, based
on elimination of degrees from 0 to 4. When specific degrees were preferred on
the basis of their $\chi^2$-values, we put its value in bold. When none of
the degree possibilities remained, we note this by '/'. A {blank} line means we
had insufficient information for mode identification.
(This table is available in its entirety in a machine-readable form in the online journal. A portion is shown here for guidance regarding its form and
content.)}
\end{table*}

Section~\ref{sect:indiv} contains the results for all the other variable
B-stars, where we followed the same approach.


\section{MODE IDENTIFICATION}
\label{sect:mode}

Successful seismic modelling involves not only detecting pulsation frequencies,
but also relies on the identification of the spherical wavenumbers $(\ell,m)$ of
the modes. Unfortunately, unambiguous identifications of detected frequencies
are rather scarce for heat-driven pulsators, given that we are not in the
asymptotic regime of the frequencies.  Our collected photometry of NGC~884 in
different colours in principle allows deducing the degree $\ell$ of the various
modes of the pulsators.  For this purpose we used the well-known method of
photometric amplitude ratios, which compares the observational values with the
theoretical ones. This technique has already been successfully applied for
B-type stars \citep[see, e.g.,][]{heynderickx, aerts_hip, handler_3betacep,
  nueri_mode, Bmercator}. As can be seen in Table~\ref{table:indiv}, {\it no
  phase differences within the observational errors} between the different
filters are detected, so we restrict to the amplitude ratios to determine the
degree $\ell$, following the method of \citet{amplrat}.

\subsection{Grid of Stellar Models and Pulsation Computations}

We made use of an extensive grid of equilibrium models calculated by the
evolutionary code \textsc{cl\'es} \citep[Code Li\'egeois d'\'Evolution
Stellaire,][]{cles}.  The models were computed using OP opacity tables
\citep{OPopac} assuming the \citet{asplund} {solar abundance} mixture. We used
{grey} atmosphere models. The reader is referred to \citet{briquet_hd46202} for
more details on the adopted input physics.  A given model is characterised by
five parameters: its mass $M$, its initial hydrogen abundance $X$, its initial
metallicity $Z$, its core convective overshoot parameter $\alpha_\mathrm{ov}$
and its age.  The grid consists of main-sequence models with masses from 2 to
20~$M_{\sun}$, with a step of 0.1~$M_{\sun}$, four values for the initial
hydrogen abundance $X=0.68, 0.70, 0.72, 0.74$, five values for the initial
metallicity $Z=0.010, 0.012, 0.014, 0.016, 0.018$ and an overshoot parameter
$\alpha_\mathrm{ov}$ between 0.00 and 0.50 in steps of 0.05. In total, this grid
contains 36\,000 evolutionary tracks from the start to the end of the central
hydrogen burning phase (ZAMS to TAMS).  It consists of over 3
million models in the core-hydrogen burning phase, the hydrogen-shell burning
phase being too short to consider in comparison with the main-sequence phase
(also known as the Hertzsprung gap).

For each of these models, the theoretical frequency spectrum of low-order
low-degree axisymmetric ($m=0$) modes was computed using the standard adiabatic
code \textsc{osc} \citep[Li\`ege Oscillation Code,][]{osc}.  The age step along
one evolutionary track is such that the median of the frequency differences of
low-order low-degree p- and g-modes for consecutive stellar models (which will
be used further in Sect.\,\ref{sect:compatib}) amounts to 0.07\,d$^{-1}$ and
0.03\,d$^{-1}$, for models of 7\,M$_\odot$ and of 20\,M$_\odot$ respectively,
and reaches a value in between those for all the masses in $[7,20]$\,M$_\odot$.
This grid was already successfully used for the seismic modelling of various
isolated massive pulsators \citep[e.g.,][for recent
applications]{12lac2,briquet_hd46202,V1449aq}.

\begin{figure}
\centering
\includegraphics[width=0.7\columnwidth]{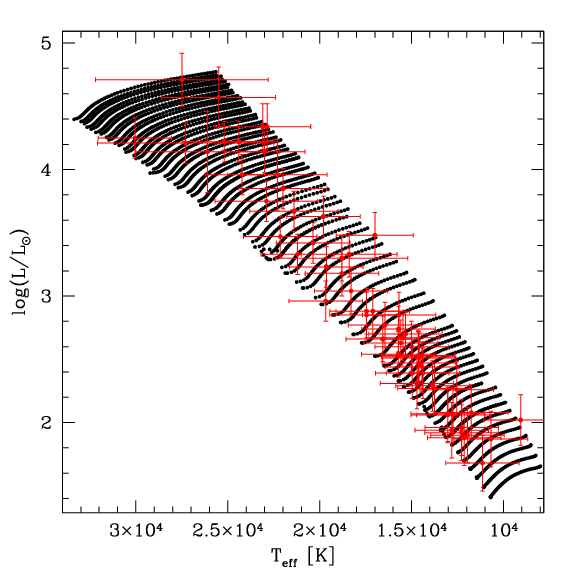}
\caption{{HR-diagram} with the model grid for
mode identification, as described in the text, denoted in black. The position
of all {treated} B-stars together with their observational $2\sigma$-error
boxes are also shown in red.\label{fig:grid}}
\end{figure}

The frequency values for B-stars in the adiabatic approximation are close to
their non-adiabatic counterparts, the difference being less than typically
0.001\,d$^{-1}$ \citep[e.g., Section 4 of\ ][]{thetaop3}. This reflects the fact
that the frequencies of g-modes and of low- to moderate-order p-modes in B-stars
are determined mostly by the internal layers, where the adiabatic approximation
is excellent \citep[see, e.g.,][for a detailed explanation]{dupretPhD}.
However, for the purpose of mode identification, we need to compute the
non-adiabatic eigenfunctions to predict the theoretical amplitude ratios. This
is much more CPU intensive than adiabatic computations and not feasible for the
grid of models discussed above. Also, an appropriate sampling of models covering
the error boxes in the HR-diagram is largely sufficient, as was demonstrated in
previous modern applications of mode identification of B-stars
\citep[e.g.,][]{decatSPB, Bmercator,thetaop1,12lac1}.  Following these previous
studies, we considered a sub-grid of the one discussed above for the mode
identification, with masses of 2.1, 2.2, 2.3, \ldots, 3.9, 4.0, 4.2, 4.5, 4.7,
5.0, ..., 15.7 and 16.0~$M_{\sun}$, restricting to $Z=0.010$, $X=0.70$ and
$\alpha_{\mathrm{ov}}=0.2$.  For all the models in this sub-grid, which is shown
in Fig.~\ref{fig:grid}, we computed the theoretical eigenfunctions and
-frequencies with the non-adiabatic pulsation code \textsc{mad}
\citep{mad}. Herein, a detailed treatment of the non-adiabatic behaviour of the
pulsations in the atmosphere is included. We considered all low-order p- and
g-modes with degrees $\ell$ between 0 and 4 as the chance to observe higher
degree modes with photometry is very small due to geometrical cancellation
effects. We computed the amplitude ratios associated to the theoretical
frequencies with respect to the $V$ filter (see \citealt{amplrat} for more
information) for grey atmosphere models and confronted these with the observed
ones in Sect.\,\ref{application}.  Before we discuss the concrete applications
of mode identification, various remarks are appropriate.

\begin{figure*}\centering
\includegraphics[width=0.7\columnwidth]{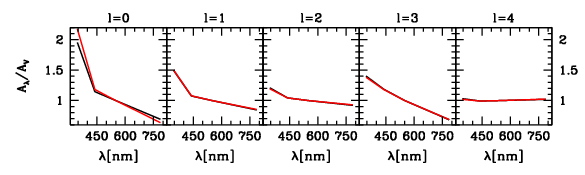}
\caption{{Comparison between the amplitude ratios for two extreme
$\beta$~Cep models at the same position in the HR-diagram. The black lines
denote the ratios for a model with initial metallicity $Z=0.01$ and
overshoot parameter $\alpha_\mathrm{ov}=0.5$, the red lines are for
$Z=0.02$ and $\alpha_\mathrm{ov}=0.0$. For both models, one theoretical
eigenfrequency was selected in the $\beta$~Cep interval with very similar
values. {In the plots for $\ell$=1, 2, 3 and 4 the red and black lines are almost indistinguishable.}}\label{fig:compMI}}
\end{figure*}

Fixing a value for $Z$ and $\alpha_\mathrm{ov}$ is commonly adopted for mode
identification of B-stars \citep[e.g.,][for a recent application]{handler2012}.
We illustrate the typical theoretical uncertainty on the amplitude ratios for
our grid of models in Fig.~\ref{fig:compMI}.  To construct this figure, we
selected two typical models for a $\beta$~Cep star with extreme values that have
the same position in the HR-diagram. On the one hand we took a model with
$Z=0.01$ and $\alpha_\mathrm{ov}=0.5$, on the other hand we selected a model
with $Z=0.02$ and $\alpha_\mathrm{ov}=0.0$. For both models and for each degree
$\ell$, we selected a frequency with very similar values in each model which
fell in the typical range for $\beta$~Cep variations and calculated its
theoretical amplitude ratios shown in Fig.~\ref{fig:compMI}. We see that there
is only little difference between the predictions for both models.  We tested
more cases, varying either only the $Z$, or only $\alpha_{\rm ov}$, or both
together as in Fig.~\ref{fig:compMI}. The general conclusion is that the
difference in amplitude ratios is usually negligible compared to the
observational errors on the observed amplitude ratios we are coping with.  This
is in agreement with the theoretical uncertainties found by \citet[][their
Fig.\,4]{thetaop1} based on independent evolution and pulsation codes.  Thus,
for the data we are dealing with here, the mode identification does not change
when changing $(Z,\alpha_{\rm ov})$ because the change in theoretical
predictions are smaller than the observed errors.

A second remark concerns the mode instability.  In our comparison between
observed and theoretical amplitude ratios from the sub-grid, we considered all
the computed eigenfrequencies, i.e., as in similar recent applications of mode
identification, we did not restrict to modes predicted to be excited
\citep[e.g.,][]{Bmercator,handler2012}.  Our argument is that recent studies
have shown there to be problems with the details of excitation computations for
various modes in main-sequence OB-pulsators, where modes predicted to be stable
were detected in the data
\citep[e.g.,][]{nu_eri3,gammapeg,briquet_hd46202,V1449aq}.  An extreme case
occurred for the O9V pulsator HD\,46202, for which none of the detected modes is
predicted to be excited. Relying on the instability predictions for mode
identifications is thus inappropriate.

Finally, we point out that the pulsational computations done with \textsc{osc}
and \textsc{mad} ignore rotational effects and only provide axisymmetric modes,
just as in \citet{balona_clusters} who considered arbitrary equatorial rotation
velocities from zero to 200\,km\,s$^{-1}$ to interpret B-type pulsators in young
open clusters.  Meanwhile, however, the theoretical study by \citet{jagoda} took
into account {the effects of rotation} on photometric mode identification from amplitude
ratios, while considering a typical $\beta\,$Cep star with an equatorial rotation
velocity of 100\,km\,s$^{-1}$. They pointed out that the effects on the
amplitude ratios and on the phase differences can be significant, depending on
the mode degree and its coupling with modes of other degrees.  Unfortunately,
this theoretical study is of no practical use in mode identification when the
equatorial rotation velocity and stellar inclination angle are not known, which
is the case for the pulsating B-stars in NGC~884. On the other hand, we note
that \citet{jagoda} found large rotational effects on the amplitude ratios to be
accompanied by large phase differences of the light curves in the different
wavebands.  We did not find any significant phase differences for the modes
detected in our data, which points out that the effects of rotation, if any, are
limited as they would otherwise result in measurable phase differences for the
amplitudes in the different filters.  Thus, as \citet{balona_clusters}, we
perform the mode identification under the assumption that the rotation can be
ignored in the computation of the theoretical amplitude ratios, realising that
this may not be optimal for all the modes in all the considered pulsators.

\subsection{Application to the B-stars in NGC~884}
\label{application}

For each pulsator, models in the sub-grid in Fig.~\ref{fig:grid} that fitted the
observed position in the {HR-diagram} within a $2\sigma$-error box, as derived
in Sect.~\ref{sect:par} and noted in Table~\ref{table:hrcalib},
Table~\ref{table:hrsimil} and the last column of Table~\ref{table:hrlit}, were
retained.  For each model and each degree $\ell$, we selected the
eigenfrequencies $f_{\mathrm{T},i}$ that match the observed frequency
$f_i~\pm~0.2$~d$^{-1}$. We take an uncertainty of 0.2~d$^{-1}$ into account
since the theoretical eigenfrequencies assume axisymmetric modes. We cannot
compute a more exact value for the rotational splitting due to unknown
inclination and rotation rate of the star, in combination with the unknown
azimuthal number $m$. Our approach is stricter than in the previous applications
mentioned above \citep{decatSPB, Bmercator,12lac1,thetaop1} where the closest
frequency in their used model grid, which was in addition much less dense than
ours, was selected, irrespective of the difference with the theoretical
frequency value. {We have therefore followed a particularly conservative approach in assigning mode identifications to our detected frequencies.}

\begin{figure*}\centering
\includegraphics[width=0.7\columnwidth]{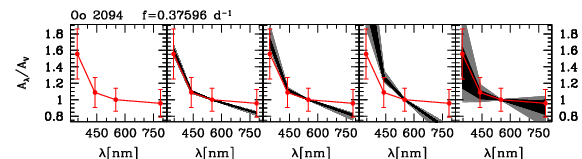}
\includegraphics[width=0.2\columnwidth]{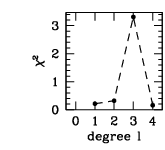}\\
\includegraphics[width=0.7\columnwidth]{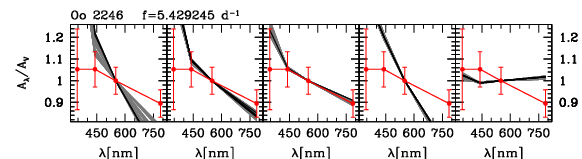}
\includegraphics[width=0.2\columnwidth]{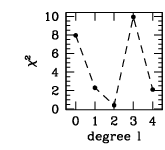}\\
\includegraphics[width=0.7\columnwidth]{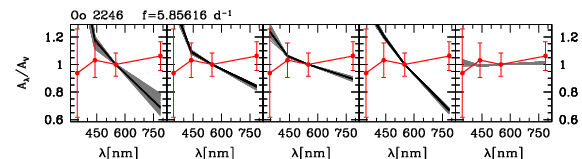}
\includegraphics[width=0.2\columnwidth]{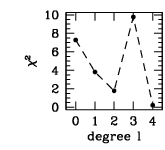}\\
\includegraphics[width=0.7\columnwidth]{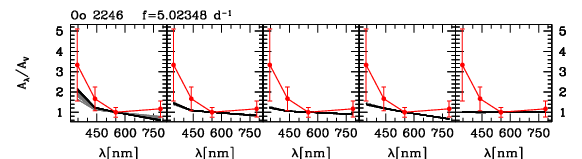}
\includegraphics[width=0.2\columnwidth]{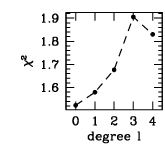}\\
\caption{{(Left) Amplitude ratios scaled to the $V$ filter for the frequency and the star denoted above the figure. The red filled circles with error bars
denote the observed amplitude ratios and their uncertainties, the grey bands indicate the theoretical predictions for these, running from $\ell=0$ at the left to $\ell=4$ at the right. The black bands denote a subsample of these theoretical models that deviate only 500~K in effective temperature and 0.05~dex in luminosity from the observed position in
the HR-diagram. (Right) Corresponding $\chi^2$-value as a function of the degree $\ell$. (The mode identification plots for all stars and frequencies are available in the electronic edition of the journal.)}
\label{fig:modeidall}}
\end{figure*}

We confronted the theoretical amplitude ratios retained in this way with those
observed.  We illustrate the results of the mode identification by two
examples. The outcome for all B-stars treated in this section can be found in
Table~\ref{table:indiv} and Fig.~\ref{fig:modeidall}. Our results rely on the selected sub-grid combined
with the elimination of certain degrees $\ell$ for the $2\sigma$-error box in
the HR-diagram of each star {as described in the previous paragraph}.

The results for Oo~2094 are based
on the observed amplitude values presented in Table~\ref{table:indiv}. The mode
identification plots with the comparison between the theoretical and
observational amplitude ratios are shown in {the top left panel of}
Fig.~\ref{fig:modeidall}. First of all, we eliminate the possibility of degree
$\ell=0$. None of the models in the grid that fall in the 2$\sigma$-error box of
Oo~2094 in the {HR-diagram} have $(\ell=0)$-modes with frequencies close enough
($\pm~0.2$~d$^{-1}$) to $f_1=0.37596$~d$^{-1}$. This is not at all surprising
since the frequency value of $f_1$ already indicated that we are probably
dealing with a g-mode. We can also eliminate
$\ell=3$, because the theoretical amplitude ratios for the $I$ filter never
match the observed one within the uncertainties. We retain $\ell=1, 2$ and 4,
but remark that the observational amplitude ratios follow best the theoretical
band for $\ell=4$. 

The mode identification for the three frequencies detected in Oo~2246, $f_1 =
5.429245$~d$^{-1}$, $f_2 = 5.85616$~d$^{-1}$ and $f_3 = 5.02348$~d$^{-1}$, is shown in the lower three left panels of Fig.~\ref{fig:modeidall}. The observed
amplitude ratios are only compatible with the theoretical models with $\ell=2$
for $f_1$, $\ell=4$ for $f_2$ and $\ell=0$ for $f_3$, although the dependence of
the observed ratios on wavelength does not always resemble the theoretical
one. We also note that, even though we are just on the limit of eliminating the
possibility of $\ell=1$ for $f_3$, there is not much reason to prefer $\ell=0$
to $\ell=1$.

We also identified the mode degree from an alternative approach, as in, e.g., 
\citet{nueri_mode,12lac1}, by calculating a $\chi^2$-value for every
frequency and every degree $\ell$, as
 \begin{equation}  
\chi^2(\ell) = \frac{1}{\sqrt{N-1}} \sum_{j=1}^N
\left(\frac{A_{j,\mathrm{th}}/A_{V,\mathrm{th}} -
A_{j,\mathrm{obs}}/A_{V,\mathrm{obs}}}{\sigma_{j,\mathrm{obs}}}\right)^2,
\end{equation}
with $N$ the total number of filters available for the star and
$\sigma_{j,\mathrm{obs}}$ the error on the observed amplitude ratio for
filter $j$ \citep[see also][]{aerts_hip,bookaerts}. The same models were
selected as described above and the $\chi^2$-value for the best fitting
model was retained for every degree, i.e., the mode with minimal $\chi^2$ is
preferred. The outcome for the stars described above is shown in the right
panels of Fig.~\ref{fig:modeidall} and the result is in
agreement with the one of the left panels of these figures. The
$\chi^2$-value indeed expresses the difference of the observed mean
amplitude ratio to the one predicted from the closest model. There is no
statistically founded acceptance criterion for such $\chi^2$-values.
However, we note that in the examples above, the elimination process relying
on the left panels often retains only models with $\chi^2<1$. We only use
this $\chi^2$-value to prefer specific degrees $\ell$ when a clear minimum
is visible, as noted in Table~\ref{table:indiv} by putting these
$\ell$-values in bold. We remark that we always included the amplitudes in
the $U$ filter, when available, because, although their error is often
larger than for the other filters, the calculated $\chi^2$-value accounts
for this.

The results for all the mode identification attempts are represented in
Fig.~\ref{fig:modeidall}. {For one star, Oo~2371, which has one frequency, we omitted the mode identification since it is an ellipsoidal binary. We securely identified the degree for 12 of the 114 detected frequencies: nine of the 64
investigated stars have at least one mode degree identified.The stars and frequencies for which we only retain one possibility for the degree $\ell$ of the modes, are the following: $f_1$\,=\,0.32811\,~d$^{-1}$ in Oo~2191 has $\ell=4$, $f_2$\,=\,3.12734\,~d$^{-1}$ in Oo~2242 has $\ell=2$, $f_1$\,=\,5.429245\,~d$^{-1}$, $f_2$\,=\,5.85616\,~d$^{-1}$ and $f_3$\,=\,5.02348\,~d$^{-1}$ in Oo~2246 have $\ell=2$, 4 and 0, respectively, $f_1$\,=\,3.07229\,~d$^{-1}$ in Oo~2262 has $\ell=4$, $f_1$\,=\,3.145150\,~d$^{-1}$ in Oo~2299 has $\ell=2$, $f_1$\,=\,0.76570\,~d$^{-1}$ in Oo~2372 has $\ell=4$, $f_1$\,=\,4.58161\,~d$^{-1}$ in Oo~2444 has $\ell=1$, $f_1$\,=\,6.16805\,~d$^{-1}$ in Oo~2488 has $\ell=2$, and $f_1$\,=\,4.41463\,~d$^{-1}$ and $f_2$\,=\,4.76330\,~d$^{-1}$ in Oo~2572 have $\ell=2$ and 1, respectively.} The identified modes
will turn out to be powerful input for the derivation of cluster properties
as we show below.  Nevertheless, the number of unambiguous mode
identifications is low, in view of the immense observational effort {put into
the observing campaign}. The lack of good photometric capabilities based on
ultraviolet filters led to too uncertain amplitudes at blue
wavelengths. This propagated to too high uncertainties for the amplitude
ratios {to be of sufficiently strong diagnostic value} to identify the majority of the
detected modes.  Future campaigns should take this into account.


\section{RESULTS FOR INDIVIDUAL STARS}
\label{sect:indiv}

In this section, we describe the results of the weighted analysis of the $V$,
$B$, $I$ and $U$ light curves for individual stars and the mode identification of the detected frequencies. For figures of the light curves, the phase diagram with the dominant frequency and the periodograms of the treated B-stars, we refer to the electronic appendix~A of \citet{saesenI}. A summary of the pulsational
properties of all these stars, such as frequencies, amplitudes, phases and
degrees $\ell$ of the modes, are noted in Table~\ref{table:indiv}. 

The manually performed frequency analysis of all these stars was done similarly
as described in Sect.~\ref{sect:freq} for Oo~2444. In the following, the
results of the non-weighted analysis of the $V$ light curves are comparable to
the weighted analysis, except if we state otherwise. When the data in a filter was too limited, or the harmonic fit was too bad, we did not note the results of
the frequency fit for that filter. Some stars need some additional remarks, they
are noted below, sorted on the star number. 

The results of the mode identification were obtained in the same way as
described in Sect.~\ref{sect:mode} for Oo~2094 and Oo~2246. We did not perform a
mode identification for harmonic and combination frequencies, since {the
probability that they are independent frequencies is very small, as we have
checked in our extensive grid of frequencies based on theoretical models,
e.g., the probability for an independent harmonic frequency is
about~2x10$^{-4}$.} In Table~\ref{table:indiv}, we eliminated degree
possibilities when no models were retained for the star and frequency of
interest {in the case of $\ell=0$} or when no theoretical amplitude ratios
matched the observed ones within the uncertainties {for the models selected
within the $2\sigma$-error box in the HR-diagram}.  {We also evaluated the
calculated $\chi^2$-value for the remaining degrees. When a specific degree
was preferred based on its $\chi^2$-value, we stated so in
Table~\ref{table:indiv} by putting its $\ell$-value in bold.}  Whenever no
possibilities for $\ell=0$ to $\ell=4$ are left, it might be caused by $\ell>4$,
as {spectroscopic studies show that} high-degree modes occur in moderately
rotating pulsating B-stars \citep[e.g.,][]{aerts_betacru, telting_omegasco,
telting_psiori, ausseloos_betacen}. {It could also be that the moderate
rotation affects the observables too much such that the mode identification is
erroneous, as pointed out by {\citet{handler2012}},} 
or that we are not dealing with
a pulsation frequency. The mode identification figures for all the stars are
shown in Fig.~\ref{fig:modeidall}. {The results of the mode identification in
Table~\ref{table:indiv} should be regarded in combination with these figures,
in order to interpret the magnitude of discrepancy between the models and
observations.}
  
{\subsection*{Oo~1898}
After optimisation of the light curves and a detailed frequency analysis, no significant frequency peaks could be found any more in star Oo~1898. We only retain a candidate frequency $f=3.648$~d$^{-1}$ with a S/N of 3.9 in the $V$ data.}

{\subsection*{Oo~1973}
After optimisation of the light curves and a detailed frequency analysis, no significant frequency peaks could be found any more in star Oo~1973.}

\subsection*{Oo~2086}
For star Oo~2086 we find frequency values around 25 and 29~d$^{-1}$,
which are not at all expected for B-type pulsators. Based on its observed
frequencies, this star is a $\delta$~Sct variable and thus not a cluster member,
as already pointed out in \citet{saesenI}.

\subsection*{Oo~2089}
The analysis of Oo~2089 revealed only significant frequencies in the
$V$ filter. The frequency value of $f_2$ is rather close to
1.003~d$^{-1}$ and so we have to be cautious since it
may not be an intrinsic frequency. 

\subsection*{Oo~2091}
After prewhitening with $f_1$ and $f_2$, we found an additional significant
frequency $f = 1.0437$~d$^{-1}$ in the weighted and non-weighted
analysis of 
the $V$ residuals. To check whether this is an intrinsic
frequency or not, we analysed the data of Bia{\l}k\'ow (Poland) and Xinglong
(China) Observatory separately. Frequency $f = 1.0437$~d$^{-1}$ only shows up in
the Bia{\l}k\'ow data and is totally absent in the Xinglong data. Therefore we
do not accept this frequency. Prewhitening with $f$ did not reveal any other
significant frequencies, so we stopped the frequency search at this point.

\subsection*{Oo~2094}
After removing a sine fit with the main frequency $f_1$ and its harmonic $f_2 =
2f_1$, we obtained the candidate frequency $f = 0.479$~d$^{-1}$. This frequency
is the highest peak in both the $B$ and $V$ periodograms of the residuals, but
its amplitude does not exceed the needed signal-to-noise ratio to accept it
formally, i.e., it has a S/N of 3.7 in $V$ and of 3.4 in $B$.

\subsection*{Oo~2110}
The two significant frequencies we accepted for Oo~2110 are $f_1$ and
$f_2$. After prewhitening, a third significant frequency, $f = 3.0231$~d$^{-1}$
showed up in the weighted and non-weighted analysis of the $V$ data. This
frequency was not present in the other filters. A separate analysis of the $V$
data of different observatories, showed $f$ in the Bia{\l}k\'ow, but not in the
Xinglong data. Therefore we did not accept this third frequency as intrinsic
for the time being.

{\subsection*{Oo~2116}
In Oo~2116 we can only detect one frequency $f_1$. In the $I$ data, the alias peak $f_1'=4.21$~d$^{-1}$ is higher, whereas in $V$ and $B$, the peak at $f_1$ itself is highest. Since the $V$ data are less suffering from aliasing, according to the spectral windows, and since in both $V$ and $B$ data $f_1$ is highest, we believe this is the right peak.}

{\subsection*{Oo~2139}
We cannot identify any significant peaks in Oo~2139, but the Fourier spectra of the data in the different filters are dominated by peaks in the range between 0.4~d$^{-1}$ and 0.6~d$^{-1}$.}

{\subsection*{Oo~2146}
Only in the $V$ data of Oo~2146, a significant frequency can be detected: $f=4.013$~d$^{-1}$ with a S/N of 4.2. However, its value is equal to four times 1.0027~d$^{-1}$ within its error bars. Since the data for Oo~2146 mainly come from one observatory, Xinglong, we cannot conduct a further analysis to decide whether this frequency could be intrinsic to the star or not. Therefore we do not include it in a harmonic fit.}

\subsection*{Oo~2185}
For this star, no frequency could be accepted in the weighted analysis.
Therefore, the parameters of the frequency fit noted in
Table~\ref{table:indiv} are the results of a non-weighted analysis. In the
weighted analysis, much importance is given to the precise Bia{\l}k\'ow data,
which are spread over the three observing seasons, whereas the Xinglong data,
spread over only four months, become important in the non-weighted analysis, due
to their overwhelming number. This may be the reason why the
non-weighted frequency search gives better results than the weighted one, since
it is possible that the frequencies of Oo~2185 are not stable over time, due to
its Be nature.

{\subsection*{Oo~2189}
After the time series optimisation, we do not find any significant frequencies any more in Oo~2189, except $f=0.10411$~d$^{-1}$ that just reaches a S/N of 4 in a non-weighted analysis of the $V$ data. This frequency, however, does not at all show up in a separate analysis of the Bia{\l}k\'ow and Xinglong data, and therefore we do not trust it.}

\subsection*{Oo~2191}
Oo~2191 was announced as candidate SPB star by \citet{ngc884_3} and the
derivation of the two frequencies $f_1$ and $f_2$ confirms this pulsational
behaviour.

\subsection*{Oo~2228}
The results from the weighted and non-weighted frequency search for Oo~2228 are
not the same. The first frequency in the weighted $V$ analysis is $f_1$ with a
S/N of 4.5 and it is also marginally present in the $I$ data. In the
non-weighted
$V$ data we first retrieved $f = 0.959$~d$^{-1}$, prewhitening for this
frequency gives $f = 2.778$~d$^{-1}$ or $f_1$. It is impossible to distinguish
between these two values, so we have to be aware that we may not have taken
the correct frequency peak. Prewhitening the weighted and non-weighted
data set with $f_1$ gave no more significant frequencies, that differ
enough from a multiple of 1~d$^{-1}$ and that are similar in both data sets.
Therefore we only accept $f_1$.

{\subsection*{Oo~2235}
After a carefully executed analysis of Oo~2235, we cannot find any significant frequencies in the different filters, although the star surely is variable.}

\subsection*{Oo~2242}
Be-star Oo~2242 was discovered to be a variable star by \citet{ngc884_1}. Their
data showed variations on a time-scale typical for $\lambda$ Eri
variables. Strong aliasing prevented them from giving the correct
frequency peak. 

With our data set, we were able to determine a first frequency peak at $f_1$.
The non-weighted $V$ analysis led to a second significant frequency $f_2$. Its
value is very close to $2f_1$. However, when we fixed the second frequency
as harmonic $2f_1$ in the sine fit, the amplitude of $2f_1$ is much lower than
the one for $f_2$ in the periodogram and after prewhitening, we again obtained
$f_2$. {Moreover, since the frequency difference between $2f_1$ and $f_2$
is nearly 4~year aliases, we also fitted ($f_1-2\mathrm{yr}^{-1}$) and its first
harmonic ($2f_1-4\mathrm{yr}^{-1}$) to the data. This time, the amplitude of
($f_1-2\mathrm{yr}^{-1}$) was much lower than the one for $f_1$ and the
amplitude of the harmonic ($2f_1-4\mathrm{yr}^{-1}$) corresponded to the one of
$f_2$. Prewhitening gave again an additional frequency $f_1$ with an amplitude
as expected.} This whole analysis points to $f_2$ being an independent
additional frequency.

We also retrieved these two frequencies when combining our data set with the
measurements of \citet{ngc884_1}. Since the time span of the data is much
longer in this way, the frequency value can be deduced much more accurately.
Therefore we adopted the non-weighted results of the combined data sets in
Table~\ref{table:indiv}.

\subsection*{Oo~2245}
The periodograms of Oo~2245 in the different filters first showed many long
periods, that were due to different zero points in the three observing seasons.
After adjusting them, the low frequencies were absent in the $V$ and $U$
periodogram, but for the $I$ and $B$ data we needed to fit one low frequency to
get to the intrinsic frequencies. After prewhitening with $f_1$ and $f_2$, only
peaks at multiples of 1~d$^{-1}$ appeared and we stopped the frequency analysis.

\subsection*{Oo~2246}
\citet{ngc884_1} discovered Oo~2246 as one of the two first $\beta$~Cep
stars in NGC~884. They could disentangle two frequencies in their data, that
correspond with our $f_1$ and $f_2$. After prewhitening, we detected a third
frequency $f_3$ in the $B$ data. We optimised all three frequencies
in the $B$ filter, after verifying that the frequency values of $f_1$ and
$f_2$ are the same as the optimised $V$ values within their
error bars. Combining our light curve with the one from \citet{ngc884_1}
reproduced the same frequencies. Due to the longer time span, the results shown
in Table~\ref{table:indiv} are thus obtained in the joined data sets.

\subsection*{Oo~2253}
Oo~2253 was reported as SPB candidate by \citet{ngc884_3} and they extracted
three significant frequencies in the data of Bia{\l}k\'ow and Xinglong
Observatory. We confirm here their first two frequencies, denoted as $f_1$ and
$f_2$ in Table~\ref{table:indiv}, but we do not detect their third frequency in
our merged data set.

\subsection*{Oo~2267}
After prewhitening for $f_1$ and $f_2$, we found a candidate frequency $f =
4.546$~d$^{-1}$ in the $I$ residuals. This frequency has a S/N of 3.3 in $B$,
3.4 in $V$ and 3.7 in $I$, which is not high enough to accept it.
Moreover, due to strong aliasing, we cannot be sure to have picked out the
correct frequency peak.

\subsection*{Oo~2299}
Oo~2299 was discovered as candidate variable star on a time scale of six hours
by \citet{percy} and was identified as second $\beta$~Cep star in NGC~884 by
\citet{ngc884_1}. They report a single periodicity at $f_1$. With our data set
we cannot detect any other frequencies within the detection threshold either.
The results noted in Table~\ref{table:indiv} originate from the combined data
set.

\subsection*{Oo~2319}
Two significant intrinsic frequencies, $f_1$ and its harmonic $f_2 = 2f_1$ were
derived for Oo~2319. Prewhitening gave alias structures of low frequencies,
which we did not accept. We searched further, but no other significant peaks
emerged. 

The phases of $f_1$ and $f_2$ are the same in the different filters. The
amplitude of $f_1$ in $U$ seems largest, and the one in $I$ smallest, which is
a typical signature for pulsations in B-stars. However, they do not differ
within their error bars, therefore we cannot formally exclude the possibility
that this star is an ellipsoidal binary.

\subsection*{Oo~2324}
Prewhitening Oo~2324 with $f_1$ and $f_2$, gives a nice peak at $f_3$ with a S/N
= 5.6 in the weighted analysis of the $V$ residuals. In the other filters,
aliasing and higher noise levels hamper the detection of other frequency
peaks. The frequency value of $f_3$ is close to 3 times 1.0027~d$^{-1}$, so we
should be careful with it. However, when analysing $V$ data of Bia{\l}k\'ow and
Xinglong Observatory separately, both sites indicate $f_3$ as significant
frequency. Therefore we are inclined to accept it as intrinsic frequency.
Further prewhitening
no longer gave significant frequency peaks.

\subsection*{Oo~2345}
The first frequency found in the data of Oo~2345 is very close to 2~d$^{-1}$.
We retrieved $f_1$ in all filters, and also in the analysis of some
observatories separately. A phase diagram folded with this frequency, as shown
in Fig.~A.37 of \citet{saesenI}, displays a clear sinusoidal variation. We are thus
inclined to accept $f_1$ as intrinsic frequency of Oo~2345. Removing a sine fit
with $f_1$ reveals another frequency $f_2$, which is present in the $B$, $V$ and
$I$ residuals. Hereafter no more frequency peaks were accepted.

{\subsection*{Oo~2371}
Oo~2371 is an ellipsoidal binary with  orbital period around 5.2~d. The binary star was first discovered as candidate by \citet{ngc884_1} and later confirmed in spectroscopic data by \citet{oo2371}. Our deduced frequency $f_1$ is, as expected, twice the orbital frequency and the amplitudes in the different filters indeed do not follow the typical relation for B-type pulsations. We also note that a frequency analysis in the different observing seasons showed evidence for a changing amplitude over time, pointing to an {interacting} binary due to a close orbit \citep{oo2371}. In Table~\ref{table:indiv} we note the frequency as found in the combined data set. We did not execute a mode identification for this star, since we are not dealing with pulsations.}

\subsection*{Oo~2406}
Frequency $f_1$ is significant in the $V$ and $I$ data of Oo~2406. Prewhitening
shows the presence of a second frequency peak, both in the $V$ and $I$
residuals, but with a different alias frequency. $I$ data suffer more from
aliasing than $V$ data and so the difference between both peaks is significantly
less than for the $V$ filter. Therefore we adopt $f_2$ given by the $V$
residuals as the correct frequency peak.

{\subsection*{Oo~2426}
After optimisation of the light curves in the different filters and a detailed frequency analysis, we do not recover any significant frequencies for star Oo~2426.}

\subsection*{Oo~2429}
For Oo~2429, the results noted in Table~\ref{table:indiv} were optimised in the
$B$ filter, since this filter led to more significant frequencies. Moreover,
optimising $f_1$ in the $B$ filter resulted in the same frequency values within
 the error as optimising in $V$.

\subsection*{Oo~2448}
The results obtained for Oo~2448 should be treated with caution: the
periodograms suffered from severe aliasing. We adopted the frequencies coming
from the $B$ periodograms, since its alias frequency was lower than for the $V$
data, as in the spectral window. The highest peaks in $V$ are $f_1^{'} =
0.497$~d$^{-1}$ and $f_2^{'} = 2.430$~d$^{-1}$ with a S/N of 4.0 and 4.8,
respectively.

\subsection*{Oo~2488}
Oo~2488 was for the first time announced as variable star with $\beta$~Cep-like
oscillations by \citet{ngc6910_2} in a preliminary report on the multi-site
campaign on NGC~884. \citet{ngc884_2} derived two significant frequencies,
$f_1$ and $ f_2$, in a frequency analysis on
the single-site data of Bia{\l}k\'ow Observatory. An analysis of the bi-site
data of Bia{\l}k\'ow and Xinglong Observatory by \citet{ngc884_3} led to the
same two frequencies. An analysis of the total data set, as carried out here,
confirms $f_1$ and $f_2$ and even permits the detection of a third frequency
$f_3$. 

{\subsection*{Oo~2462}
As for star Oo~2185, no frequencies could be accepted in a weighted analysis, so that the results in Table~\ref{table:indiv} come from a non-weighted frequency analysis.}

{\subsection*{Oo~2562}
After prewhitening with significant frequencies $f_1$ and $f_2$, a third frequency $f_3$ is retrieved in the $I$ data with a S/N of 4.6. Inclusion of this frequency and optimisation in the $I$ filter do not change the harmonic fit for the two first frequencies in any of the filters. Given the large errors, the parameters of the harmonic fit for this third frequency as noted in table~\ref{table:indiv} cannot be trusted.}

{\subsection*{Oo~2566}
\citet{ngc6910_2} and \citet{ngc884_2} already mentioned Oo~2566 as candidate $\beta$~Cep star based on (a part of) the first season Bia{\l}k\'ow data. A frequency analysis on the whole campaign data set is suffering from long-term variations caused by the Be-character of the star. Frequency $f_1$, however, recurs in the $V$, $B$ an $U$ data after removing some arbitrary low frequencies, but is only significant in the $V$ data. Since the chaotic behaviour of this Be-star hampers our time series analysis, the results in Table~\ref{table:indiv} should be taken with caution.}

\subsection*{Oo~2572}
Oo~2572 was discovered and analysed in the same preliminary reports on the
multi-site campaign as for Oo~2488. \citet{ngc884_3} remarked that the analysis
of the bi-site data already was important in order to pinpoint the correct
frequency peak, in comparison with the single site data. They derived two
frequencies, $f_1$ and $f_2$, and reported that there is still power present in
the residuals, but that no other frequency value could be
determined. 

We confirm these two frequencies in the frequency search in the
whole campaign data set and retrieve a third frequency peak $f_3$. After
prewhitening with $f_1$, $f_2$ and $f_3$, no other frequencies can be accepted
for Oo~2572. Only in the non-weighted analysis of the $V$ data, we find evidence
for a candidate frequency $f = 4.389$~d$^{-1}$ with a signal-to-noise level of
3.9.

{\subsection*{Oo~2579}
The results noted in Table~\ref{table:indiv} for the $I$ data do not include the ones coming from Bia{\l}k\'ow Observatory, since they suffered severely from long-term effects hampering the detection of $f_1$. For the other filters, Bia{\l}k\'ow data could be included.}

{\subsection*{Oo~2616}
After prewhitening the $B$ data of Oo~2616 with $f_1$, a second frequency $f_2$ is found on the limit of acceptance, i.e., with a S/N-level of 4.0. This frequency was also visible in the first periodogram of the $V$ and $I$ data and therefore we accept it and include it in the harmonic fit. The parameters of the harmonic fit were optimised in the $B$ filter, but we have no high confidence in them due to their high error bars.}

{\subsection*{Oo~2622}
Oo~2622 is definitely a variable star, as clearly seen in the $V$, $B$, and $I$ light curves. However, residual trends hamper a clear and coherent frequency determination in the different filters. Therefore we do not attempt a harmonic fit for this star.}

\subsection*{Oo~2649}
We can only accept one frequency in the Be-star Oo~2649. However, the phase
diagram folded with this frequency \citep[see Fig.~A.59 in][]{saesenI} clearly shows
additional variations. A possible candidate frequency is $f =
2.264$~d$^{-1}$, which is clearly the highest peak in the $V$ periodogram of
the residuals, but its amplitude does not fulfill the signal-to-noise criteria
to accept it.

\subsection*{Oo~2694}
For Oo~2694 only one frequency, $f_1$ can be accepted, but surely more
variation is still hidden in the residuals. The residuals in the different
filters pointed towards a different frequency peak, the one in the $V$ data is
most significant, at $f = 1.502$~d$^{-1}$ or one of its aliases, with a S/N =
3.8. We cannot accept this frequency, but prewhitening with it even reveals
more candidate frequencies. However, we cannot retrieve a trustable value for
them, so we stopped the frequency search at this point.

{\subsection*{Oo~2752}
No significant frequencies can be found in Oo~2752 after optimisation of the light curves. We only retain one candidate frequency $f=1.811$~d$^{-1}$ that has a S/N of 3.9 in the $V$ data, which is not high enough to formally accept it.}

{\subsection*{Oo~2752}
Only one frequency just reaches the 4.0 S/N-level in the $V$ data of Oo~2752, namely $f=5.991$~d$^{-1}$. However, because of its closeness to 6~d$^{-1}$, the S/N-level of the peak and the absence of proof in the other filters, we do not accept this frequency.}


\section{ADDITIONAL SEISMIC RESULTS}
\label{sect:seismic}
\subsection{Relations between Pulsation and Basic Stellar Parameters}
\label{sect:relation}

In this section, we do not consider star Oo~2371, since it is an ellipsoidal
binary and not a pulsating star. We also exclude Oo~2086 and Oo~2633, which are,
according to their frequency spectra, $\delta$~Sct stars rather than members of
the cluster.  This also means that their deduced position in the HR-diagram of
Sect.~\ref{sect:par} cannot be trusted. All other 62~B-stars that have at least
one accepted frequency in Table~\ref{table:indiv} can be considered {members} of NGC~884 based on the photometric diagrams presented
in Appendix~A of \citet{saesenI}.  \citet{slesnick}, \citet{uribe} and
\citet{currie} also investigated the membership of (some) of these periodic
B-stars in detail, and all of them are believed to be a member star of NGC~884
by at least one of these sources.

\begin{figure}[ht!]
\centering
\includegraphics[width=\columnwidth]{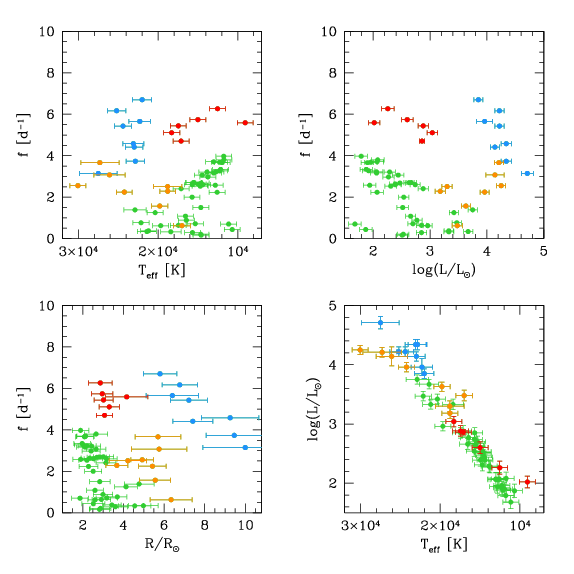}
\caption{{Dominant frequencies for all B-stars with
respect to the basic stellar parameters $T_\mathrm{eff}$, $\log(L/L_{\sun})$ and
$R$,  with their $1\sigma$-error. The error on the frequency is smaller than
  the used symbol.
Different colours are attributed to different apparent groups in the
$(R,f)$-diagram (see Sect.~\ref{sect:relation}). 
The bottom right figure shows all these stars in the
HR-diagram.}\label{fig:paramobs}}
\end{figure}

\begin{figure}[t!]
\centering
\includegraphics[width=\columnwidth]{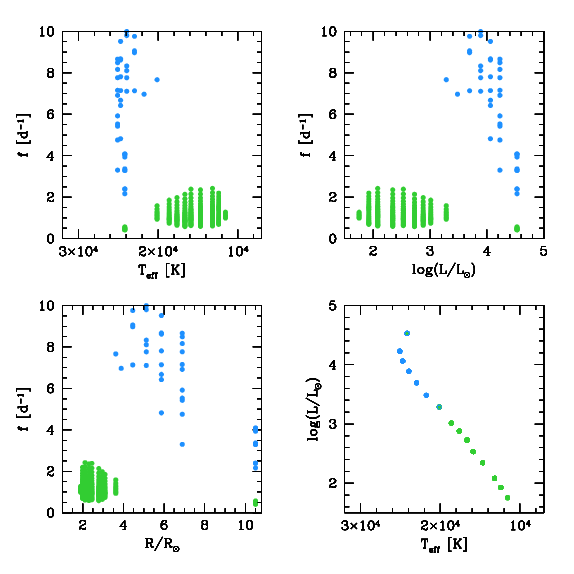}
\caption{{Same as Fig.~\ref{fig:paramobs}, but for the excited
theoretical eigenfrequencies of  axisymmetric modes of 
degree $\ell=0, 1$ and 2 calculated for
\textsc{cl\'es} models with $X=0.70$, $Z=0.02$, $\alpha_\mathrm{ov}=0.2$ and
$\log(\mathrm{age/yr})=7.1$.} The green points denote the g-modes, the blue
points the
p-modes.\label{fig:paramth}}
\end{figure}

In Fig.~\ref{fig:paramobs} we show the frequency with the highest amplitude for
every star with respect to the basic stellar parameters, $\log(L/L_{\sun})$,
$T_\mathrm{eff}$ and the radius $R$ of the star. The latter was determined as
$(R/R_{\sun})^2 = (L/L_{\sun}) (T_\mathrm{eff}/T_{\mathrm{eff},\sun})^{-4}$. We
attributed different colours to different apparent groups in these diagrams,
especially on the basis of the radius.  We used these figures to make a
selection of the most appropriate stars for asteroseismology by visual
inspection.  Fig.\,\ref{fig:paramth} shows the theoretical analogue of the
observational Fig.\,\ref{fig:paramobs} for axisymmetric modes and is based on
\textsc{cl\'es} models with parameters $X=0.70$, $Z=0.02$,
$\alpha_\mathrm{ov}=0.2$ and $\log(\mathrm{age/yr})=7.1$
\citep{southworth_chiper}. We selected all theoretical eigenfrequencies with
degrees $\ell=0, 1$ and 2 for p- and g-modes that are excited (a lower $Z$ would
result in less excited frequencies). In this way, we identify the blue
observational group as p-mode pulsators and the green one as g-mode
pulsators. The group in between, denoted in orange, contains all Be-pulsators
amongst the studied B-stars.  The stars denoted in red also reside between the
p- and g-modes, but these are not known as Be-stars nor do we know their
rotational velocities.  On the basis of this comparison between observational
and theoretical frequencies, we identify eight $\beta$~Cep pulsators in the
cluster: Oo~2246, Oo~2299, Oo~2444, Oo~2488, Oo~2520, Oo~2572, Oo~2601 and
Oo~2694.


\begin{figure}
\centering
\includegraphics[width=0.68\columnwidth]{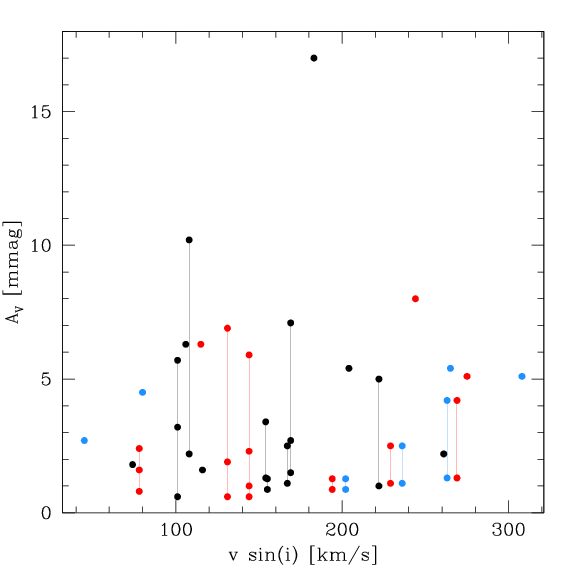}
\caption{Amplitude in the $V$ filter of all observed frequencies
with respect to the projected rotational velocities of \citet{strom} (black
points), \citet{huang} (red points), and \citet{marsh2012} (blue points)
 for all stars for which these data are
available. {The amplitudes originating from the same star are connected.}
\label{fig:amprot}}
\end{figure}

\begin{figure}
\centering
\includegraphics[width=0.68\columnwidth]{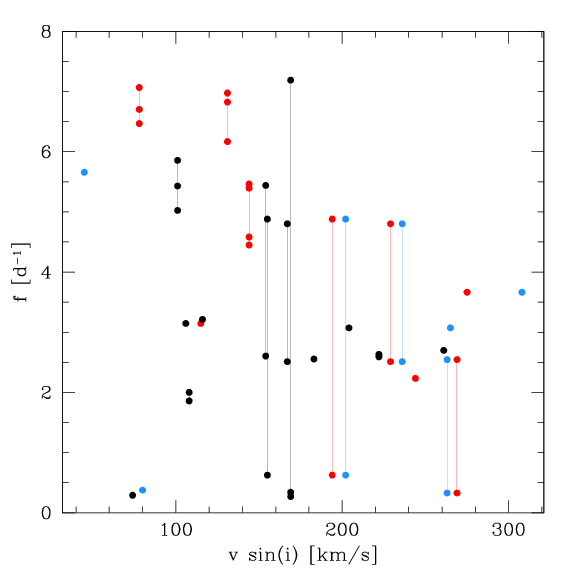}
\caption{{All observed frequency values with respect to the projected rotational
    velocities of \citet{strom} (black points), \citet{huang} (red points), and
\citet{marsh2012} (blue points) 
    for all stars for which these data are available. The frequencies observed
    in the same star are connected.}
\label{fig:freqrot}}
\end{figure}

\begin{figure}
\centering
\includegraphics[width=0.68\columnwidth]{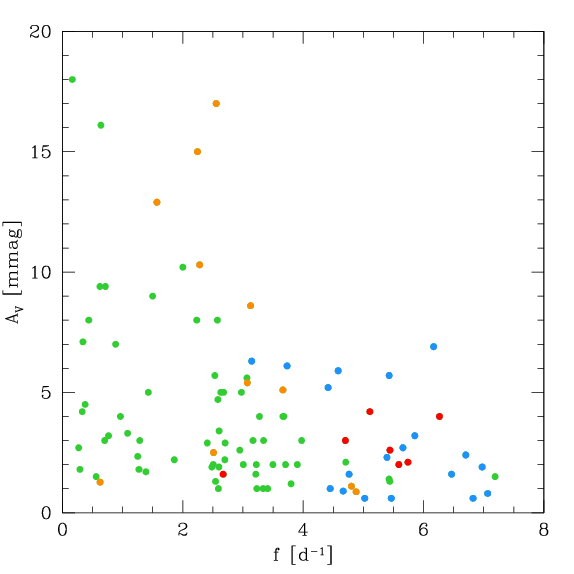}
\caption{{All observed frequency values with respect to their $V$
    amplitude. {For clarity we left out two points situated at
      $(f,A_V)=(0.20871~\textrm{d}^{-1},56~\textrm{mmag})$ and
      $(f,A_V)=(11.951~\textrm{d}^{-1},1~\textrm{mmag})$.} The used colour code
    is the same as for Fig.~\ref{fig:paramobs} and is explained in
    Sect.~\ref{sect:relation}.}
\label{fig:freqamp}}
\end{figure}

In Fig.~\ref{fig:amprot} we show the amplitudes in the $V$ filter of all
observed frequencies {for all B-stars} with respect to the projected rotational
velocities of \citet{strom}, \citet{huang}, and \citet{marsh2012} for all stars
for which these data are available. {\citet{catalog_betacep} found that
slowly-rotating Galactic $\beta$~Cep stars tend to have higher pulsation
amplitudes, lending support to the hypothesis that rotation would act as an
amplitude limiting mechanism. We do not find a clear connection between the
observed rotation velocities and the mode amplitudes. We also note that our
observed amplitudes are in general lower than the ones of the stars treated in
\citet{catalog_betacep}. We do not find any dependency {either} between the
projected rotational velocities and observed frequencies, as shown in
Fig.~\ref{fig:freqrot}, but the $V$ amplitude seems to get lower for higher
frequencies, as shown in Fig.~\ref{fig:freqamp}.}

\subsection{Asteroseismology of the B-stars in  the Cluster}
\label{sect:compatib} 

We subsequently conducted a comparison between the observed frequencies of the
eight selected $\beta$~Cep stars and those that occur for our dense full grid of
stellar models, to see whether we can find a consistent cluster solution, in the
sense of demanding a similar chemical composition and age for the eight
$\beta\,$Cep stars without specifying any values for these quantities.  In this
way, we present a proof-of-concept for asteroseismology of the cluster without
imposing information already deduced in the literature. This is useful as an
independent cluster analysis because {\it none\/} of the previous studies
allowed for a broad range in metallicity and core overshooting in their models
due to which the age uncertainties reported for the cluster in the literature
are serious underestimations, as already pointed out by
\citet{southworth_chiper}. {We have to restrict our asteroseismic analysis to the p-mode pulsators ($\beta$~Cep stars) of the cluster and are not able to include the g-mode pulsators (SPB stars). The reason for this restriction is the dense theoretical oscillation spectrum for their g-modes which we cannot exploit as long as mode identification is lacking. Indeed, given the rotation rate of the cluster stars, the predicted g-mode multiplets of the SPB stars are merged and we can always assign several of them to the observed frequencies. Hence, they do not contribute to a better restriction of the cluster age, unless we can pinpoint a unique $(\ell,m)$-value to each of the detected frequencies. This is also the reason why seismic modelling of SPB stars is essentially lacking in the literature, with the notable exception of the star HD\,50230 in \citet{spacinggmodes} for which period spacings of detected g-modes in this very slow rotator could be exploited  in terms of input physics of the models.}

In a first step, we searched for agreement between the observed and theoretical
frequency spectrum on a star-to-star basis. For each of the eight $\beta\,$Cep
stars, we retained the grid models within the observed position in the
HR-diagram within a 2$\sigma$-error box, as derived in Sect.~\ref{sect:par}. For
these models, we made a further downselection by demanding that each observed
frequency must be present in the theoretical frequency spectrum.  To do this
selection, we took an allowed difference between the theoretical frequency and
the observed one of $\Delta f=0.15$~d$^{-1}$ for the radial modes, because this
value is half of the maximum difference in frequency for the same mode of
subsequent models on the considered evolutionary tracks.  For the non-radial
modes, the allowed difference between the observed frequency and the theoretical
one is the maximum of $\Delta f=0.15$~d$^{-1}$ or the one based on rotational
splitting in the first-order approximation.  Since the theoretical frequency
spectra only contain axisymmetric modes, we took as {the maximum allowed frequency}
difference for the non-radial modes of degree $\ell$ a value of $\Delta f = \ell
\, (\beta_{n,l} v \sin i) / (2 \pi R \sin i)$, with $\beta_{n,l}$ connected with
the Ledoux constant and $R$ the radius of the star (both of which are known for
each model) and $v \sin i$ the projected rotational velocity. For the latter, we
used the literature values given by \citet{strom}, \citet{huang} or
\citet{marsh2012} if available (ranging from 45 to 144~km~s$^{-1}$), or its mean
value, if more than one value is available. For two stars we did not have any
estimated projected rotational velocity, and we adopted 100~km~s$^{-1}$ to
compute $\Delta f$, but we repeated the model selection experiment taking $v
\sin i = 50$~km~s$^{-1}$ and for 150~km~s$^{-1}$ to evaluate this choice for the
two stars on the outcome.  We performed the downselection twice, for two values
of the inclination angle: $i=30^\circ$ and $i=60^\circ$, and we restricted {the analysis} to
radial modes and non-radial modes of $\ell=1$ or 2 in each case.

Since the mode identification performed in Sect.~\ref{sect:mode} was not
  conclusive for all the 19~detected frequencies in the eight selected stars, we
  proceeded in two different ways.

  In a first scenario, the model selection done after eliminating all models
  outside the $2\sigma$-error boxes in the HR-diagram only made use of the
  frequencies corresponding to a unique identification of $\ell$ from the
  amplitude ratios in Sect.~\ref{sect:mode}, restricting to frequencies of
  $\ell=0,1,2$ modes.  This type of model selection with one unique $\ell\leq2$
  per observed frequency, allowing rotational splitting, concerns seven
  frequencies in five $\beta$~Cep stars: Oo~2246 with $\ell=2$ for $f_1$ and a
  radial mode for $f_3$, Oo~2299 with $\ell=2$ for $f_1$, Oo~2444 with $\ell=1$
  for $f_1$, Oo~2488 with $\ell=2$ for $f_1$ and Oo~2572 with $\ell=2$ for $f_1$
  and $\ell=1$ for $f_2$.  Since the cluster stars must have similar chemical
  composition and age, we only retained models with equal $(X,Z,{\rm age})$ for
  the five stars.  For each $(X,Z)$ combination, a series of models indeed
  survives these requirements.  These selected models have an overall age range
  of $\log(\mathrm{age/yr}) = 7.12-7.28$, where the lowest values occur for
  models with $Z=0.018$ and the highest values for those with $Z=0.010$, per
  fixed $X$ (see Table\,\ref{clustersolution}).  It is noteworthy that
  \citet{southworth_chiper} cite an age problem, i.e., they deduced a too low
  range of $\log(\mathrm{age/yr}) = 7.09-7.11$ from models with fixed $Z=0.02$
  and without overshoot, while they also quote the observed metallicity to be
  $Z\simeq 0.01$. We consider $Z<0.014$ as unlikely for the cluster, since no
  modes are predicted to be excited at such low metallicity for a solar mixture,
  but we refrain from imposing this constraint to narrow down the allowed age
  range at present.  We looked whether the model elimination implied a
  restriction on the overshoot parameter for some of the five $\beta$~Cep stars,
  by taking their observed frequency spectrum and the obtained cluster age range
  into account.  The only restriction on the overshoot parameters found in this
  way is $\alpha_\mathrm{ov}\geq 0.20$ for Oo~2299. We have presently
  insufficient observational constraints to disentangle the different physical
  phenomena causing near-core extra mixing, such as rotational, diffusive or
  convective mixing \citep[for a discussion of those phenomena, see][]{maeder2009}.  Hence, despite the fact that rotational mixing leaves a
  different signature than plain overshoot in $\beta\,$Cep stars
  \citep{montalban2008,miglio2008}, our estimation of $\alpha_{\rm ov}$ is not
  sufficient to unravel the physical nature of the mixing processes at this
  stage.

\begin{table*}
\footnotesize
  \caption{Results of the elimination of models from the grid, for scenario 1
    (defined in the text).\label{clustersolution}} 
\centering
\tabcolsep=4pt
\begin{tabular}{cc|cc|cc|cc}
\tableline
&&\multicolumn{2}{c|}{$v\sin i=50\,$km\,s$^{-1}$ for Oo~2572} &
\multicolumn{2}{c|}{$v\sin i=100\,$km\,s$^{-1}$ for Oo~2572} &
\multicolumn{2}{c}{$v\sin i=150\,$km\,s$^{-1}$ for Oo~2572}\\
\tableline
&&{$i=30^\circ$} & {$i=60^\circ$} &
{$i=30^\circ$} & {$i=60^\circ$} &
{$i=30^\circ$} & {$i=60^\circ$} \\
$X$ & $Z$ & age range & age range & age range & age range & age range & age range \\
\tableline
0.68&0.010& \nodata & \nodata &16.32--17.84& \nodata &15.20--17.84&16.32--16.88\\
0.68&0.012&17.44--17.52& \nodata &15.28--17.52& \nodata &14.72--17.52&15.36--16.64\\
0.68&0.014&16.40--17.20& \nodata &14.72--17.20&15.84--16.32&13.84--17.20&15.20--16.32\\
0.68&0.016&15.76--17.12& \nodata &13.84--17.12&15.20--16.00&13.36--17.12&14.56--16.00\\
0.68&0.018&15.12--17.28& \nodata &13.68--17.28&15.04--15.84&13.20--17.28&13.76--15.84\\
0.70&0.010& \nodata &\nodata &16.80--18.24& \nodata &15.60--18.24&16.80--17.28\\
0.70&0.012&17.44--17.84& \nodata &15.84--17.84& \nodata &14.72--17.84&15.84--16.72\\
0.70&0.014&16.72--17.68& \nodata &15.12--17.68&16.32--16.64&14.32--17.68&15.20--16.64\\
0.70&0.016&15.76--17.92& \nodata &14.32--17.92&15.68--16.32&13.76--17.92&15.04--16.32\\
0.70&0.018&15.60--17.60&16.56--16.64&14.08--17.60&15.04--16.64&13.28--17.60&14.16--16.64\\
0.72&0.010& \nodata & \nodata &16.80--18.72& \nodata &16.08--18.72&17.28--17.60\\
0.72&0.012&18.24& \nodata &16.32--18.24&17.28&15.20--18.24&16.32--17.28\\
0.72&0.014&17.20--18.48& \nodata &15.60--18.48& \nodata &14.56--18.48&15.60--16.80\\
0.72&0.016&16.24--18.48& \nodata &15.04--18.48&16.16--16.64&13.92--18.48&15.04--16.64\\
0.72&0.018&16.08--18.16&17.28--17.44&14.24--18.16&15.52--17.44&13.68--18.16&14.64--17.44\\
0.74&0.010& \nodata &\nodata&17.28--19.12& \nodata &16.56--19.12&17.76--18.00\\
0.74&0.012&18.80& \nodata &16.56--18.80& \nodata &15.68--18.80&17.04--17.68\\
0.74&0.014&17.68--19.20& \nodata &15.68--19.20&17.12--17.20&14.80--19.20&16.16--17.20\\
0.74&0.016&16.96--18.88& \nodata &15.44--18.88&16.64--17.44&14.24--18.88&15.52--17.44\\
0.74&0.018&16.08--18.72&17.76--18.24&14.64--18.72&16.00--18.24&14.08--18.72&15.04--18.24\\
\tableline
\end{tabular}
\tablecomments{We tabulate the ranges of stellar parameters that survive the elimination process when     considering three values for $v\sin i$ for the star Oo~2572, while     using the measured $v\sin i$ for the other four pulsators. The acceptable    age ranges are indicated in $10^6$yr. We considered the cases of    $i=30^\circ$, respectively $60^\circ$, for all five pulsators. {Blank} entries imply that none of the models in the grid fulfill the mode identification of the detected frequencies.}
\end{table*}

In a second scenario, we did not make any model selection based on the
identification of the degree $\ell$ of the oscillation modes as obtained from
the amplitude ratios.  Rather, we kept all models fulfilling the requirement of
equal $(X,Z,{\rm age})$ while having $\ell=0,1,2$ modes whose frequencies differ
less than $\Delta f$ as determined above for the 19 detected ones in the eight
stars.  Again, a series of models survives these requirements, having an age
range of $\log(\mathrm{age/yr}) = 7.09-7.30$, i.e., this is less demanding on
the model selection than using the seven identified frequencies as in
scenario\,1.  The comparison between the results of both scenarios is an
illustration of the advantage of having a unique identification of the mode
degree, even if only for a few modes, compared to the case where no such
identification is available.


\section{Summary and Discussion}
\label{sect:summ}

In this paper, we investigated in detail {75 periodic} B-stars in NGC~884.  The
estimates for the basic stellar parameters can be found in
Tables~\ref{table:hrcalib}, \ref{table:hrsimil} and \ref{table:hrlit}.  From a
detailed frequency analysis, we deduced 65~periodic B-stars of which we
classified 36~as mono-periodic, {16} as bi-periodic, {10} as tri-periodic, and 2
as quadru-periodic, while one star showed 9~independent frequencies. An overview
of the different frequency values, together with their amplitudes and phases
from a multi-frequency fit, can be found in Table~\ref{table:indiv}.

For all the detected frequencies we performed a photometric mode identification
based on the comparison between the observed and the theoretical amplitude
ratios,  ignoring the rotational effects on the amplitude ratios.  Given
the input physics of the models, we could eliminate some degree
possibilities. However, the large uncertainties on the observed amplitude ratios
hampered a unique mode identification for the majority of detected
frequencies. Table~\ref{table:indiv} contains the results for the mode
identification in its last column. Twelve modes of nine pulsators could be
securely identified.

We did not find a correlation between the projected rotational velocity and the
amplitude or frequency of the modes.  The amplitudes of the oscillations,
however, decrease as the frequency values increase. A comparison between the
observed and theoretical frequency-radius relation allowed us to select eight
$\beta$~Cep stars in our sample, namely Oo~2246, Oo~2299, Oo~2444, Oo~2488,
Oo~2520, Oo~2572, Oo~2601 and Oo~2694.  The seismic properties of these
$\beta$~Cep stars were confronted with those predicted from an extensive grid of
stellar models by requesting common cluster parameters, i.e., equal age and
initial chemical composition.  Elimination of models in the considered grid led
to a global cluster age of $\log(\mathrm{age/yr}) =7.12-7.28$ by imposing the
pulsation characteristics {of five of the eight} $\beta\,$Cep stars with in total seven
identified modes. This is fully compatible with the age estimate obtained from
modelling of an eclipsing binary in the cluster \citep[$>12.6$~Myr,\
][]{southworth_chiper} and illustrates that the concept of cluster
asteroseismology offers a valuable alternative to isochrone fitting or eclipsing
binary modelling, particularly if one keeps in mind that our seismic approach
considered a much broader range of $(X,Z,\alpha_{\rm ov},{\rm age})$ than the
previous methods.

{If we would have had several more identified p-mode frequencies per star, then the elimination
of models would become much more powerful. Although we have many more pulsating B-stars with
high-order g-modes at hand, they are presently not useful to restrict the seismic cluster age further, as already explained above. Ideally, one would want to reach the stage of having very detailed information on several modes, as in, e.g., \citet{nu_eri3,thetaop3,V1449aq}, but then for several cluster pulsators. This requires clear detections of multiplets or high-resolution spectroscopy in addition to multi-colour photometry to identify the modes. If such a stage could be reached, then the input physics of the models can be improved and a very precise cluster age could be deduced. In our present study, we only reached the stage of finding an age consistent with the one reported in the literature.}

{Future improvements of our present cluster modelling can be achieved by
collecting high-precision spectroscopy for the cluster pulsators, in an attempt
to derive usable spectroscopic limits on the metallicity. Additionally, the
gathering of a time-series of high-resolution high-S/N spectra of the brightest
high-amplitude p-mode pulsators in the cluster may open the door for
spectroscopic mode identificaton of some of the oscillation modes through an
interpretation of the line-profile variability \citep[e.g., Chapter~6
in][]{bookaerts}.}

In addition to the conclusion we reached for the particular case of NGC\,884,
several conclusions can be drawn as best practices for further studies of this
kind. A prime concern remains the overcoming of frequency aliasing and the
accompanying high demands on the duty cycle of the data, particularly when the
amplitudes of the cluster pulsators are of order mmag or lower.  Secondly, the
need occurs to have more precise amplitude determinations for secure
identification of mode degrees, which was, in our study, severely hampered by
the lack of a high-precision amplitude in the UV domain. This wavelength range had deserved more attention in the planning of the observing
runs than we gave it.

\acknowledgements
  The authors are much indebted to Prof.\ A.\ Pigulski and Dr.\ J.\ De Ridder
  for valuable comments on an earlier version of this study, in the framework of
  their PhD jury evaluation task. We thank R.~Scuflaire and M.-A.~Dupret for the
  use of their software. The research leading to these results has received
  funding from the European Research Council under the European Community's
  Seventh Framework Programme (FP7/2007--2013)/ERC grant agreement
  n$^\circ$227224 (PROSPERITY). This research has made use of the WEBDA
  database, operated at the Institute for Astronomy of the University of Vienna,
  as well as NASA's Astrophysics Data System, and the SIMBAD database and VizieR
  catalogue access tool, both operated at CDS, Strasbourg, France.


\bibliographystyle{aa}
\bibliography{paper}

\begin{thebibliography}{76}
\expandafter\ifx\csname natexlab\endcsname\relax\def\natexlab#1{#1}\fi

\bibitem[{{Aerts}(2000)}]{aerts_hip}
{Aerts}, C. 2000, \aap, 361, 245

\bibitem[{{Aerts} {et~al.}(2011){Aerts}, {Briquet}, {Degroote}, {Thoul}, \&
  {van Hoolst}}]{V1449aq}
{Aerts}, C., {Briquet}, M., {Degroote}, P., {Thoul}, A., \& {van Hoolst}, T.
  2011, \aap, 534, A98

\bibitem[{{Aerts} {et~al.}(2010){Aerts}, {Christensen-Dalsgaard}, \&
  {Kurtz}}]{bookaerts}
{Aerts}, C., {Christensen-Dalsgaard}, J., \& {Kurtz}, D.~W. 2010,
  {Asteroseismology} ({Heidelberg: Springer})

\bibitem[{{Aerts} {et~al.}(1998){Aerts}, {De Cat}, {Cuypers}, {Becker},
  {Mathias}, {De Mey}, {Gillet}, \& {Waelkens}}]{aerts_betacru}
{Aerts}, C., {De Cat}, P., {Cuypers}, J., {et~al.} 1998, \aap, 329, 137

\bibitem[{{Aerts} {et~al.}(2004){Aerts}, {Lamers}, \&
  {Molenberghs}}]{aertslamers}
{Aerts}, C., {Lamers}, H.~J.~G.~L.~M., \& {Molenberghs}, G. 2004, \aap, 418,
  639

\bibitem[{{Arenou} {et~al.}(1992){Arenou}, {Grenon}, \& {Gomez}}]{arenou}
{Arenou}, F., {Grenon}, M., \& {Gomez}, A. 1992, \aap, 258, 104

\bibitem[{{Asplund} {et~al.}(2005){Asplund}, {Grevesse}, \& {Sauval}}]{asplund}
{Asplund}, M., {Grevesse}, N., \& {Sauval}, A.~J. 2005, in Astronomical Society
  of the Pacific Conference Series, Vol. 336, Cosmic Abundances as Records of
  Stellar Evolution and Nucleosynthesis, ed. {T.~G.~Barnes III \& F.~N.~Bash},
  25

\bibitem[{{Ausseloos} {et~al.}(2006){Ausseloos}, {Aerts}, {Lefever}, {Davis},
  \& {Harmanec}}]{ausseloos_betacen}
{Ausseloos}, M., {Aerts}, C., {Lefever}, K., {Davis}, J., \& {Harmanec}, P.
  2006, \aap, 455, 259

\bibitem[{{Balona} {et~al.}(1997){Balona}, {Dziembowski}, \&
  {Pamyatnykh}}]{balona_clusters}
{Balona}, L.~A., {Dziembowski}, W.~A., \& {Pamyatnykh}, A. 1997, \mnras, 289,
  25

\bibitem[{{Balona} {et~al.}(2011){Balona}, {Pigulski}, {Cat}, {Handler},
  {Guti{\'e}rrez-Soto}, {Engelbrecht}, {Frescura}, {Briquet}, {Cuypers},
  {Daszy{\'n}ska-Daszkiewicz}, {Degroote}, {Dukes}, {Garcia}, {Green}, {Heber},
  {Kawaler}, {Lehmann}, {Leroy}, {Molenda-{\.Z}aaowicz}, {Neiner}, {Noels},
  {Nuspl}, {{\O}stensen}, {Pricopi}, {Roxburgh}, {Salmon}, {Smith},
  {Su{\'a}rez}, {Suran}, {Szab{\'o}}, {Uytterhoeven}, {Christensen-Dalsgaard},
  {Kjeldsen}, {Caldwell}, {Girouard}, \& {Sanderfer}}]{balona_kepler}
{Balona}, L.~A., {Pigulski}, A., {Cat}, P.~D., {et~al.} 2011, \mnras, 413, 2403

\bibitem[{{Breger} {et~al.}(1993){Breger}, {Stich}, {Garrido}, {Martin},
  {Jiang}, {Li}, {Hube}, {Ostermann}, {Paparo}, \& {Scheck}}]{breger_sn}
{Breger}, M., {Stich}, J., {Garrido}, R., {et~al.} 1993, \aap, 271, 482

\bibitem[{{Briquet} {et~al.}(2011){Briquet}, {Aerts}, {Baglin}, {Nieva},
  {Degroote}, {Przybilla}, {Noels}, {Schiller}, {Vu{\v c}kovi{\'c}}, {Oreiro},
  {Smolders}, {Auvergne}, {Baudin}, {Catala}, {Michel}, \&
  {Samadi}}]{briquet_hd46202}
{Briquet}, M., {Aerts}, C., {Baglin}, A., {et~al.} 2011, \aap, 527, A112

\bibitem[{{Briquet} {et~al.}(2007){Briquet}, {Morel}, {Thoul}, {Scuflaire},
  {Miglio}, {Montalb{\'a}n}, {Dupret}, \& {Aerts}}]{thetaop3}
{Briquet}, M., {Morel}, T., {Thoul}, A., {et~al.} 2007, \mnras, 381, 1482

\bibitem[{{Briquet} {et~al.}(2009){Briquet}, {Uytterhoeven}, {Morel}, {Aerts},
  {De Cat}, {Mathias}, {Lefever}, {Miglio}, {Poretti}, {Mart{\'{\i}}n-Ruiz},
  {Papar{\'o}}, {Rainer}, {Carrier}, {Guti{\'e}rrez-Soto}, {Valtier}, {Benk{\H
  o}}, {Bogn{\'a}r}, {Niemczura}, {Amado}, {Su{\'a}rez}, {Moya},
  {Rodr{\'{\i}}guez-L{\'o}pez}, \& {Garrido}}]{briquetcorot}
{Briquet}, M., {Uytterhoeven}, K., {Morel}, T., {et~al.} 2009, \aap, 506, 269

\bibitem[{{Currie} {et~al.}(2010){Currie}, {Hernandez}, {Irwin}, {Kenyon},
  {Tokarz}, {Balog}, {Bragg}, {Berlind}, \& {Calkins}}]{currie}
{Currie}, T., {Hernandez}, J., {Irwin}, J., {et~al.} 2010, \apjs, 186, 191

\bibitem[{{Daszy{\'n}ska-Daszkiewicz}
  {et~al.}(2002){Daszy{\'n}ska-Daszkiewicz}, {Dziembowski}, {Pamyatnykh}, \&
  {Goupil}}]{jagoda}
{Daszy{\'n}ska-Daszkiewicz}, J., {Dziembowski}, W.~A., {Pamyatnykh}, A.~A., \&
  {Goupil}, M.-J. 2002, \aap, 392, 151

\bibitem[{{De Cat} {et~al.}(2007){De Cat}, {Briquet}, {Aerts}, {Goossens},
  {Saesen}, {Cuypers}, {Yakut}, {Scuflaire}, {Dupret}, {Uytterhoeven}, {van
  Winckel}, {Raskin}, {Davignon}, {Le Guillou}, {van Malderen}, {Reyniers},
  {Acke}, {de Meester}, {Vanautgaerden}, {Vandenbussche}, {Verhoelst},
  {Waelkens}, {Deroo}, {Reyniers}, {Ausseloos}, {Broeders},
  {Daszy{\'n}ska-Daszkiewicz}, {Debosscher}, {de Ruyter}, {Lefever}, {Decin},
  {Kolenberg}, {Mazumdar}, {van Kerckhoven}, {de Ridder}, {Drummond}, {Barban},
  {Vanhollebeke}, {Maas}, \& {Decin}}]{Bmercator}
{De Cat}, P., {Briquet}, M., {Aerts}, C., {et~al.} 2007, \aap, 463, 243

\bibitem[{{De Cat} {et~al.}(2005){De Cat}, {Briquet},
  {Daszy{\'n}ska-Daszkiewicz}, {Dupret}, {De Ridder}, {Scuflaire}, \&
  {Aerts}}]{decatSPB}
{De Cat}, P., {Briquet}, M., {Daszy{\'n}ska-Daszkiewicz}, J., {et~al.} 2005,
  \aap, 432, 1013

\bibitem[{{De Ridder} {et~al.}(2004){De Ridder}, {Telting}, {Balona},
  {Handler}, {Briquet}, {Daszy{\'n}ska-Daszkiewicz}, {Lefever}, {Korn},
  {Heiter}, \& {Aerts}}]{nueri_mode}
{De Ridder}, J., {Telting}, J.~H., {Balona}, L.~A., {et~al.} 2004, \mnras, 351,
  324

\bibitem[{{Degroote} {et~al.}(2010){Degroote}, {Aerts}, {Baglin}, {Miglio},
  {Briquet}, {Noels}, {Niemczura}, {Montalban}, {Bloemen}, {Oreiro}, {Vu{\v
  c}kovi{\'c}}, {Smolders}, {Auvergne}, {Baudin}, {Catala}, \&
  {Michel}}]{spacinggmodes}
{Degroote}, P., {Aerts}, C., {Baglin}, A., {et~al.} 2010, \nat, 464, 259

\bibitem[{{Degroote} {et~al.}(2009){Degroote}, {Briquet}, {Catala},
  {Uytterhoeven}, {Lefever}, {Morel}, {Aerts}, {Carrier}, {Auvergne}, {Baglin},
  \& {Michel}}]{betacepcorot}
{Degroote}, P., {Briquet}, M., {Catala}, C., {et~al.} 2009, \aap, 506, 111

\bibitem[{{Desmet} {et~al.}(2009){Desmet}, {Briquet}, {Thoul}, {Zima}, {De
  Cat}, {Handler}, {Ilyin}, {Kambe}, {Krzesi\'nski}, {Lehmann}, {Masuda},
  {Mathias}, {Mkrtichian}, {Telting}, {Uytterhoeven}, {Yang}, \&
  {Aerts}}]{12lac2}
{Desmet}, M., {Briquet}, M., {Thoul}, A., {et~al.} 2009, \mnras, 396, 1460

\bibitem[{{Dupret}(2002)}]{dupretPhD}
{Dupret}, M.-A. 2002, PhD thesis, Bulletin de la Soci\'et\'e des Sciences de
  Li\`ege, Vol.\ 71, 5--6, pp.\ 249 -- 445

\bibitem[{{Dupret} {et~al.}(2003){Dupret}, {De Ridder}, {De Cat}, {Aerts},
  {Scuflaire}, {Noels}, \& {Thoul}}]{amplrat}
{Dupret}, M.-A., {De Ridder}, J., {De Cat}, P., {et~al.} 2003, \aap, 398, 677

\bibitem[{{Dupret} {et~al.}(2002){Dupret}, {De Ridder}, {Neuforge}, {Aerts}, \&
  {Scuflaire}}]{mad}
{Dupret}, M.-A., {De Ridder}, J., {Neuforge}, C., {Aerts}, C., \& {Scuflaire},
  R. 2002, \aap, 385, 563

\bibitem[{{Dziembowski} \& {Pamyatnykh}(2008)}]{nu_eri3}
{Dziembowski}, W.~A., \& {Pamyatnykh}, A.~A. 2008, \mnras, 385, 2061

\bibitem[{{Flower}(1996)}]{flower}
{Flower}, P.~J. 1996, \apj, 469, 355

\bibitem[{{Handler} {et~al.}(2006){Handler}, {Jerzykiewicz},
  {Rodr{\'{\i}}guez}, {Uytterhoeven}, {Amado}, {Dorokhova}, {Dorokhov},
  {Poretti}, {Sareyan}, {Parrao}, {Lorenz}, {Zsuffa}, {Drummond},
  {Daszy{\'n}ska-Daszkiewicz}, {Verhoelst}, {De Ridder}, {Acke}, {Bourge},
  {Movchan}, {Garrido}, {Papar{\'o}}, {Sahin}, {Antoci}, {Udovichenko},
  {Csorba}, {Crowe}, {Berkey}, {Stewart}, {Terry}, {Mkrtichian}, \&
  {Aerts}}]{12lac1}
{Handler}, G., {Jerzykiewicz}, M., {Rodr{\'{\i}}guez}, E., {et~al.} 2006,
  \mnras, 365, 327

\bibitem[{{Handler} {et~al.}(2009){Handler}, {Matthews}, {Eaton},
  {Daszy{\'n}ska-Daszkiewicz}, {Kuschnig}, {Lehmann}, {Rodr{\'{\i}}guez},
  {Pamyatnykh}, {Zdravkov}, {Lenz}, {Costa}, {D{\'{\i}}az-Fraile}, {Sota},
  {Kwiatkowski}, {Schwarzenberg-Czerny}, {Borczyk}, {Dimitrov}, {Fagas},
  {Kami{\'n}ski}, {Ro{\.z}ek}, {van Wyk}, {Pollard}, {Kilmartin}, {Weiss},
  {Guenther}, {Moffat}, {Rucinski}, {Sasselov}, \& {Walker}}]{gammapeg}
{Handler}, G., {Matthews}, J.~M., {Eaton}, J.~A., {et~al.} 2009, \apjl, 698,
  L56

\bibitem[{{Handler} \& {Meingast}(2011)}]{handler_2011}
{Handler}, G., \& {Meingast}, S. 2011, \aap, 533, A70

\bibitem[{{Handler} {et~al.}(2004){Handler}, {Shobbrook}, {Jerzykiewicz},
  {Krisciunas}, {Tshenye}, {Rodr{\'{\i}}guez}, {Costa}, {Zhou}, {Medupe},
  {Phorah}, {Garrido}, {Amado}, {Papar{\'o}}, {Zsuffa}, {Ramokgali}, {Crowe},
  {Purves}, {Avila}, {Knight}, {Brassfield}, {Kilmartin}, \&
  {Cottrell}}]{nueri_phot}
{Handler}, G., {Shobbrook}, R.~R., {Jerzykiewicz}, M., {et~al.} 2004, \mnras,
  347, 454

\bibitem[{{Handler} {et~al.}(2005){Handler}, {Shobbrook}, \&
  {Mokgwetsi}}]{thetaop1}
{Handler}, G., {Shobbrook}, R.~R., \& {Mokgwetsi}, T. 2005, \mnras, 362, 612

\bibitem[{{Handler} {et~al.}(2012){Handler}, {Shobbrook}, {Uytterhoeven},
  {Briquet}, {Neiner}, {Tshenye}, {Ngwato}, {van Winckel}, {Guggenberger},
  {Raskin}, {Rodr{\'{\i}}guez}, {Mazumdar}, {Barban}, {Lorenz},
  {Vandenbussche}, {{\c S}ahin}, {Medupe}, \& {Aerts}}]{handler2012}
{Handler}, G., {Shobbrook}, R.~R., {Uytterhoeven}, K., {et~al.} 2012, \mnras,
  424, 2380

\bibitem[{{Handler} {et~al.}(2003){Handler}, {Shobbrook}, {Vuthela}, {Balona},
  {Rodler}, \& {Tshenye}}]{handler_3betacep}
{Handler}, G., {Shobbrook}, R.~R., {Vuthela}, F.~F., {et~al.} 2003, \mnras,
  341, 1005

\bibitem[{{Handler} {et~al.}(2008){Handler}, {Tuvikene}, {Lorenz}, {Shobbrook},
  {Saesen}, {Provencal}, {Pagani}, {Quint}, {Desmet}, {Sterken}, {Kanaan}, \&
  {Aerts}}]{handler_2008}
{Handler}, G., {Tuvikene}, T., {Lorenz}, D., {et~al.} 2008, Communications in
  Asteroseismology, 157, 315

\bibitem[{{Hekker} {et~al.}(2011){Hekker}, {Basu}, {Stello}, {Kallinger},
  {Grundahl}, {Mathur}, {Garc{\'{\i}}a}, {Mosser}, {Huber}, {Bedding},
  {Szab{\'o}}, {De Ridder}, {Chaplin}, {Elsworth}, {Hale},
  {Christensen-Dalsgaard}, {Gilliland}, {Still}, {McCauliff}, \&
  {Quintana}}]{hekker_clusters}
{Hekker}, S., {Basu}, S., {Stello}, D., {et~al.} 2011, \aap, 530, A100

\bibitem[{{Heynderickx} {et~al.}(1994){Heynderickx}, {Waelkens}, \&
  {Smeyers}}]{heynderickx}
{Heynderickx}, D., {Waelkens}, C., \& {Smeyers}, P. 1994, \aaps, 105, 447

\bibitem[{{Huang} \& {Gies}(2006)}]{huang}
{Huang}, W., \& {Gies}, D.~R. 2006, \apj, 648, 591

\bibitem[{{Jerzykiewicz} {et~al.}(2011){Jerzykiewicz}, {Kopacki}, {Pigulski},
  {Ko{\l}aczkowski}, \& {Kim}}]{jerzykiewicz}
{Jerzykiewicz}, M., {Kopacki}, G., {Pigulski}, A., {Ko{\l}aczkowski}, Z., \&
  {Kim}, S.-L. 2011, \actaa, 61, 247

\bibitem[{{Krzesi\'nski}(1998)}]{krzesinski1}
{Krzesi\'nski}, J. 1998, in Astronomical Society of the Pacific Conference
  Series, Vol. 135, A Half Century of Stellar Pulsation Interpretation, ed.
  P.~A. {Bradley} \& J.~A. {Guzik}, 157

\bibitem[{{Krzesi\'nski} \& {Pigulski}(1997)}]{ngc884_1}
{Krzesi\'nski}, J., \& {Pigulski}, A. 1997, \aap, 325, 987

\bibitem[{{Kunzli} {et~al.}(1997){Kunzli}, {North}, {Kurucz}, \&
  {Nicolet}}]{kunzli}
{Kunzli}, M., {North}, P., {Kurucz}, R.~L., \& {Nicolet}, B. 1997, \aaps, 122,
  51

\bibitem[{{Lenz} \& {Breger}(2005)}]{period04}
{Lenz}, P., \& {Breger}, M. 2005, Communications in Asteroseismology, 146, 53

\bibitem[{{Maeder}(2009)}]{maeder2009}
{Maeder}, A. 2009, {Physics, Formation and Evolution of Rotating Stars}
  ({Heidelberg: Springer})

\bibitem[{{Majewska} {et~al.}(2008){Majewska}, {Pigulski}, \&
  {Rucinski}}]{majewska}
{Majewska}, A., {Pigulski}, A., \& {Rucinski}, S.~M. 2008, Communications in
  Asteroseismology, 157, 338

\bibitem[{{Malchenko}(2007)}]{oo2371}
{Malchenko}, S.~L. 2007, in 14th Young Scientists Conference on Astronomy and
  Space Physics, ed. G.~{Ivashchenko} \& A.~{Golovin}, 59--63

\bibitem[{{Marsh Boyer} {et~al.}(2012){Marsh Boyer}, {McSwain}, {Aragona}, \&
  {Ou-Yang}}]{marsh2012}
{Marsh Boyer}, A.~N., {McSwain}, M.~V., {Aragona}, C., \& {Ou-Yang}, B. 2012,
  \aj, 144, 158

\bibitem[{{Meylan} \& {Hauck}(1981)}]{meylan}
{Meylan}, G., \& {Hauck}, B. 1981, \aaps, 46, 281

\bibitem[{{Michalska} {et~al.}(2009){Michalska}, {Pigulski}, {St{\c e}licki},
  \& {Narwid}}]{michalska}
{Michalska}, G., {Pigulski}, A., {St{\c e}licki}, M., \& {Narwid}, A. 2009,
  \actaa, 59, 349

\bibitem[{{Miglio} {et~al.}(2012){Miglio}, {Brogaard}, {Stello}, {Chaplin},
  {D'Antona}, {Montalb{\'a}n}, {Basu}, {Bressan}, {Grundahl}, {Pinsonneault},
  {Serenelli}, {Elsworth}, {Hekker}, {Kallinger}, {Mosser}, {Ventura},
  {Bonanno}, {Noels}, {Silva Aguirre}, {Szabo}, {Li}, {McCauliff}, {Middour},
  \& {Kjeldsen}}]{miglio_clusters}
{Miglio}, A., {Brogaard}, K., {Stello}, D., {et~al.} 2012, \mnras, 419, 2077

\bibitem[{{Miglio} {et~al.}(2008){Miglio}, {Montalb{\'a}n}, {Eggenberger}, \&
  {Noels}}]{miglio2008}
{Miglio}, A., {Montalb{\'a}n}, J., {Eggenberger}, P., \& {Noels}, A. 2008,
  Astronomische Nachrichten, 329, 529

\bibitem[{{Montalb{\'a}n} {et~al.}(2008){Montalb{\'a}n}, {Miglio},
  {Eggenberger}, \& {Noels}}]{montalban2008}
{Montalb{\'a}n}, J., {Miglio}, A., {Eggenberger}, P., \& {Noels}, A. 2008,
  Astronomische Nachrichten, 329, 535

\bibitem[{{Montgomery} \& {O'Donoghue}(1999)}]{montgomery}
{Montgomery}, M.~H., \& {O'Donoghue}, D. 1999, Delta Scuti Star Newsletter, 13,
  28

\bibitem[{{Morel} {et~al.}(2006){Morel}, {Butler}, {Aerts}, {Neiner}, \&
  {Briquet}}]{morel}
{Morel}, T., {Butler}, K., {Aerts}, C., {Neiner}, C., \& {Briquet}, M. 2006,
  \aap, 457, 651

\bibitem[{{Percy}(1972)}]{percy}
{Percy}, J.~R. 1972, \pasp, 84, 420

\bibitem[{{Pigulski} {et~al.}(2007){Pigulski}, {Handler}, {Michalska},
  {Ko{\l}aczkowski}, {Kopacki}, {Narwid}, {Vanhollebeke}, {St\c{e}{\'s}licki},
  {Lefever}, {Gazeas}, {de Meester}, {Vanautgaerden}, {Leitner}, {de Ridder},
  {van Helshoecht}, {Gielen}, {Vandenbussche}, {Saesen}, {Reed}, {Eggen},
  {Gelven}, {Desmet}, {Puga Antol{\'{\i}}n}, {Aerts}, {Schmidt}, {Huygen},
  {Lorenz}, {Vu{\v c}kovi{\'c}}, {Broeders}, {Bauwens}, {Verhoelst}, {Deroo},
  {Lenz}, {Dehaes}, {Ladjal}, {Steininger}, {Davignon}, {Beck}, {Yakut},
  {Drummond}, {Fu}, {Jiang}, {Zhang}, {Provencal}, \& {Decin}}]{ngc6910_2}
{Pigulski}, A., {Handler}, G., {Michalska}, G., {et~al.} 2007, Communications
  in Asteroseismology, 150, 191

\bibitem[{{Poretti} \& {Zerbi}(1993)}]{redspm}
{Poretti}, E., \& {Zerbi}, F. 1993, \aap, 268, 369

\bibitem[{{Rufener}(1964)}]{redp7_1}
{Rufener}, F. 1964, Publ.~Obs.~Gen{\`e}ve, Ser.~A, 66, 413

\bibitem[{{Rufener}(1985)}]{redp7_2}
{Rufener}, F. 1985, in IAU Symposium, Vol. 111, Calibration of Fundamental
  Stellar Quantities, ed. D.~S. {Hayes}, L.~E. {Pasinetti}, \& A.~G.~D.
  {Philip}, 253--268

\bibitem[{{Saesen} {et~al.}(2009){Saesen}, {Carrier}, {Pigulski}, {Aerts},
  {Handler}, {Narwid}, {Fu}, {Zhang}, {Jiang}, {Kopacki}, {Vanautgaerden},
  {Ste{\'s}licki}, {Acke}, {Poretti}, {Uytterhoeven}, {de Meester}, {Reed},
  {Ko{\l}aczkowski}, {Michalska}, {Schmidt}, {{\"O}stensen}, {Gielen}, {Yakut},
  {Leitner}, {Kalomeni}, {Prins}, {van Helshoecht}, {Zima}, {Huygen},
  {Vandenbussche}, {Lenz}, {Ladjal}, {Puga Antol{\'{\i}}n}, {Verhoelst}, {de
  Ridder}, {Niarchos}, {Liakos}, {Lorenz}, {Dehaes}, {Reyniers}, {Davignon},
  {Kim}, {Kim}, {Lee}, {Lee}, {Kwon}, {Broeders}, {van Winckel},
  {Vanhollebeke}, {Raskin}, {Blom}, {Eggen}, {Beck}, {Puschnig},
  {Schmitzberger}, {Gelven}, {Steininger}, \& {Drummond}}]{ngc884_3}
{Saesen}, S., {Carrier}, F., {Pigulski}, A., {et~al.} 2009, Communications in
  Asteroseismology, 158, 179

\bibitem[{{Saesen} {et~al.}(2010){Saesen}, {Carrier}, {Pigulski}, {Aerts},
  {Handler}, {Narwid}, {Fu}, {Zhang}, {Jiang}, {Vanautgaerden}, {Kopacki},
  {St{\c e}{\'s}licki}, {Acke}, {Poretti}, {Uytterhoeven}, {Gielen},
  {{\O}stensen}, {De Meester}, {Reed}, {Ko{\l}aczkowski}, {Michalska},
  {Schmidt}, {Yakut}, {Leitner}, {Kalomeni}, {Cherix}, {Spano}, {Prins}, {van
  Helshoecht}, {Zima}, {Huygen}, {Vandenbussche}, {Lenz}, {Ladjal}, {Puga
  Antol{\'{\i}}n}, {Verhoelst}, {De Ridder}, {Niarchos}, {Liakos}, {Lorenz},
  {Dehaes}, {Reyniers}, {Davignon}, {Kim}, {Kim}, {Lee}, {Lee}, {Kwon},
  {Broeders}, {van Winckel}, {Vanhollebeke}, {Waelkens}, {Raskin}, {Blom},
  {Eggen}, {Degroote}, {Beck}, {Puschnig}, {Schmitzberger}, {Gelven},
  {Steininger}, {Blommaert}, {Drummond}, {Briquet}, \& {Debosscher}}]{saesenI}
{Saesen}, S., {Carrier}, F., {Pigulski}, A., {et~al.} 2010, \aap, 515, A16

\bibitem[{{Saesen} {et~al.}(2008){Saesen}, {Pigulski}, {Carrier}, {DeRidder},
  {Aerts}, {Handler}, {Narwid}, {Fu}, {Zhang}, {Jiang}, {Kopacki},
  {Vanautgaerden}, {St{\c e}{\'s}licki}, {Acke}, {Poretti}, {Uytterhoeven},
  {DeMeester}, {Reed}, {Ko{\l}aczkowski}, {Michalska}, {Schmidt},
  {{\O}stensen}, {Gielen}, {Yakut}, {Leitner}, {Kalomeni}, {Prins}, {Van
  Helshoecht}, {Zima}, {Huygen}, {Vandenbussche}, {Lenz}, {Ladjal}, {Puga
  Antol{\'{\i}}n}, {Verhoelst}, {Niarchos}, {Liakos}, {Lorenz}, {Dehaes},
  {Reyniers}, {Davignon}, {Kim}, {Kim}, {Lee}, {Lee}, {Kwon}, {Broeders}, {Van
  Winckel}, {Vanhollebeke}, {Raskin}, {Blom}, {Eggen}, {Beck}, {Puschnig},
  {Schmitt}, {Gelven}, {Steiniger}, \& {Drummond}}]{ngc884_2}
{Saesen}, S., {Pigulski}, A., {Carrier}, F., {et~al.} 2008, Journal of Physics
  Conference Series, 118, 012071

\bibitem[{{Schwarzenberg-Czerny}(1991)}]{schwarzenberg}
{Schwarzenberg-Czerny}, A. 1991, \mnras, 253, 198

\bibitem[{{Scuflaire} {et~al.}(2008{\natexlab{a}}){Scuflaire}, {Montalb{\'a}n},
  {Th{\'e}ado}, {Bourge}, {Miglio}, {Godart}, {Thoul}, \& {Noels}}]{osc}
{Scuflaire}, R., {Montalb{\'a}n}, J., {Th{\'e}ado}, S., {et~al.}
  2008{\natexlab{a}}, \apss, 316, 149

\bibitem[{{Scuflaire} {et~al.}(2008{\natexlab{b}}){Scuflaire}, {Th{\'e}ado},
  {Montalb{\'a}n}, {Miglio}, {Bourge}, {Godart}, {Thoul}, \& {Noels}}]{cles}
{Scuflaire}, R., {Th{\'e}ado}, S., {Montalb{\'a}n}, J., {et~al.}
  2008{\natexlab{b}}, \apss, 316, 83

\bibitem[{{Seaton}(2005)}]{OPopac}
{Seaton}, M.~J. 2005, \mnras, 362, L1

\bibitem[{{Slesnick} {et~al.}(2002){Slesnick}, {Hillenbrand}, \&
  {Massey}}]{slesnick}
{Slesnick}, C.~L., {Hillenbrand}, L.~A., \& {Massey}, P. 2002, \apj, 576, 880

\bibitem[{{Southworth} {et~al.}(2004{\natexlab{a}}){Southworth}, {Maxted}, \&
  {Smalley}}]{southworth_hper}
{Southworth}, J., {Maxted}, P.~F.~L., \& {Smalley}, B. 2004{\natexlab{a}},
  \mnras, 349, 547

\bibitem[{{Southworth} {et~al.}(2004{\natexlab{b}}){Southworth}, {Zucker},
  {Maxted}, \& {Smalley}}]{southworth_chiper}
{Southworth}, J., {Zucker}, S., {Maxted}, P.~F.~L., \& {Smalley}, B.
  2004{\natexlab{b}}, \mnras, 355, 986

\bibitem[{{Stankov} \& {Handler}(2005)}]{catalog_betacep}
{Stankov}, A., \& {Handler}, G. 2005, \apjs, 158, 193

\bibitem[{{Stello} {et~al.}(2011){Stello}, {Huber}, {Kallinger}, {Basu},
  {Mosser}, {Hekker}, {Mathur}, {Garc{\'{\i}}a}, {Bedding}, {Kjeldsen},
  {Gilliland}, {Verner}, {Chaplin}, {Benomar}, {Meibom}, {Grundahl},
  {Elsworth}, {Molenda-{\.Z}akowicz}, {Szab{\'o}}, {Christensen-Dalsgaard},
  {Tenenbaum}, {Twicken}, \& {Uddin}}]{Stello2011}
{Stello}, D., {Huber}, D., {Kallinger}, T., {et~al.} 2011, \apjl, 737, L10

\bibitem[{{Strom} {et~al.}(2005){Strom}, {Wolff}, \& {Dror}}]{strom}
{Strom}, S.~E., {Wolff}, S.~C., \& {Dror}, D.~H.~A. 2005, \aj, 129, 809

\bibitem[{{Telting} {et~al.}(2001){Telting}, {Abbott}, \&
  {Schrijvers}}]{telting_psiori}
{Telting}, J.~H., {Abbott}, J.~B., \& {Schrijvers}, C. 2001, \aap, 377, 104

\bibitem[{{Telting} \& {Schrijvers}(1998)}]{telting_omegasco}
{Telting}, J.~H., \& {Schrijvers}, C. 1998, \aap, 339, 150

\bibitem[{{Torres}(2010)}]{torres}
{Torres}, G. 2010, \aj, 140, 1158

\bibitem[{{Uribe} {et~al.}(2002){Uribe}, {Garc{\'{\i}}a-Varela},
  {Sabogal-Mart{\'{\i}}nez}, {Higuera G.}, \& {Brieva}}]{uribe}
{Uribe}, A., {Garc{\'{\i}}a-Varela}, J.-A., {Sabogal-Mart{\'{\i}}nez}, B.-E.,
  {Higuera G.}, M.~A., \& {Brieva}, E. 2002, \pasp, 114, 233

\bibitem[{{Waelkens} {et~al.}(1990){Waelkens}, {Lampens}, {Heynderickx},
  {Cuypers}, {Degryse}, {Poedts}, {Polfliet}, {Denoyelle}, {van den Abeele},
  {Rufener}, \& {Smeyers}}]{waelkens_ngc884}
{Waelkens}, C., {Lampens}, P., {Heynderickx}, D., {et~al.} 1990, \aaps, 83, 11

\end{thebibliography}

\newpage
\section{Online Material}
\tabcolsep=4pt
\addtocounter{table}{-2}
\begin{table*}
\footnotesize
\caption{Results of the multi-frequency fit to the $U$, $B$, $V$ and $I$ light
curves of all B-stars.} 
\centering\begin{tabular}{l || lll | lll | lll| lll | c}
\tableline
~ & \multicolumn{3}{c|}{$U$} & \multicolumn{3}{c|}{$B$} & \multicolumn{3}{c|}{$V$} & \multicolumn{3}{c|}{$I$} &  \\
$f_i$ & $A_i$ & $\phi_i$ & S/N & $A_i$ & $\phi_i$ & S/N & $A_i$ & $\phi_i$ & S/N & $A_i$ & $\phi_i$ & S/N & $\ell$ \\
(d$^{-1}$) & (mmag) & (rad) & ~ & (mmag) & (rad) & ~ & (mmag) & (rad) & ~ & (mmag) & (rad) & ~ &  \\
\tableline
\multicolumn{14}{l}{}\\

\multicolumn{14}{l}{\textbf{{Oo~1898}}}\\
/ & \nodata & \nodata & \nodata & \nodata & \nodata & \nodata & \nodata & \nodata & \nodata & \nodata & \nodata & \nodata & \nodata \\
\multicolumn{14}{l}{}\\

\multicolumn{14}{l}{\textbf{{Oo~1973}}}\\
/ & \nodata & \nodata & \nodata & \nodata & \nodata & \nodata & \nodata & \nodata & \nodata & \nodata & \nodata & \nodata & \nodata \\
\multicolumn{14}{l}{}\\

\multicolumn{14}{l}{\textbf{{Oo~1980}}}\\
$f_1 = 6.2656(2)$ & \nodata & \nodata & \nodata & 3(8) & 0.9(4) & 3.1 & 4(1) & 0.91(6) & 7.3 & 4(7) & 0.9(3) & 3.9 & 0,1,2,3,\textbf{4}\\
\multicolumn{14}{l}{}\\

\multicolumn{14}{l}{\textbf{{Oo~1990}}}\\
$f_1 = 2.9742(2)$ & \nodata & \nodata & \nodata & 7(8) & 0.0(2) & 3.2 & 5(1) & 0.01(4) & 7.1 & 4(8) & 0.0(3) & 3.4 & 1,\textbf{2},3,\textbf{4}\\
\multicolumn{14}{l}{}\\

\multicolumn{14}{l}{\textbf{{Oo~2006}}}\\
$f_1 = 2.7032(2)$ & \nodata & \nodata & \nodata & 3(3) & 0.6(2) & \nodata & 2.9(6) & 0.52(4) & 4.3 & 2(3) & 0.7(3) & \nodata & \textbf{1},2,\textbf{3},4\\
\multicolumn{14}{l}{}\\

\multicolumn{14}{l}{\textbf{{Oo~2019}}}\\
$f_1 = 5.1088(7)$ & \nodata & \nodata & \nodata & 4(5) & 0.7(2) & 3.7 & 4.2(8) & 0.65(3) & 4.9 & 3(6) & 0.7(3) & \nodata & \textbf{2},3,4\\
\multicolumn{14}{l}{}\\

\multicolumn{14}{l}{\textbf{{Oo~2037}}}\\
$f_1 = 0.16371(4)$ & \nodata & \nodata & \nodata & 20(5) & 0.90(4) & 3.0 & 18(1) & 0.94(1) & 4.5 & 18(4) & 0.92(4) & \nodata & 1,2,\textbf{4}\\
\multicolumn{14}{l}{}\\

\multicolumn{14}{l}{\textbf{Oo~2086}}\\
$f_1 = 25.7444(9)$ & \nodata & \nodata & \nodata & \nodata & \nodata & \nodata & 2.5(7) & 0.34(4) & 4.8 & \nodata & \nodata & \nodata & \nodata \\
$f_2 = 25.2747(9)$ & \nodata & \nodata & \nodata & \nodata & \nodata & \nodata & 2.5(7) & 0.09(4) & 6.0 & \nodata & \nodata & \nodata & \nodata \\
$f_3 = 29.448(1)  $ & \nodata & \nodata & \nodata & \nodata & \nodata & \nodata & 1.7(7) & 0.07(6) & 5.3 & \nodata & \nodata & \nodata & \nodata \\
\multicolumn{14}{l}{}\\

\multicolumn{14}{l}{\textbf{Oo~2089}}\\
$f_1 = 0.8832(6)$ & \nodata & \nodata & \nodata & 8(6) & 0.5(1) & 3.0 & 7(1) & 0.56(3) & 4.0 & 3(6) &
0.5(3) & \nodata & 1,2,3,\textbf{4}\\
$f_2 = 0.963(1)$  & \nodata & \nodata & \nodata & 5(7) & 0.8(2) &    \nodata   & 4(1) & 0.87(5) & 4.1 & 4(7) &
0.8(2) & \nodata & \textbf{1},\textbf{2},\textbf{3},4\\
$f_3 = 1.284(1)$  & \nodata & \nodata & \nodata & 5(6) & 1.0(2) &    \nodata   & 3(1) & 0.87(6) & 4.9 & 2(5) &
0.1(4) & \nodata & 1,\textbf{2},\textbf{3},\textbf{4}\\
\multicolumn{14}{l}{}\\

\multicolumn{14}{l}{\textbf{Oo~2091}}\\
$f_1 = 2.51102(6)$ & \nodata & \nodata & \nodata & 3.1(6) & 0.46(3) & 3.6 & 2.5(2) & 0.40(1) & 5.1 & 
1.9(3) & 0.42(2) & 4.4 & \textbf{1},2,\textbf{3},4\\
$f_2 = 4.8018(1)$   & \nodata & \nodata & \nodata & 1.8(6) & 0.41(6) & 3.8 & 1.1(2) & 0.38(3) & 4.4 & 
1.0(3) & 0.38(5) & 3.7 & 0,\textbf{1},\textbf{2},3,4\\
\multicolumn{14}{l}{}\\

\tableline
\end{tabular}
\tablecomments{The signal is written as $C+\sum_{i=1}^N A_i \sin(2\pi f_i
(t-t_0) +2 \pi \phi_i)$ with $t_0$=HJD 2453000. $A_i$ stands for amplitude and
is expressed in millimag, $2 \pi \phi_i$ is the phase expressed in radians.
Uncertainties in units of the last given digit on the amplitudes and phases are
denoted in brackets. We also indicate the signal-to-noise {level} of the peak,
if it is larger than 3. In the last column we note the mode identification possibilities of the degree $\ell$, based
on elimination of degrees from 0 to 4. When specific degrees were preferred on
the basis of their $\chi^2$-values, we put its value in bold. When none of
the degree possibilities remained, we note this by '/'. A {blank} line means we
had insufficient information for mode identification.}
\end{table*}


\addtocounter{table}{-1}
\begin{table*}
\footnotesize
\caption{Continued.}
\centering\begin{tabular}{l || lll | lll | lll| lll | c}
\tableline
~ & \multicolumn{3}{c|}{$U$} & \multicolumn{3}{c|}{$B$} & \multicolumn{3}{c|}{$V$} & \multicolumn{3}{c|}{$I$} &  \\
$f_i$ & $A_i$ & $\phi_i$ & S/N & $A_i$ & $\phi_i$ & S/N & $A_i$ & $\phi_i$ & S/N & $A_i$ & $\phi_i$ & S/N & $\ell$ \\
(d$^{-1}$) & (mmag) & (rad) & ~ & (mmag) & (rad) & ~ & (mmag) & (rad) & ~ & (mmag) & (rad) & ~ &  \\
\tableline
\multicolumn{14}{l}{}\\

\multicolumn{14}{l}{\textbf{Oo~2094}}\\
$f_1 = 0.37596(4)$ & 7(1) & 0.49(3) & 3.7 & 4.9(5) & 0.46(2) & 5.0 & 4.5(2) &
0.439(9) & 4.1 & 4.3(5) & 0.46(2) & 4.8 &1,2,{4}\\
$f_2 = 2f_1$       & 2(1) & 0.4(1)  &   \nodata    & 2.2(5) & 0.47(4) & 3.9 & 1.3(2) &
0.24(3)  & 3.2 & 1.6(5) & 0.37(5) &    \nodata   & \nodata \\
\multicolumn{14}{l}{}\\

\multicolumn{14}{l}{\textbf{Oo~2110}}\\
$f_1 = 2.6004(2)$ & 3(2) & 1.0(1) & 3.6 & 1.9(9) & 0.91(7) & 4.3 & 1.9(4) &
0.92(4) & 4.5 & 2.0(7) & 0.90(5) & 3.9 & 1,\textbf{2},3,4\\
$f_2 = 5.4289(2)$ & 1(2) & 0.1(3) &   \nodata    & 1.3(9) & 0.1(1)  & 4.4 & 1.4(4) &
0.16(5) & 3.9 & 1.3(6) & 0.16(8) & 4.3 & 1,\textbf{2},3,\textbf{4}\\
\multicolumn{14}{l}{}\\

\multicolumn{14}{l}{\textbf{Oo~2114}}\\
$f_1 = 0.34023(2)$ & 6.9(8) & 0.16(2) & 4.5 & 9.1(4) & 0.260(7) & 5.3 & 7.1(2) &
0.231(4) & 5.7 & 6.2(3) & 0.256(7) & 5.8 & /\\
$f_2 = 7.18922(9)$ & 1.3(8) & 0.4(1)   & 3.1 & 1.6(4) & 0.36(4)   & 5.9 &
1.5(2) & 0.35(2)   & 8.0 & 1.1(3) & 0.35(4)   & 4.8 & \textbf{2},3,4\\
$f_3 = 0.26925(5)$ & 2.3(8) & 0.09(6) &   \nodata      & 3.7(4) & 0.15(2)   & 3.9 &
2.7(2) & 0.14(1)   & 4.1 & 2.0(3) & 0.15(2)   & 3.5 & /\\
\multicolumn{14}{l}{}\\

\multicolumn{14}{l}{\textbf{{Oo~2116}}}\\
$f_1 = 3.2130(3)$ & 2(5) & 0.3(5) & 5.4 & 2(2) & 0.3(1) & 5.4 & 1.6(8) & 0.23(8) & 6.1 & 1(1) & 0.2(1) & 4.0 & 1,2,\textbf{3},4\\
\multicolumn{14}{l}{}\\
 
\multicolumn{14}{l}{\textbf{{Oo~2139}}}\\
/ & \nodata & \nodata & \nodata & \nodata & \nodata & \nodata & \nodata & \nodata & \nodata & \nodata  & \nodata & \nodata & \nodata\\
\multicolumn{14}{l}{}\\

\multicolumn{14}{l}{\textbf{{Oo~2141}}}\\
$f_1 = 0.20871(9)$ & \nodata & \nodata & \nodata & 52(7) & 0.54(3) & 3.4 & 56(2) & 0.519(6) & 5.1 & 41(7) & 0.50(3) & 3.4 & 1,2,4\\
$f_2 = 2 f_1$            & \nodata & \nodata & \nodata & 11(7) & 0.1(1)    &     \nodata    & 12(2) & 0.17(3)   & 5.8 & 5(7) & 0.0(2) & \nodata & \nodata \\
\multicolumn{14}{l}{}\\

\multicolumn{14}{l}{\textbf{{Oo~2146}}}\\
/ & \nodata & \nodata & \nodata & \nodata & \nodata & \nodata & \nodata & \nodata & \nodata & \nodata  & \nodata & \nodata & \nodata\\
\multicolumn{14}{l}{}\\

\multicolumn{14}{l}{\textbf{{Oo~2151}}}\\
$f_1 = 5.590(3)$ & \nodata & \nodata & \nodata & 2(13) & 0.0(1) & \nodata & 2(2) & 0.6(2) & 4.8 & 4(9) & 0.5(3) & \nodata & 1,2,3,\textbf{4}\\
\multicolumn{14}{l}{}\\

\multicolumn{14}{l}{\textbf{Oo~2185}}\\
$f_1 = 0.62466(3)$ & 0.08(13) & 0.8(3)   & \nodata & 1.58(8) & 0.593(8) &  \nodata       &
1.27(3) & 0.675(5) & 4.2 & 0.50(6) & 0.59(2) & \nodata & /\\
$f_2 = 4.87975(4)$ & 0.3(1)     & 0.75(7) & \nodata & 1.35(8) & 0.936(9) & 3.1 & 0.87(4)
& 0.867(6) & 4.6 & 0.22(6) & 0.79(4) & \nodata & /\\
\multicolumn{14}{l}{}\\

\multicolumn{14}{l}{\textbf{{Oo~2189}}}\\
/ & \nodata & \nodata & \nodata & \nodata & \nodata & \nodata & \nodata & \nodata & \nodata & \nodata  & \nodata & \nodata & \nodata\\
\multicolumn{14}{l}{}\\

\multicolumn{14}{l}{\textbf{Oo~2191}}\\
$f_1 = 0.32811(4)$ & \nodata & \nodata & \nodata & 4.6(7) & 0.16(2) & 4.1 & 4.2(2) & 0.158(8) & 4.5 &
4.6(3) & 0.17(1) & 4.8 & 4\\
$f_2 = 2.5447(1) $  & \nodata & \nodata & \nodata & 1.0(7) & 0.3(1)   &     \nodata    & 1.3(2) & 0.27(3)   & 4.0
& 0.8(3) & 0.22(6) &    \nodata     & {1},2,\textbf{3}\\
\multicolumn{14}{l}{}\\

\multicolumn{14}{l}{\textbf{Oo~2228}}\\
$f_1 = 3.7993(5)$ & \nodata & \nodata & \nodata & \nodata & \nodata & \nodata & 1.2(9) & 0.0(1) & 4.5 & 1(1) & 0.0(2) & 3.6
& {1},\textbf{2},\textbf{3},4\\
\multicolumn{14}{l}{}\\

\tableline
\end{tabular}\end{table*}

\addtocounter{table}{-1}
\begin{table*}
\footnotesize
\caption{Continued.}
\centering\begin{tabular}{l || lll | lll | lll| lll | c}
\tableline
~ & \multicolumn{3}{c|}{$U$} & \multicolumn{3}{c|}{$B$} & \multicolumn{3}{c|}{$V$} & \multicolumn{3}{c|}{$I$} &  \\
$f_i$ & $A_i$ & $\phi_i$ & S/N & $A_i$ & $\phi_i$ & S/N & $A_i$ & $\phi_i$ & S/N & $A_i$ & $\phi_i$ & S/N & $\ell$ \\
(d$^{-1}$) & (mmag) & (rad) & ~ & (mmag) & (rad) & ~ & (mmag) & (rad) & ~ & (mmag) & (rad) & ~ &  \\
\tableline
\multicolumn{14}{l}{}\\

\multicolumn{14}{l}{\textbf{{Oo~2235}}}\\
/ & \nodata & \nodata & \nodata & \nodata & \nodata & \nodata & \nodata & \nodata & \nodata & \nodata  & \nodata & \nodata & \nodata\\
\multicolumn{14}{l}{}\\

\multicolumn{14}{l}{\textbf{Oo~2242}}\\
$f_1 = 1.56904(1)$ & 17(2) & 0.76(2) & \nodata & 16(1) & 0.78(1) & 4.6 & 12.9(6) &
0.773(8) & 4.8 & 8.8(3) & 0.836(6) &   \nodata      & 1,\textbf{3}\\
$f_2 = 3.12734(2)$ & 10(2) & 0.49(3) & \nodata & 7(1)   & 0.55(4) & 3.0 & 8.6(6)   &
0.57(1)   & 4.1 & 6.5(3) & 0.561(8) & 3.1 & 2\\
\multicolumn{14}{l}{}\\

\multicolumn{14}{l}{\textbf{Oo~2245}}\\
$f_1 = 5.4416(1)$ & 2(2) & 0.6(2) & 3.5 & 2.7(9) & 0.53(5) & 5.5 & 2.6(4) &
0.55(3) & 5.5 & 2.0(9) & 0.58(7) & 4.9 & \textbf{2},3,4\\
$f_2 = 2.6701(1)$ & 3(2) & 0.5(1) & 4.0 & 2.4(9) & 0.50(6) & 3.7 & 1.6(4) &
0.44(3) & 4.4 & 1.8(9) & 0.49(8) & 3.5 & 1,\textbf{2},3,\textbf{4}\\
\multicolumn{14}{l}{}\\

\multicolumn{14}{l}{\textbf{Oo~2246}}\\
$f_1 = 5.429245(7)$ & 6(1) & 0.17(3) & 5.5 & 6.0(3) & 0.169(9) & 8.6 & 5.7(1) &
0.168(4) & 8.9 & 5.1(2) & 0.166(7) & 7.3 & 2\\
$f_2 = 5.85616(1)$   & 3(1) & 0.01(8) & 3.6 & 3.3(3) & 0.07(2)   & 8.7 & 3.2(1)
& 0.078(7) & 8.9 & 3.4(2) & 0.07(1)   & 8.9 & 4\\
$f_3 = 5.02348(4)$   & 2(1) & 0.5(1)   & 3.2 & 1.0(3) & 0.51(5)   & 4.2 & 0.6(1)
& 0.51(4)   & 3.7 & 0.7(2) & 0.53(5)   & 3.4 & 0\\
\multicolumn{14}{l}{}\\

\multicolumn{14}{l}{\textbf{Oo~2253}}\\
$f_1 = 0.71442(3)$ & 16(2) & 0.31(2) & 5.9 & 18.4(8) & 0.999(7) & 8.6 & 9.4(4) &
0.054(8) & 6.9 & 6.8(6) & 0.30(1) & 6.2 & /\\
$f_2 = 2.9481(1)$   & 5(2)   & 0.50(7) & 5.2 & 3.6(8)   & 0.56(4)   & 7.6 &
2.6(4) & 0.56(3)   & 5.3 & 2.4(6) & 0.54(4) & 6.7 & \textbf{1},2,3,4\\
\multicolumn{14}{l}{}\\

\multicolumn{14}{l}{\textbf{{Oo~2262}}}\\
$f_1 = 3.07229(5)$ & \nodata & \nodata & \nodata & 4.4(7) & 0.67(3) & 3.3 & 5.4(3) & 0.69(1) & 4.7 & 6.3(6) & 0.67(2) & 4.5 & 4\\
\multicolumn{14}{l}{}\\

\multicolumn{14}{l}{\textbf{Oo~2267}}\\
$f_1 = 2.6035(1)$ & 5(3) & 0.55(8) & 3.6 & 4(1) & 0.55(6) & 7.0 & 3.4(6) &
0.55(3) & 8.0 & 2.8(9) & 0.56(5) & 6.4 & \textbf{1},\textbf{2},{3},4\\
$f_2 = 5.4396(3)$ & 1(3) & 0.4(5)   &      \nodata   & 1(1) & 0.3(2)   & 4.4 & 1.3(6) &
0.35(8) & 6.4 & 1.3(9) & 0.4(1)   & 5.9 & 1,{2},3,\textbf{4}\\
\multicolumn{14}{l}{}\\

\multicolumn{14}{l}{\textbf{{Oo~2285}}}\\
$f_1 = 0.62307(3) $ & \nodata & \nodata & \nodata & 11(1) & 0.90(2) & 5.7 & 9.4(4) & 0.886(8) & 6.2 & 11.0(8) & 0.89(1) & 5.3 & /\\
$f_2 = 2 f_1 $ & \nodata & \nodata & \nodata & 3(1) & 0.47(5) & 5.0 & 2.8(4) & 0.43(3) & 4.1 & 2.7(7) & 0.48(4) & 4.4 & \nodata \\
\multicolumn{14}{l}{}\\

\multicolumn{14}{l}{\textbf{Oo~2299}}\\
$f_1 = 3.145150(7)$  & 8(1) & 0.41(2) & 6.9 & 6.8(4) & 0.398(9) & 9.4 & 6.3(2)
& 0.375(5) & 9.2 & 6.2(3) & 0.386(7) & 8.6 & 2\\
\multicolumn{14}{l}{}\\

\multicolumn{14}{l}{\textbf{{Oo~2309}}}\\
$f_1 = 5.7408(2)$ & 1(3) & 0.3(3) & \nodata & 2(1) & 0.24(9) & 4.4 & 2.1(5) & 0.28(4) & 6.8 & 1.5(7) & 0.27(8) & 5.1 & 1,2,\textbf{3},4\\
\multicolumn{14}{l}{}\\

\multicolumn{14}{l}{\textbf{Oo~2319}}\\
$f_1 = 3.06471(9)$ &  8(3) & 0.77(5) & 6.9 & 5(1) & 0.75(4) & 7.7 & 5.6(7) &
0.76 (2) & 8.9 & 4(1) & 0.77(4) & 8.0 & \textbf{1},2,3,4\\
$f_2 = 2f_1$            &  1(3) & 0.9(6)   &       & 1(1) & 0.9(2)   &     \nodata    &
1.0(7) & 0.9(1)    & 4.1 & 1(1) & 0.9(1)   & 4.6 & \nodata \\
\multicolumn{14}{l}{}\\

\tableline
\end{tabular}\end{table*}

\addtocounter{table}{-1}
\begin{table*}
\footnotesize
\caption{Continued.}
\centering\begin{tabular}{l || lll | lll | lll| lll | c}
\tableline
~ & \multicolumn{3}{c|}{$U$} & \multicolumn{3}{c|}{$B$} &
\multicolumn{3}{c|}{$V$} & \multicolumn{3}{c|}{$I$} &  \\
$f_i$ & $A_i$ & $\phi_i$ & S/N & $A_i$ & $\phi_i$ & S/N & $A_i$ & $\phi_i$ & S/N
& $A_i$ & $\phi_i$ & S/N & $\ell$ \\
(d$^{-1}$) & (mmag) & (rad) & ~ & (mmag) & (rad) & ~ & (mmag) & (rad) & ~ &
(mmag) & (rad) & ~ &  \\
\tableline
\multicolumn{14}{l}{}\\

\multicolumn{14}{l}{\textbf{Oo~2323}}\\
$f_1 = 3.9740(5)$ & \nodata & \nodata & \nodata & \nodata & \nodata & \nodata & 3(2) & 0.1(1) & 6.6 & \nodata & \nodata & \nodata & \nodata \\
$f_2 = 11.951(1)$ & \nodata & \nodata & \nodata & \nodata & \nodata & \nodata & 1(2) & 0.7(4) & 4.2 & \nodata & \nodata & \nodata & \nodata \\
\multicolumn{14}{l}{}\\

\multicolumn{14}{l}{\textbf{Oo~2324}}\\
$f_1 = 3.6795(2)$ & 8(8) & 1.0(2) & 5.6 & 5(3) & 0.9(1) & 5.5 & 4(1) & 0.94(5) &
6.8 & 4(2) & 0.96(8) & 6.0 & \textbf{1},\textbf{2},3,4\\
$f_2 = 3.9041(4)$ & 5(8) & 0.6(3) & 4.4 & 3(3) & 0.7(2) & 4.3 & 2(1) & 0.7(1)  
& 5.0 & 2(2) & 0.6(1)   & 5.5 & \textbf{1},{2},3,4\\
$f_3 = 3.0066(5)$ & 2(9) & 1.0(7) &   \nodata    & 2(3) & 1.0(2) & 3.5 & 2(1) & 0.9(1) &
5.6 & 0.4(1.9) & 0.9(9) & \nodata   &{1},{2},\textbf{3},4 \\
\multicolumn{14}{l}{}\\

\multicolumn{14}{l}{\textbf{{Oo~2342}}}\\
$f_1 = 2.678(1)$ & \nodata & \nodata & \nodata & 5(9) & 0.2(3) & 3.3 & 5(2) & 0.25(7) & 6.1 & 5(7) & 0.3(2) & \nodata & 1,\textbf{2},3,4\\
\multicolumn{14}{l}{}\\

\multicolumn{14}{l}{\textbf{Oo~2345}}\\
$f_1 = 1.99990(4)$ & 9(3) & 0.11(7) & 7.2 & 12(1) & 0.08(2) & 8.8 & 10.2(6) &
0.05(1) & 8.5 & 10.8(9) & 0.06(2) & 8.0 & \textbf{2},3,4\\
$f_2 = 1.8589(2)$   & 1(3) & 0.9(4)   &      \nodata   & 3(1)   & 0.91(7) & 4.4 & 2.2(6) 
 & 0.99(5) & 4.8 & 2.2(9)   & 0.97(7) & 4.4 & 1,\textbf{2},\textbf{3},\textbf{4}\\
\multicolumn{14}{l}{}\\

\multicolumn{14}{l}{\textbf{Oo~2349}}\\
$f_1 = 2.6986(1)$ & 3(2) & 0.11(9) & 6.0 & 2.7(8) & 0.10(5) & 7.2 & 2.2(4) &
0.10(3) & 6.4 & 2.1(7) & 0.08(5) & 5.5 & {1},\textbf{2},3,4\\
\multicolumn{14}{l}{}\\

\multicolumn{14}{l}{\textbf{{Oo~2350}}}\\
$f_1 = 2.6320(2)$ & 10(4) & 0.47(7) & 7.6 & 6(2) & 0.48(6) & 8.8 & 5(1) & 0.47(4) & 8.1 & 4(1) & 0.47(5) & 7.4 & \textbf{1},2,3,4\\
$f_2 = 2.5918(7)$ & 2(4)   & 0.6(3)    & 3.6 & 1(2) & 0.5(3)   & 3.8 & 1(1) & 0.5(2)   & 3.4 & 2(1) & 0.5(2) & 4.4 & 1,2,3,\textbf{4}\\
\multicolumn{14}{l}{}\\

\multicolumn{14}{l}{\textbf{{Oo~2352}}}\\
$f_1 = 0.2909(2)$ & 2(1) & 0.88(8) & 4.3 & 1.8(7) & 0.91(6) & 4.8 & 1.8(4) & 0.87(4) & 4.2 & 1.4(5) & 0.91(6) & \nodata & 1,2,\textbf{4}\\
\multicolumn{14}{l}{}\\

\multicolumn{14}{l}{\textbf{{Oo~2370}}}\\
$f_1 = 3.2744(2)$ & 9(8) & 0.1(1) & 7.2 & 5(3) & 0.1(1) & 7.1 & 4(1) & 0.11(6) & 9.7 & 3(2) & 0.1(1) & 5.9 & \textbf{1},2,3,4\\
\multicolumn{14}{l}{}\\

\multicolumn{14}{l}{\textbf{{Oo~2371}}}\\
$f_1 = 0.383573(5)$ & 24(3) & 0.22(2) & 4.6 & 24(3) & 0.231(7) & 6.8 & 20.5(4) & 0.257(4) & 5.5 & 26(1) & 0.252(5) & 5.7 & \nodata \\
\multicolumn{14}{l}{}\\

\multicolumn{14}{l}{\textbf{{Oo~2372}}}\\
$f_1=0.76570(4)$ & 4(4) & 0.8(2) & 4.4 & 3.4(5) & 0.82(2) & 6.1 & 3.2(2) & 0.80(1) & 5.3 & 3.5(3) & 0.82(1) & 5.3 & 4\\
\multicolumn{14}{l}{}\\

\multicolumn{14}{l}{\textbf{{Oo~2377}}}\\
$f_1 = 1.38760(6)$ & 9(3) & 0.2(6) & 3.3 & 2.0(3) & 0.22(3) & 4.9 & 1.7(1) & 0.27(1) & 5.3 & 1.5(2) & 0.24(2) & 4.7 & /\\
\multicolumn{14}{l}{}\\

\multicolumn{14}{l}{\textbf{Oo~2406}}\\
$f_1 = 3.7066(5)$ & \nodata & \nodata & \nodata & \nodata & \nodata & \nodata & 2(1) & 0.5(1) & 4.5 & 1(2) & 0.5(2) & 3.6 &
{1},{2},\textbf{3},4\\
$f_2 = 3.4981(6)$ & \nodata & \nodata & \nodata & \nodata & \nodata & \nodata & 2(1) & 0.5(1) & 4.0 & 1(2) & 0.5(3) & 3.8 &
{1},{2},\textbf{3},\textbf{4}\\
\multicolumn{14}{l}{}\\

\tableline
\end{tabular}\end{table*}


\addtocounter{table}{-1}
\begin{table*}
\footnotesize
\caption{Continued.}
\centering\begin{tabular}{l || lll | lll | lll| lll | c}
\tableline
~ & \multicolumn{3}{c|}{$U$} & \multicolumn{3}{c|}{$B$} &
\multicolumn{3}{c|}{$V$} & \multicolumn{3}{c|}{$I$} &  \\
$f_i$ & $A_i$ & $\phi_i$ & S/N & $A_i$ & $\phi_i$ & S/N & $A_i$ & $\phi_i$ & S/N
& $A_i$ & $\phi_i$ & S/N & $\ell$ \\
(d$^{-1}$) & (mmag) & (rad) & ~ & (mmag) & (rad) & ~ & (mmag) & (rad) & ~ &
(mmag) & (rad) & ~ &  \\
\tableline
\multicolumn{14}{l}{}\\

\multicolumn{14}{l}{\textbf{Oo~2410}}\\
$f_1 = 3.3420(4)$ & 8(10) & 0.3(2) & 7.0 & 3(4) & 0.2(2) & 6.2 & 3(2) & 0.25(8)
& 7.1 & 3(2) & 0.2(1) & 6.1 & 1,\textbf{2},\textbf{3},4\\
$f_2 = 3.2303(7)$ & 4(10) & 0.2(4) & 3.8 & 2(4) & 0.2(3) & 3.4 & 1(2) & 0.2(2)
& 4.5 & 1(2) & 0.2(3) & 3.0 & 1,{2},3,\textbf{4}\\
$f_3 = 3.4099(9)$ & 3(10) & 1.0(5) & 4.0 & 2(4) & 0.0(3) & 4.1 & 1(2) & 0.0(2)
& 4.5 & 1(2) & 0.0(4) & \nodata & 1,2,3,\textbf{4}\\
\multicolumn{14}{l}{}\\

\multicolumn{14}{l}{\textbf{{Oo~2414}}}\\
$f_1 = 0.43731(9)$ & 24(8) & 0.92(6) & 7.4 & 7(3) & 0.01(7) & 3.5 & 8(1) & 0.96(2) & 6.7 & 11(2) & 0.94(2) & 6.1 & 1,2,\textbf{3},4\\
\multicolumn{14}{l}{}\\

\multicolumn{14}{l}{\textbf{{Oo~2426}}}\\
/ & \nodata & \nodata & \nodata & \nodata & \nodata & \nodata & \nodata & \nodata & \nodata & \nodata  & \nodata & \nodata & \nodata\\
\multicolumn{14}{l}{}\\

\multicolumn{14}{l}{\textbf{Oo~2429}}\\
$f_1 = 1.0827(2)$ & 6(2) & 0.14(6) & 4.2 & 3.9(9) & 0.14(4) & 5.0 & 3.3(5) &
0.13(2) & 4.0 & 2.9(7) & 0.10(4) & 4.5 & \textbf{1},{2},{3},{4}\\
$f_2 = 0.5593(3)$ & 3(2) & 0.6(1)  & 3.4 & 2.8(9) & 0.58(6) & 4.3 & 1.5(5) &
0.58(5) &   \nodata    & 2.1(7) & 0.57(6) & 3.8 & \textbf{1},\textbf{2},{3},\textbf{4}\\
$f_3 = 2.4821(2)$ & 4(2) & 0.23(8) & 4.8 & 2.5(9) & 0.25(6) & 4.6 & 1.9(5) &
0.15(4) & 3.1 & 1.7(7) & 0.20(7) & 3.7 & {1},\textbf{2},3,4\\
$f_4 = f_3-f_1$   & 3(2) & 0.8(1)  & 4.2 & 1.9(9) & 0.87(8) & 3.7 & 2.5(5) &
0.86(3) & 3.6 & 1.4(8) & 0.87(8) & 3.1 & \nodata \\
$f_5 = 1.2710(4)$ & 3(2) & 0.4(1)  & 4.4 & 1.8(9) & 0.39(8) & 3.7 & 1.8(5) &
0.41(4) & 3.5 & 1.5(7) & 0.43(8) & 3.4 & {1},{2},\textbf{3},4\\
\multicolumn{14}{l}{}\\

\multicolumn{14}{l}{\textbf{Oo~2444}}\\
$f_1 = 4.58161(2)$  & 7(2) & 0.63(5) & 7.2 & 6.1(3) & 0.622(7) & 9.5 & 5.9(1) &
0.624(4) & 8.9 & 4.8(2) & 0.624(8) & 7.8 & 1\\
$f_2 = 5.39333(5)$  & 2(2) & 0.3(2)   &    \nodata    & 1.9(3) & 0.30(2)   & 6.0 & 2.3(1)
& 0.297(9) & 7.9 & 1.8(2) & 0.29(2)   & 5.5 & /\\
$f_3 = 2f_1$            & 1(2) & 0.5(2)    & 3.3 & 1.2(3) & 0.59(4)  & 6.4 & 1.2(1) & 0.57(2)   & 7.1 & 0.9(2) & 0.57(4)   & 5.0 & \nodata \\
$f_4 = 4.4494(1)$    & 1(2) & 0.2(4)   &     \nodata   & 1.2(3) & 0.13(4)   & 4.7 &
1.0(1) & 0.12(2)   & 4.6 & 0.8(2) & 0.15(4)  & 3.3 &
0,\textbf{1},\textbf{2},3,4\\
$f_5 = 5.4643(3)$    & 1(2) & 0.4(3)   &   \nodata     & 1.0(3) & 0.44(5)   & 4.2 &
0.6(1) & 0.49(3)   & 3.4 & 0.8(2) & 0.44(5)  & 3.7 & /\\
\multicolumn{14}{l}{}\\

\multicolumn{14}{l}{\textbf{Oo~2448}}\\
$f_1 = 1.4983(7)$ & \nodata & \nodata & \nodata & 7(7) & 0.5(2) & 3.2 & 9(2) & 0.50(3) & 5.2 & 3(7) &
0.4(3) & \nodata & \textbf{1},{2},{3},\textbf{4}\\
$f_2 = 1.427(1)$ & \nodata & \nodata & \nodata & 9(7) & 0.8(1) & 3.2 & 5(2) & 0.77(5) & 3.6 & 5(7) &
0.8(2) & \nodata & 1,\textbf{2},\textbf{3},4\\
\multicolumn{14}{l}{}\\

\multicolumn{14}{l}{\textbf{{Oo~2455}}}\\
$f_1 = 0.63915(2)$ & 23(8) & 0.83(5) & 4.2 & 21(1) & 0.84(1) & 6.4 & 16.1(5) &  0.831(6) & 6.0 & 14.8(7) & 0.827(7) & 5.2 & 1,2,\textbf{4}\\
$f_2 = 2 f_1$ & 9(7) & 0.5(1) & 3.8 & 10(1) & 0.50(2) & 5.6 & 7.7(5) & 0.49(1) & 6.1 & 6.9(7) & 0.47(2) & 6.0 & \nodata \\
\multicolumn{14}{l}{}\\

\multicolumn{14}{l}{\textbf{{Oo~2462}}}\\
$f_1 = 1.25142(2)$ & 2.4(3) & 0.51(2) & \nodata & 1.86(9) & 0.478(7) & 3.2 & 2.34(5) & 0.471(4) & 4.5 & 1.45(9) & 0.450(9) &  \nodata  & /\\
\multicolumn{14}{l}{}\\

\multicolumn{14}{l}{\textbf{{Oo~2482}}}\\
$f_1 = 0.7014(4)$ & 10(23) & 0.9(3) & 4.2 & 6(6) & 0.9(2) & 4.9 & 3(2) & 0.9(1) & 5.0 & 3(3) & 0.9(2) & 4.1 & 1,{2},\textbf{3},4\\
\multicolumn{14}{l}{}\\

\multicolumn{14}{l}{\textbf{Oo~2488}}\\
$f_1 = 6.16805(1)$ & 7(3) & 0.61(6) & 6.8 & 7.2(3) & 0.608(6) & 10.7 & 6.9(1) &
0.612(3) & 10.8 & 6.4(2) & 0.610(5) & 9.8 & {2}\\
$f_2 = 6.97476(5)$ & 2(3) & 0.3(3)   & 3.3 & 1.5(3) & 0.30(3)   & 6.7   & 1.9(1)
& 0.29(1)   & 8.3   & 1.7(2) & 0.28(2)   & 7.7 & /\\
$f_3 = 6.8239(2)$   & 1(3) & 0.6(3)   &     \nodata    & 0.4(3) & 0.6(1)     &    \nodata       &
0.6(1) & 0.62(3)   & 4.0   & 0.2(2) & 0.6(2)     &     \nodata    & /\\
\multicolumn{14}{l}{}\\

\tableline
\end{tabular}\end{table*}

\addtocounter{table}{-1}
\begin{table*}
\footnotesize
\caption{Continued.}
\centering\begin{tabular}{l || lll | lll | lll| lll | c}
\tableline
~ & \multicolumn{3}{c|}{$U$} & \multicolumn{3}{c|}{$B$} &
\multicolumn{3}{c|}{$V$} & \multicolumn{3}{c|}{$I$} &  \\
$f_i$ & $A_i$ & $\phi_i$ & S/N & $A_i$ & $\phi_i$ & S/N & $A_i$ & $\phi_i$ & S/N
& $A_i$ & $\phi_i$ & S/N & $\ell$ \\
(d$^{-1}$) & (mmag) & (rad) & ~ & (mmag) & (rad) & ~ & (mmag) & (rad) & ~ &
(mmag) & (rad) & ~ &  \\
\tableline
\multicolumn{14}{l}{}\\

\multicolumn{14}{l}{\textbf{Oo~2507}}\\
$f_1 = 4.7075(1)$  & 2(4) & 0.6(4) &    \nodata   & 2(1) & 0.53(7) & 5.2 & 2.1(4) &
0.50(3) & 5.3 & 1.7(6) & 0.48(6) & 5.5 & {1},\textbf{2},3,4\\
$f_2 = 2.40720(9)$ & 2(4) & 0.4(3) & 3.6 & 3(1) & 0.49(6) & 4.3 & 2.9(4) &
0.52(2) & 5.4 & 2.0(6) & 0.52(5) & 4.1 & {1},{2},\textbf{3}\\
$f_3 = f_1 - f_2$  & 2(4) & 0.9(4) &   \nodata    & 2(1) & 0.98(9) & 3.6 & 2.1(4) &
0.92(3) & 4.6 & 1.8(6) & 0.93(6) & 4.9 & \nodata \\
\multicolumn{14}{l}{}\\

\multicolumn{14}{l}{\textbf{Oo~2520}}\\
$f_1 = 5.6571(4)$ & \nodata & \nodata & \nodata & \nodata & \nodata & \nodata & 2.7(3) & 0.60(2) & 6.1 & 2(1) & 0.73(8) & 3.2 &
\textbf{0},\textbf{1},{2},\textbf{3},4\\
\multicolumn{14}{l}{}\\

\multicolumn{14}{l}{\textbf{{Oo~2524}}}\\
$f_1 = 3.2217(5)$ & 4(10) & 0.8(4) & 4.2 & 2(3) & 0.8(3) & 3.7 & 2(1) & 0.8(1) & 5.7 & 1(2) & 0.7(2) & 3.6 & 1,2,\textbf{3},4\\
\multicolumn{14}{l}{}\\

\multicolumn{14}{l}{\textbf{Oo~2531}}\\
$f_1 = 3.6644(5)$ & 5(14) & 0.7(5) & 3.9 & 4(9) & 0.7(4) & 4.1 & 4(2) & 0.7(1)
& 7.3 & 3(3) & 0.7(1) & 4.9 & {1},{2},\textbf{3},4\\
\multicolumn{14}{l}{}\\

\multicolumn{14}{l}{\textbf{{Oo~2562}}}\\
$f_1 = 2.5740(3)$ & 19(20) & 0.6(2) & 7.3 & 9(6) & 0.6(1) & 7.0 & 8(2) & 0.57(3) & 7.2 & 7(2) & 0.60(5) & 6.7 & \textbf{1},2,3,4\\
$f_2 = 2.5068(7)$ & 4(21)   & 0.0(7) &    \nodata     & 4(6)  & 0.0(3) & 4.4 & 2(2) & 0.9(1)   & 3.9 & 3(2) & 0.9(1) & 4.5 & 1,2,3,\textbf{4}\\
$f_3 = 12.649(2)$ & 1(20)   & 0(3)    &    \nodata     & 0(6)  & 0(4)     &    \nodata     & 0(2) & 0(1)      &     \nodata      & 1(2) & 0.1(3) & 4.6 & \nodata \\
\multicolumn{14}{l}{}\\

\multicolumn{14}{l}{\textbf{{Oo~2563}}}\\
$f_1 = 3.6637(4)$ & \nodata & \nodata & \nodata & 6(2) & 0.64(6) & 3.1 & 5.1(6) & 0.65(2) & 4.3 & 
3(2) & 0.6(1) & \nodata & \textbf{0},1,2,\textbf{3},4\\
\multicolumn{14}{l}{}\\

\multicolumn{14}{l}{\textbf{{Oo~2566}}}\\
$f_1 = 2.55480(6)$ & 6(10) & 0.8(3) & 3.3 & 11(2)& 0.88(3) & 3.3 & 17(1) & 
0.87(1) & 4.2 & 9(4) & 0.01(7) & \nodata & /\\
\multicolumn{14}{l}{}\\

\multicolumn{14}{l}{\textbf{Oo~2572}}\\
$f_1 = 4.41463(2)$ & 6(3) & 0.3(1) & 5.5 & 5.4(3) & 0.298(8) & 8.4 & 5.2(1) &
0.295(4) & 10.1 & 4.8(2) & 0.299(7) & 7.7 & {2}\\
$f_2 = 4.76330(6)$ & 2(3) & 0.4(2) & 3.0 & 2.1(3) & 0.43(2)   & 7.0 & 1.6(1) &
0.42(1)   & 6.7   & 1.5(2) & 0.44(2)   & 6.0 & 1\\
$f_3 = 4.6648(1)$   & 0(3) & 0(2)    &    \nodata     & 1.1(3) & 0.94(4)   & 4.4 & 0.9(1)
& 0.95(2)   & 4.3   & 0.9(2) & 0.97(4)   & 3.8 & {0},1,\textbf{2},4 \\
\multicolumn{14}{l}{}\\

\multicolumn{14}{l}{\textbf{{Oo~2579}}}\\
$f_1 = 2.28187(5)$ & 13(11) & 0.6(1) & 4.7 & 9(1) & 0.62(2) & 5.5 & 10.3(6) & 0.60(1) & 5.8 & 5.6(2) & 0.643(7) & 3.7 & /\\
\multicolumn{14}{l}{}\\

\multicolumn{14}{l}{\textbf{Oo~2601}}\\
$f_1 = 6.7020(2)$ & 3(3) & 0.8(1) & 3.5 & 2.9(5) & 0.79(3) & 6.4 & 2.4(2) &
0.79(1) & 7.4 & 2.0(3) & 0.78(2) & 5.4 &
{0},\textbf{1},\textbf{2},{3},4\\
$f_2 = 6.4667(3)$ & 3(3) & 0.7(2) & 3.3 & 2.1(5) & 0.67(4) & 6.5 & 1.6(2) &
0.69(2) & 7.0 & 1.3(3) & 0.67(3) & 4.6 &
\textbf{0},\textbf{1},{2},{3},4\\
$f_3 = 7.0670(6)$ & 1(3) & 0.1(8) &   \nodata    & 0.5(5) & 0.8(1)  &   \nodata    & 0.8(2) &
0.82(4) & 4.5 & 0.5(3) & 0.87(8) &  \nodata  &
\textbf{0},\textbf{1},{2},3,4\\
\multicolumn{14}{l}{}\\

\multicolumn{14}{l}{\textbf{{Oo~2611}}}\\
$f_1 = 4.702(2)$ & \nodata & \nodata & \nodata & \nodata &   \nodata & \nodata & 3(1) & 0.90(7) & 4.3 & \nodata & \nodata & \nodata & \nodata \\
\multicolumn{14}{l}{}\\

\tableline
\end{tabular}\end{table*}

\addtocounter{table}{-1}
\begin{table*}
\footnotesize
\caption{Continued.}
\centering\begin{tabular}{l || lll | lll | lll| lll | c}
\tableline
~ & \multicolumn{3}{c|}{$U$} & \multicolumn{3}{c|}{$B$} &
\multicolumn{3}{c|}{$V$} & \multicolumn{3}{c|}{$I$} &  \\
$f_i$ & $A_i$ & $\phi_i$ & S/N & $A_i$ & $\phi_i$ & S/N & $A_i$ & $\phi_i$ & S/N
& $A_i$ & $\phi_i$ & S/N & $\ell$ \\
(d$^{-1}$) & (mmag) & (rad) & ~ & (mmag) & (rad) & ~ & (mmag) & (rad) & ~ &
(mmag) & (rad) & ~ &  \\
\tableline
\multicolumn{14}{l}{}\\

\multicolumn{14}{l}{\textbf{{Oo~2616}}}\\
$f_1 = 3.166(2)$ & \nodata & \nodata & \nodata & 3(4) & 0.5(2) & 4.0 & 3(2) & 0.5(1) & 5.9 & 2(2) & 0.5(1) & 4.2 & 1,2,\textbf{3},4\\
$f_2 = 3.335(2)$ & \nodata & \nodata & \nodata & 3(4) & 0.6(3) & 4.0 & 1(2) & 0.6(2) & 3.4 & 2(2) & 0.6(2) & 3.4 & 1,2,3,\textbf{4}\\
\multicolumn{14}{l}{}\\

\multicolumn{14}{l}{\textbf{{Oo~2622}}}\\
/ & \nodata & \nodata & \nodata & \nodata & \nodata & \nodata & \nodata & \nodata & \nodata & \nodata & \nodata & \nodata & \nodata \\
\multicolumn{14}{l}{}\\

\multicolumn{14}{l}{\textbf{Oo~2633}}\\
$f_1 = 8.2357(3)$  & \nodata & \nodata & \nodata & 2.5(8) & 0.91(5) & 5.9 & 2.4(4) & 0.92(2) & 6.5 &
1.4(3) & 0.93(4) & 5.4 & 0,1,\textbf{3}\\
$f_2 = 5.8847(3)$  & \nodata & \nodata & \nodata & 2.9(8) & 0.20(5) & 6.0 & 2.3(4) & 0.23(3) & 6.1 &
1.5(3) & 0.24(4) & 4.9 & 1,2,\textbf{3}\\
$f_3 = 10.3578(4)$ & \nodata & \nodata & \nodata & 1.7(8) & 0.27(7) & 5.4 & 1.6(4) & 0.24(4) & 6.0 &
0.8(3) & 0.26(7) & 4.2 & {0},\textbf{3}\\
$f_4 = 5.9007(4)$  & \nodata & \nodata & \nodata & 1.8(8) & 0.84(7) & 5.0 & 1.8(4) & 0.86(4) & 5.0 &
1.1(3) & 0.92(5) & 4.8 & 1,2,\textbf{3}\\
$f_5 = 8.1236(8)$  & \nodata & \nodata & \nodata & 1.4(8) & 0.14(8) & 5.1 & 0.8(3) & 0.07(7) & 3.9 &
0.8(3) & 0.13(7) & 4.5 & 0,1,\textbf{2},3,4\\
$f_6 = 15.6889(9)$ & \nodata & \nodata & \nodata & 0.8(7) & 0.4(1)  & 5.2 & 0.8(3) & 0.36(7) & 4.6 &
0.4(3) & 0.3(1)  & 4.2 & 0,{1},2,\textbf{3}\\
$f_7 = 34.502(2)$  & \nodata & \nodata & \nodata & 0.4(7) & 0.3(3)  & 3.1 & 0.4(3) & 0.4(1)  & 4.2 &
0.4(3) & 0.4(1)  & 4.1 & 0,1,2,{3},\textbf{4}\\
$f_8 = 12.301(1)$  & \nodata & \nodata & \nodata & 0.9(7) & 0.2(1)  & 4.2 & 0.6(3) & 0.06(9) & 3.3 &
0.3(3) & 0.1(2)  &  \nodata     & 0,1,2,\textbf{3},4\\
$f_9 = 19.182(1)$  & \nodata & \nodata & \nodata & 0.6(7) & 0.2(2) & 4.2 & 0.2(3) & 0.2(3) &   \nodata      &
0.2(3) & 0.2(2) & \nodata & \textbf{0},\textbf{1},\textbf{2},3,4\\ 
\multicolumn{14}{l}{}\\

\multicolumn{14}{l}{\textbf{Oo~2649}}\\
$f_1 = 2.24380(8)$ & \nodata & \nodata & \nodata & 11.4(8) & 0.65(1) & 5.8 & 15.0(6) & 0.710(6) & 6.3 &
10.9(8) & 0.65(1) & 5.0 & /\\
\multicolumn{14}{l}{}\\

\multicolumn{14}{l}{\textbf{Oo~2694}}\\
$f_1 = 3.7333(3)$ & \nodata & \nodata & \nodata & 6(1) & 0.81(3) & 5.1 & 6.1(5) & 0.76(1) & 5.1 &
4.8(7) & 0.81(2) & 5.6 & {0},\textbf{1},{2},{3},4\\
\multicolumn{14}{l}{}\\

\multicolumn{14}{l}{\textbf{{Oo~2725}}}\\
/ & \nodata & \nodata & \nodata & \nodata & \nodata & \nodata & \nodata & \nodata & \nodata & \nodata & \nodata & \nodata & \nodata \\
\multicolumn{14}{l}{}\\

\multicolumn{14}{l}{\textbf{{Oo~2752}}}\\
/ & \nodata & \nodata & \nodata & \nodata & \nodata & \nodata & \nodata & \nodata & \nodata & \nodata & \nodata & \nodata & \nodata \\
\multicolumn{14}{l}{}\\

\multicolumn{14}{l}{\textbf{Oo~2753}}\\
$f_1 = 2.233(2)$ & \nodata & \nodata & \nodata & \nodata & \nodata & \nodata & 8(4) & 0.13(8) & 4.7 & \nodata & \nodata & \nodata & \nodata \\
\multicolumn{14}{l}{}\\

\tableline
\end{tabular}\end{table*}


\FloatBarrier

\addtocounter{figure}{-6}
\begin{figure*}\centering
\includegraphics[width=0.7\columnwidth]{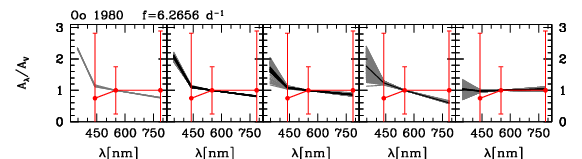}
\includegraphics[width=0.2\columnwidth]{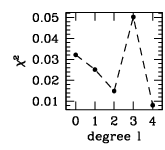}\\
\includegraphics[width=0.7\columnwidth]{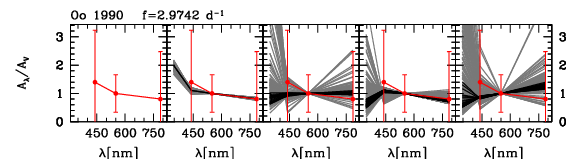}
\includegraphics[width=0.2\columnwidth]{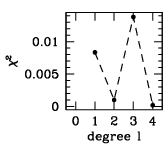}\\
\includegraphics[width=0.7\columnwidth]{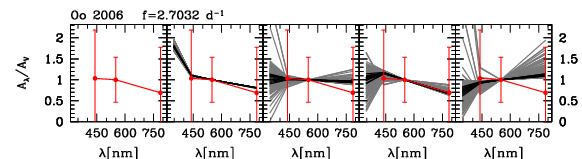}
\includegraphics[width=0.2\columnwidth]{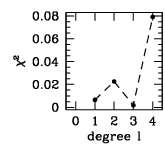}\\
\includegraphics[width=0.7\columnwidth]{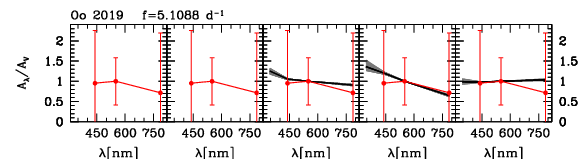}
\includegraphics[width=0.2\columnwidth]{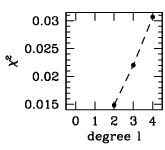}\\
\includegraphics[width=0.7\columnwidth]{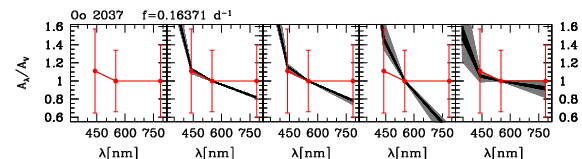}
\includegraphics[width=0.2\columnwidth]{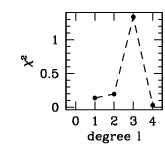}\\
\includegraphics[width=0.7\columnwidth]{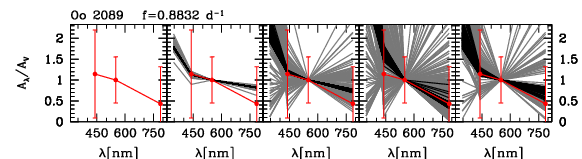}
\includegraphics[width=0.2\columnwidth]{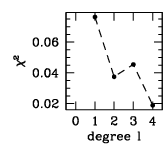}\\
\caption{{(Left) Amplitude ratios scaled to the $V$ filter for the frequency and the star denoted above the figure. The red filled circles with error bars
denote the observed amplitude ratios and their uncertainties, the grey bands indicate the theoretical predictions for these, running from $\ell=0$ at the left to $\ell=4$ at the right. The black bands denote a subsample of these theoretical models that deviate only 500~K in effective temperature and 0.05~dex in luminosity from the observed position in
the HR-diagram. (Right) Corresponding $\chi^2$-value as a function of the degree $\ell$.}}
\end{figure*}
\addtocounter{figure}{-1}\begin{figure*}\centering
\includegraphics[width=0.7\columnwidth]{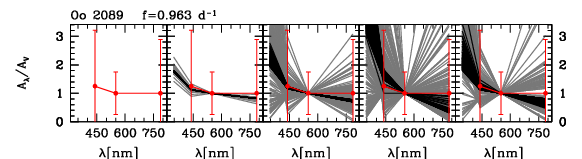}
\includegraphics[width=0.2\columnwidth]{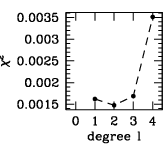}\\
\includegraphics[width=0.7\columnwidth]{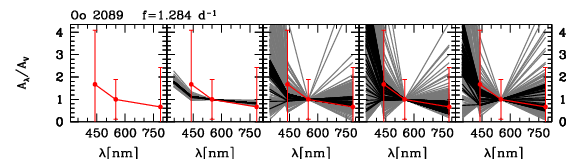}
\includegraphics[width=0.2\columnwidth]{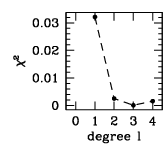}\\
\includegraphics[width=0.7\columnwidth]{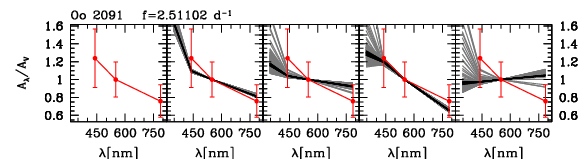}
\includegraphics[width=0.2\columnwidth]{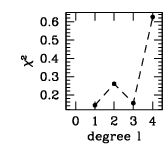}\\
\includegraphics[width=0.7\columnwidth]{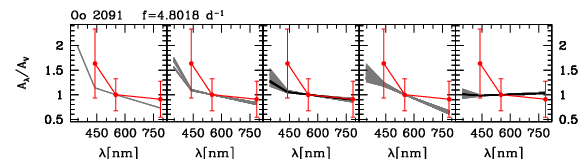}
\includegraphics[width=0.2\columnwidth]{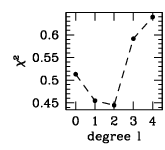}\\
\includegraphics[width=0.7\columnwidth]{fig04_11a.png}
\includegraphics[width=0.2\columnwidth]{fig04_11b.png}\\
\includegraphics[width=0.7\columnwidth]{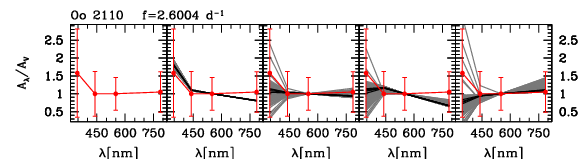}
\includegraphics[width=0.2\columnwidth]{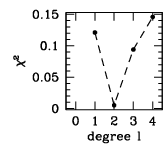}\\
\caption{Continued.}\end{figure*}
\FloatBarrier
\addtocounter{figure}{-1}\begin{figure*}
\centering
\includegraphics[width=0.7\columnwidth]{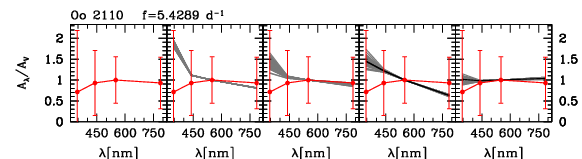}
\includegraphics[width=0.2\columnwidth]{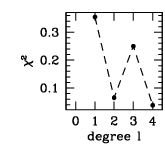}\\
\includegraphics[width=0.7\columnwidth]{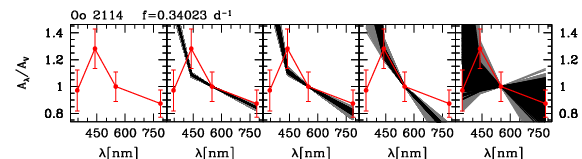}
\includegraphics[width=0.2\columnwidth]{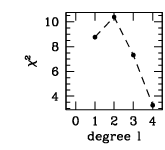}\\
\includegraphics[width=0.7\columnwidth]{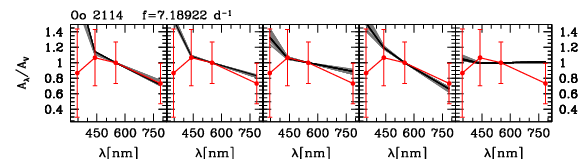}
\includegraphics[width=0.2\columnwidth]{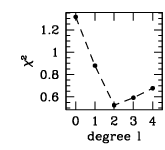}\\
\includegraphics[width=0.7\columnwidth]{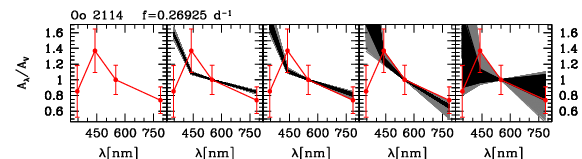}
\includegraphics[width=0.2\columnwidth]{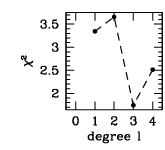}\\
\includegraphics[width=0.7\columnwidth]{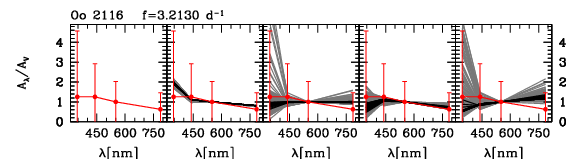}
\includegraphics[width=0.2\columnwidth]{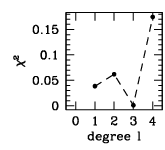}\\
\includegraphics[width=0.7\columnwidth]{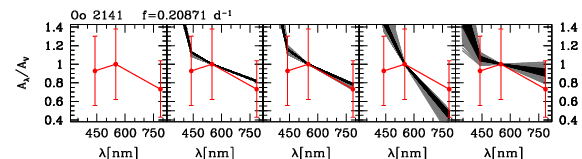}
\includegraphics[width=0.2\columnwidth]{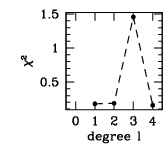}\\
\caption{Continued.}\end{figure*}\addtocounter{figure}{-1}\begin{figure*}
\centering
\includegraphics[width=0.7\columnwidth]{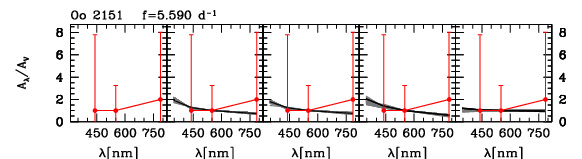}
\includegraphics[width=0.2\columnwidth]{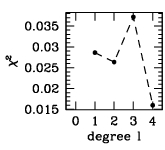}\\
\includegraphics[width=0.7\columnwidth]{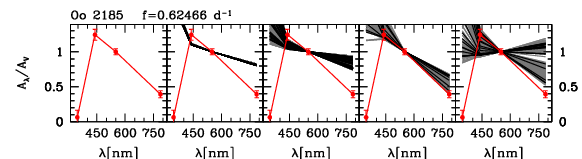}
\includegraphics[width=0.2\columnwidth]{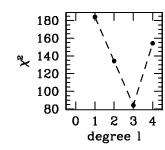}\\
\includegraphics[width=0.7\columnwidth]{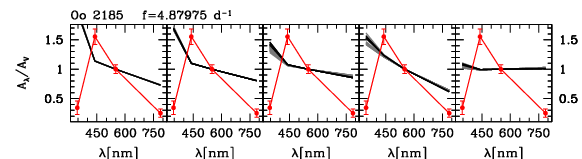}
\includegraphics[width=0.2\columnwidth]{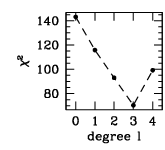}\\
\includegraphics[width=0.7\columnwidth]{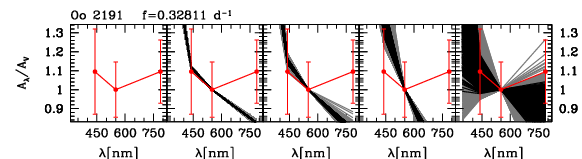}
\includegraphics[width=0.2\columnwidth]{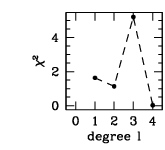}\\
\includegraphics[width=0.7\columnwidth]{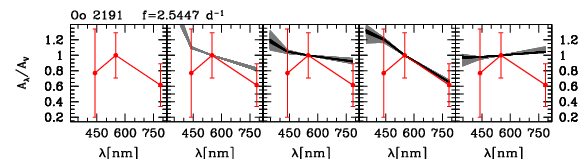}
\includegraphics[width=0.2\columnwidth]{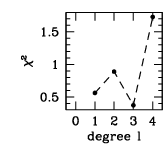}\\
\includegraphics[width=0.7\columnwidth]{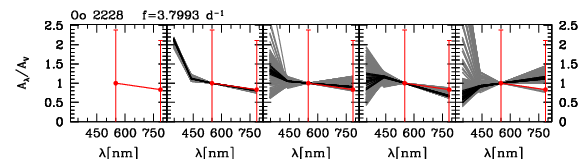}
\includegraphics[width=0.2\columnwidth]{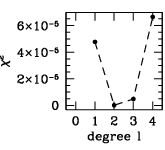}\\
\caption{Continued.}\end{figure*}
\FloatBarrier
\addtocounter{figure}{-1}\begin{figure*}
\centering
\includegraphics[width=0.7\columnwidth]{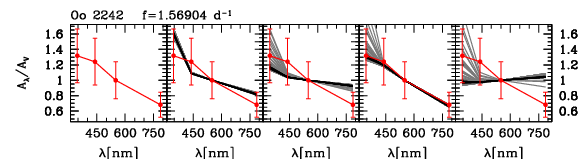}
\includegraphics[width=0.2\columnwidth]{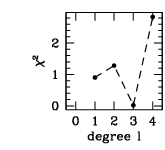}\\
\includegraphics[width=0.7\columnwidth]{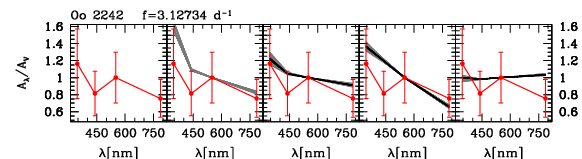}
\includegraphics[width=0.2\columnwidth]{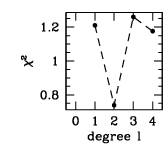}\\
\includegraphics[width=0.7\columnwidth]{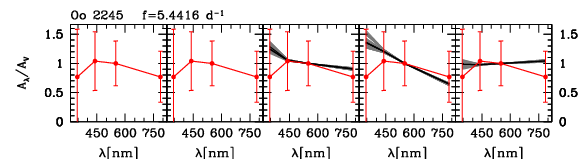}
\includegraphics[width=0.2\columnwidth]{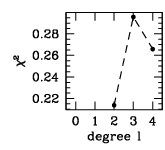}\\
\includegraphics[width=0.7\columnwidth]{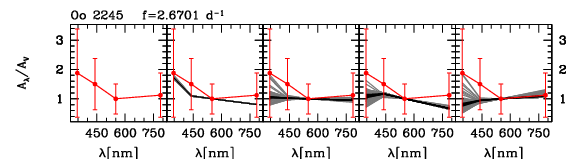}
\includegraphics[width=0.2\columnwidth]{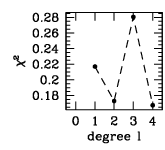}\\
\includegraphics[width=0.7\columnwidth]{fig04_29a.png}
\includegraphics[width=0.2\columnwidth]{fig04_29b.png}\\
\includegraphics[width=0.7\columnwidth]{fig04_30a.png}
\includegraphics[width=0.2\columnwidth]{fig04_30b.png}\\
\caption{Continued.}\end{figure*}\addtocounter{figure}{-1}\begin{figure*}
\centering
\includegraphics[width=0.7\columnwidth]{fig04_31a.png}
\includegraphics[width=0.2\columnwidth]{fig04_31b.png}\\
\includegraphics[width=0.7\columnwidth]{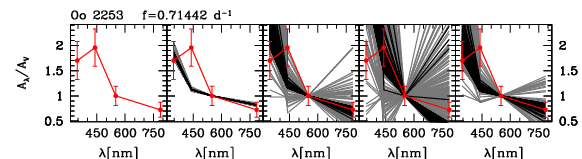}
\includegraphics[width=0.2\columnwidth]{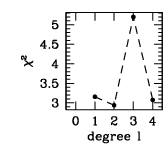}\\
\includegraphics[width=0.7\columnwidth]{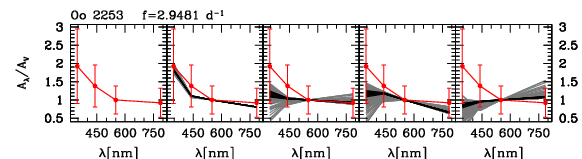}
\includegraphics[width=0.2\columnwidth]{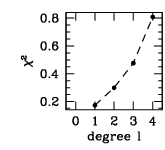}\\
\includegraphics[width=0.7\columnwidth]{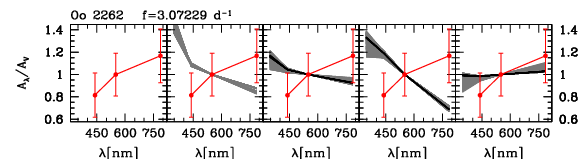}
\includegraphics[width=0.2\columnwidth]{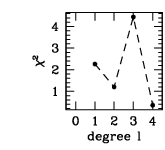}\\
\includegraphics[width=0.7\columnwidth]{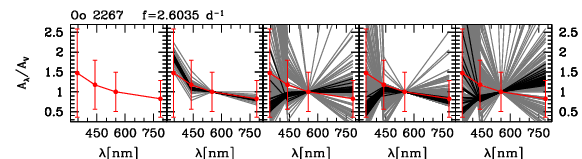}
\includegraphics[width=0.2\columnwidth]{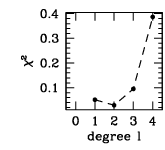}\\
\includegraphics[width=0.7\columnwidth]{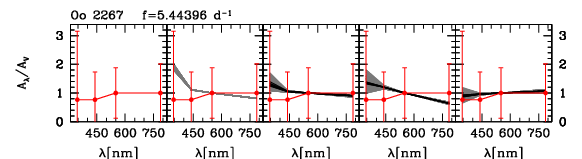}
\includegraphics[width=0.2\columnwidth]{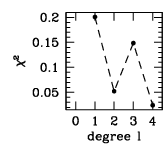}\\
\caption{Continued.}\end{figure*}\addtocounter{figure}{-1}\begin{figure*}
\centering
\includegraphics[width=0.7\columnwidth]{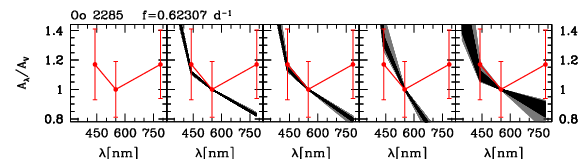}
\includegraphics[width=0.2\columnwidth]{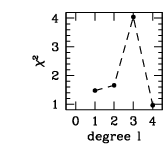}\\
\includegraphics[width=0.7\columnwidth]{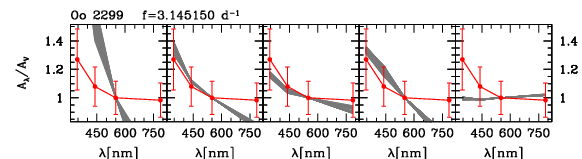}
\includegraphics[width=0.2\columnwidth]{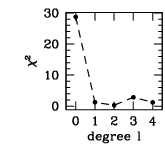}\\
\includegraphics[width=0.7\columnwidth]{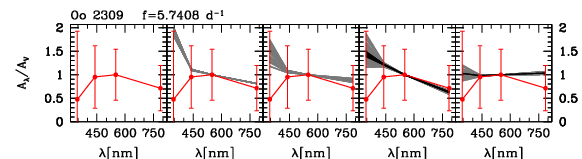}
\includegraphics[width=0.2\columnwidth]{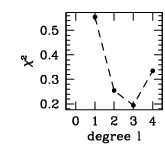}\\
\includegraphics[width=0.7\columnwidth]{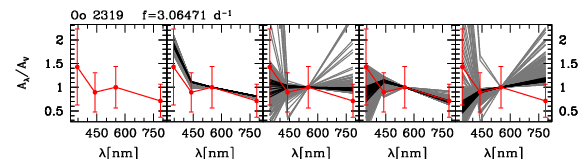}
\includegraphics[width=0.2\columnwidth]{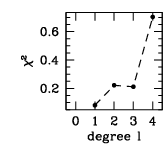}\\
\includegraphics[width=0.7\columnwidth]{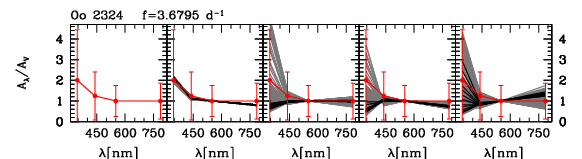}
\includegraphics[width=0.2\columnwidth]{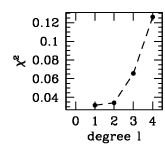}\\
\includegraphics[width=0.7\columnwidth]{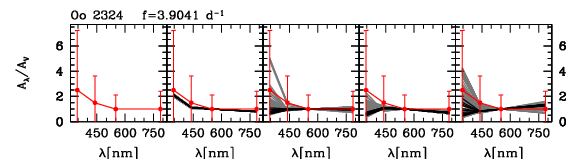}
\includegraphics[width=0.2\columnwidth]{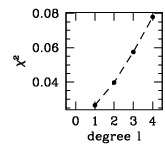}\\
\caption{Continued.}\end{figure*}\FloatBarrier
\addtocounter{figure}{-1}\begin{figure*}
\centering
\includegraphics[width=0.7\columnwidth]{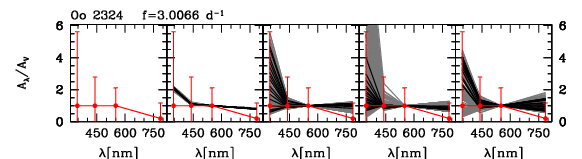}
\includegraphics[width=0.2\columnwidth]{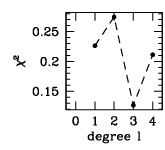}\\
\includegraphics[width=0.7\columnwidth]{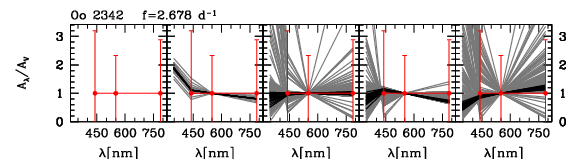}
\includegraphics[width=0.2\columnwidth]{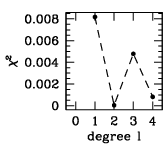}\\
\includegraphics[width=0.7\columnwidth]{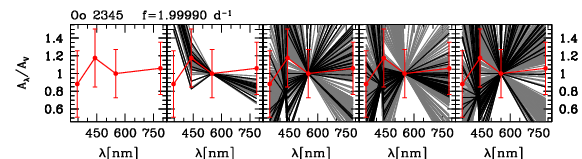}
\includegraphics[width=0.2\columnwidth]{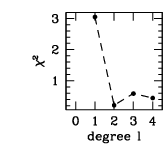}\\
\includegraphics[width=0.7\columnwidth]{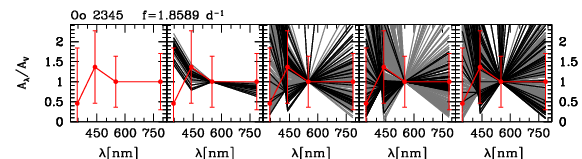}
\includegraphics[width=0.2\columnwidth]{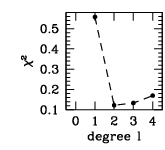}\\
\includegraphics[width=0.7\columnwidth]{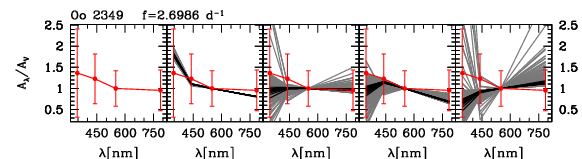}
\includegraphics[width=0.2\columnwidth]{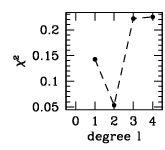}\\
\includegraphics[width=0.7\columnwidth]{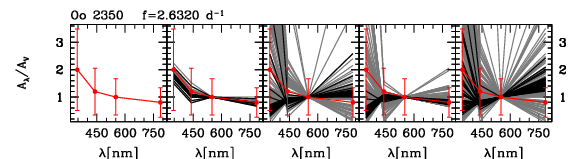}
\includegraphics[width=0.2\columnwidth]{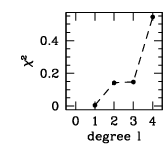}\\
\caption{Continued.}\end{figure*}\addtocounter{figure}{-1}\begin{figure*}
\centering
\includegraphics[width=0.7\columnwidth]{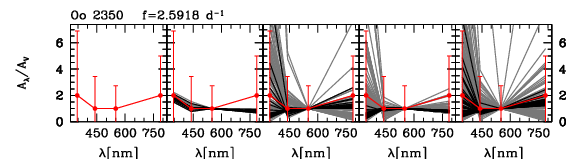}
\includegraphics[width=0.2\columnwidth]{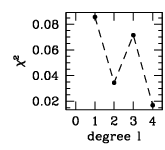}\\
\includegraphics[width=0.7\columnwidth]{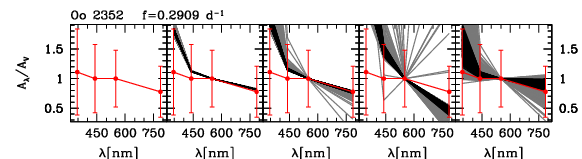}
\includegraphics[width=0.2\columnwidth]{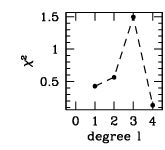}\\
\includegraphics[width=0.7\columnwidth]{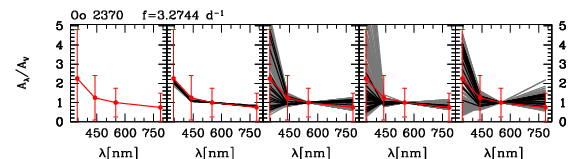}
\includegraphics[width=0.2\columnwidth]{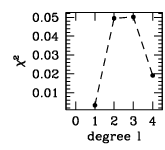}\\
\includegraphics[width=0.7\columnwidth]{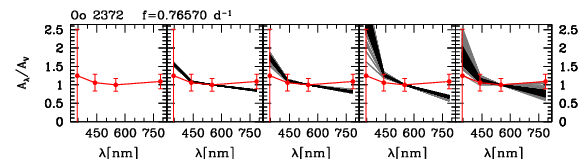}
\includegraphics[width=0.2\columnwidth]{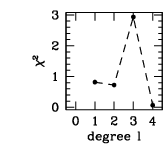}\\
\includegraphics[width=0.7\columnwidth]{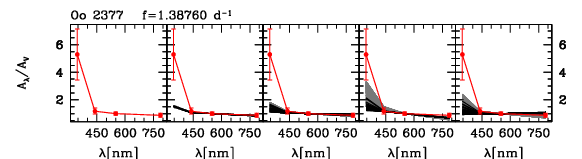}
\includegraphics[width=0.2\columnwidth]{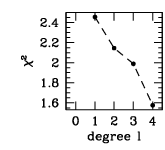}\\
\includegraphics[width=0.7\columnwidth]{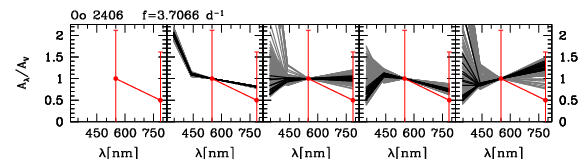}
\includegraphics[width=0.2\columnwidth]{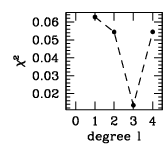}\\
\caption{Continued.}\end{figure*}\FloatBarrier
\addtocounter{figure}{-1}\begin{figure*}
\centering
\includegraphics[width=0.7\columnwidth]{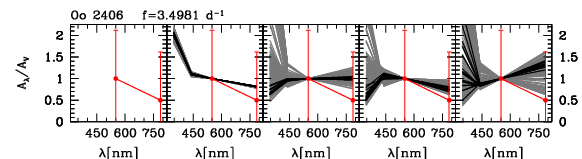}
\includegraphics[width=0.2\columnwidth]{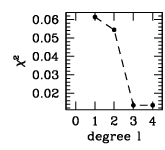}\\
\includegraphics[width=0.7\columnwidth]{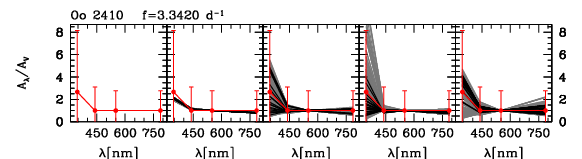}
\includegraphics[width=0.2\columnwidth]{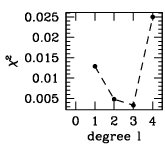}\\
\includegraphics[width=0.7\columnwidth]{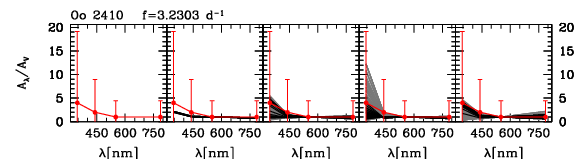}
\includegraphics[width=0.2\columnwidth]{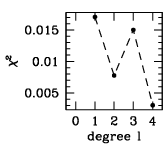}\\
\includegraphics[width=0.7\columnwidth]{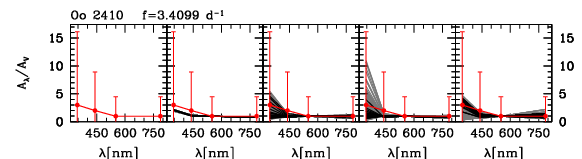}
\includegraphics[width=0.2\columnwidth]{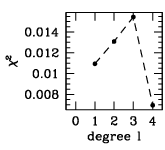}\\
\includegraphics[width=0.7\columnwidth]{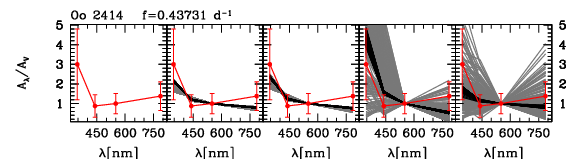}
\includegraphics[width=0.2\columnwidth]{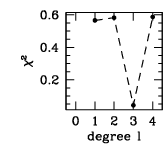}\\
\includegraphics[width=0.7\columnwidth]{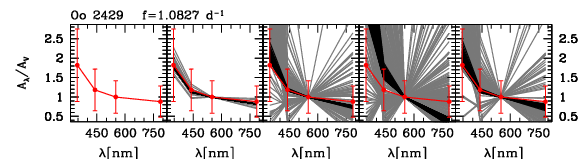}
\includegraphics[width=0.2\columnwidth]{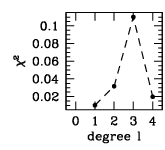}\\
\caption{Continued.}\end{figure*}
\addtocounter{figure}{-1}\begin{figure*}
\centering
\includegraphics[width=0.7\columnwidth]{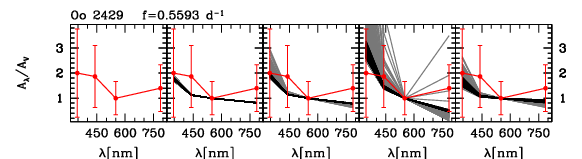}
\includegraphics[width=0.2\columnwidth]{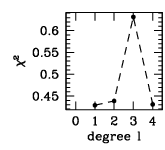}\\
\includegraphics[width=0.7\columnwidth]{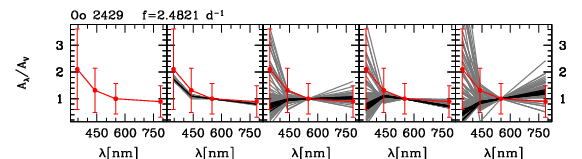}
\includegraphics[width=0.2\columnwidth]{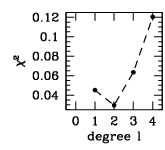}\\
\includegraphics[width=0.7\columnwidth]{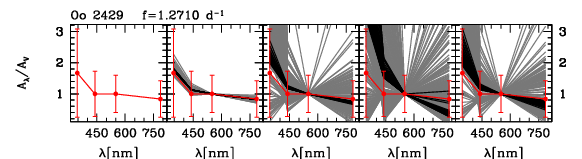}
\includegraphics[width=0.2\columnwidth]{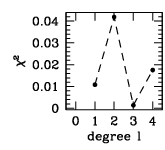}\\
\includegraphics[width=0.7\columnwidth]{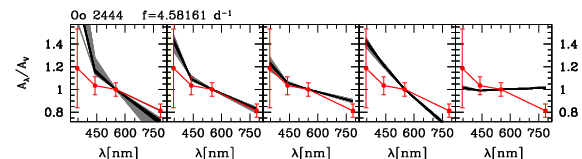}
\includegraphics[width=0.2\columnwidth]{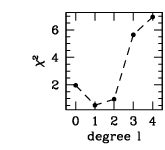}\\
\includegraphics[width=0.7\columnwidth]{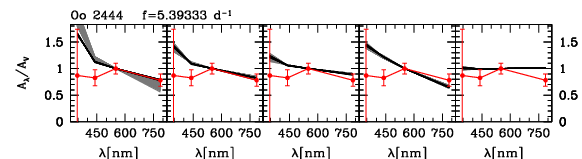}
\includegraphics[width=0.2\columnwidth]{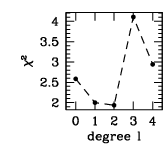}\\
\includegraphics[width=0.7\columnwidth]{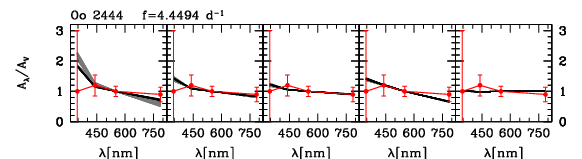}
\includegraphics[width=0.2\columnwidth]{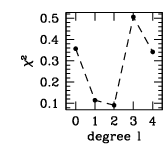}\\
\caption{Continued.}\end{figure*}\FloatBarrier
\addtocounter{figure}{-1}\begin{figure*}
\centering
\includegraphics[width=0.7\columnwidth]{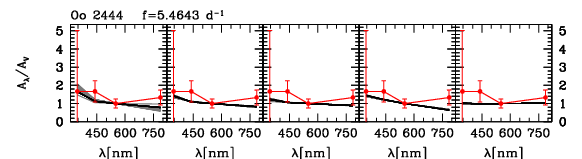}
\includegraphics[width=0.2\columnwidth]{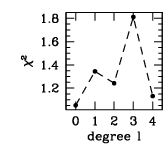}\\
\includegraphics[width=0.7\columnwidth]{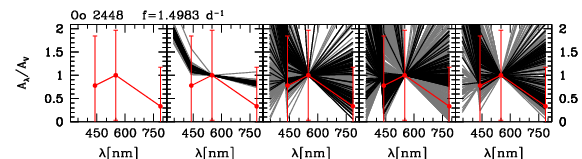}
\includegraphics[width=0.2\columnwidth]{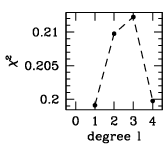}\\
\includegraphics[width=0.7\columnwidth]{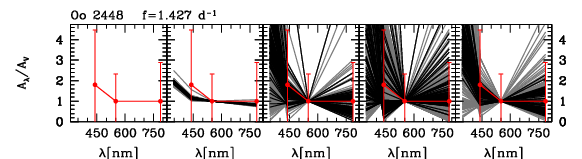}
\includegraphics[width=0.2\columnwidth]{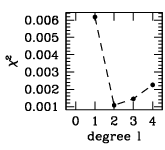}\\
\includegraphics[width=0.7\columnwidth]{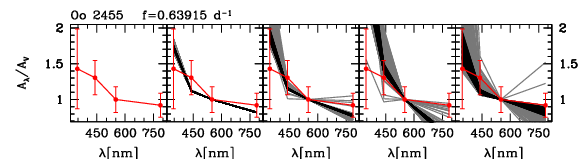}
\includegraphics[width=0.2\columnwidth]{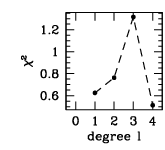}\\
\includegraphics[width=0.7\columnwidth]{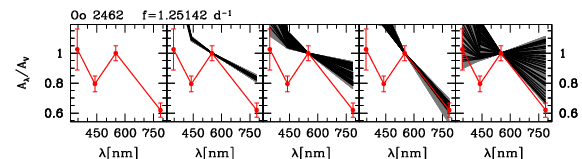}
\includegraphics[width=0.2\columnwidth]{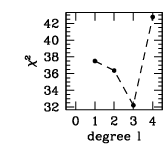}\\
\includegraphics[width=0.7\columnwidth]{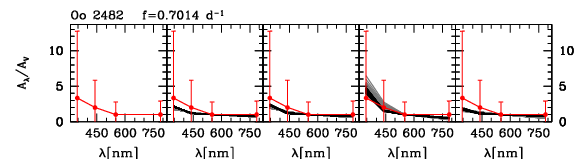}
\includegraphics[width=0.2\columnwidth]{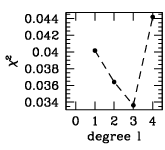}\\
\caption{Continued.}\end{figure*}\addtocounter{figure}{-1}\begin{figure*}
\centering
\includegraphics[width=0.7\columnwidth]{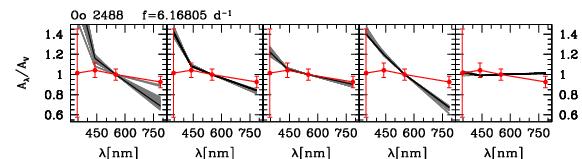}
\includegraphics[width=0.2\columnwidth]{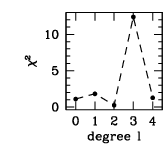}\\
\includegraphics[width=0.7\columnwidth]{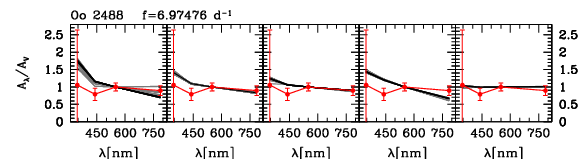}
\includegraphics[width=0.2\columnwidth]{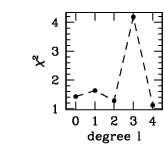}\\
\includegraphics[width=0.7\columnwidth]{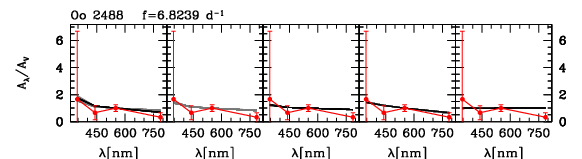}
\includegraphics[width=0.2\columnwidth]{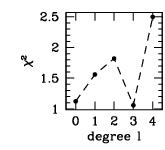}\\
\includegraphics[width=0.7\columnwidth]{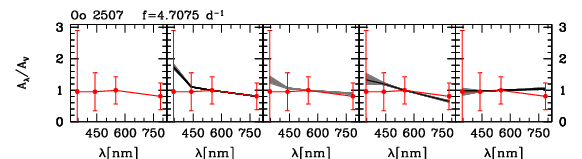}
\includegraphics[width=0.2\columnwidth]{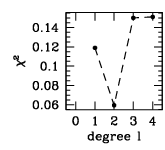}\\
\includegraphics[width=0.7\columnwidth]{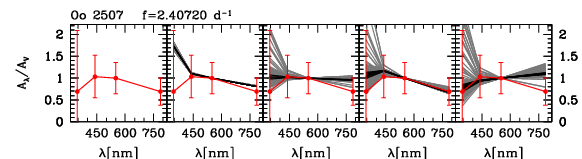}
\includegraphics[width=0.2\columnwidth]{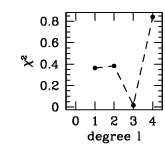}\\
\includegraphics[width=0.7\columnwidth]{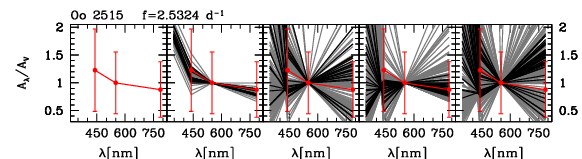}
\includegraphics[width=0.2\columnwidth]{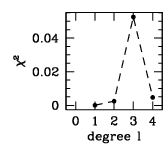}\\
\caption{Continued.}\end{figure*}\FloatBarrier
\addtocounter{figure}{-1}\begin{figure*}
\centering
\includegraphics[width=0.7\columnwidth]{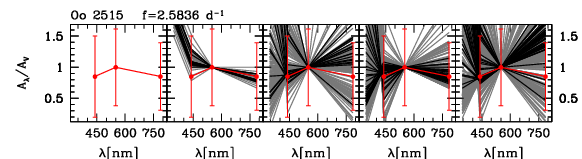}
\includegraphics[width=0.2\columnwidth]{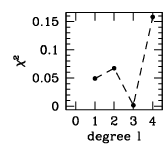}\\
\includegraphics[width=0.7\columnwidth]{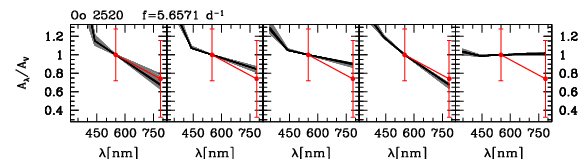}
\includegraphics[width=0.2\columnwidth]{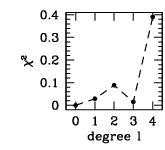}\\
\includegraphics[width=0.7\columnwidth]{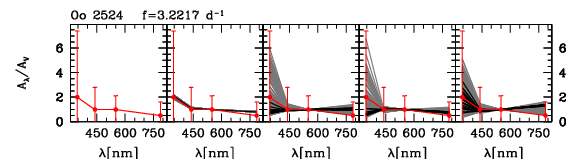}
\includegraphics[width=0.2\columnwidth]{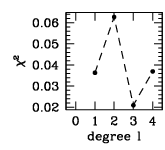}\\
\includegraphics[width=0.7\columnwidth]{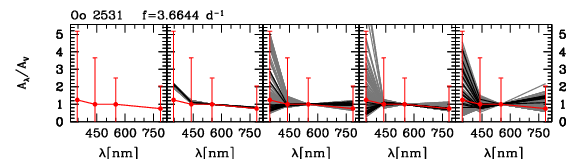}
\includegraphics[width=0.2\columnwidth]{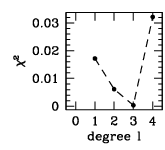}\\
\includegraphics[width=0.7\columnwidth]{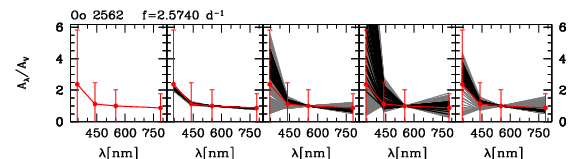}
\includegraphics[width=0.2\columnwidth]{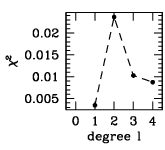}\\
\includegraphics[width=0.7\columnwidth]{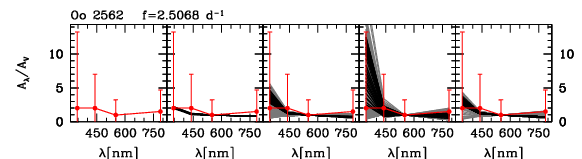}
\includegraphics[width=0.2\columnwidth]{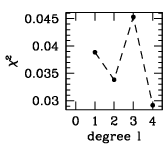}\\
\caption{Continued.}\end{figure*}\FloatBarrier
\addtocounter{figure}{-1}\begin{figure*}
\centering
\includegraphics[width=0.7\columnwidth]{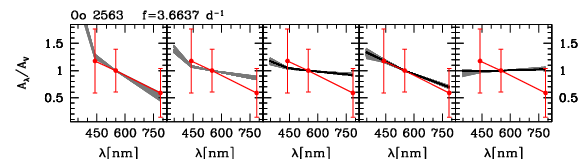}
\includegraphics[width=0.2\columnwidth]{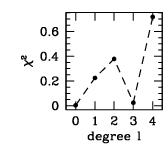}\\
\includegraphics[width=0.7\columnwidth]{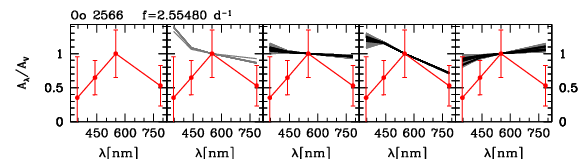}
\includegraphics[width=0.2\columnwidth]{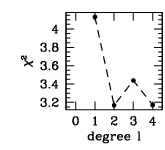}\\
\includegraphics[width=0.7\columnwidth]{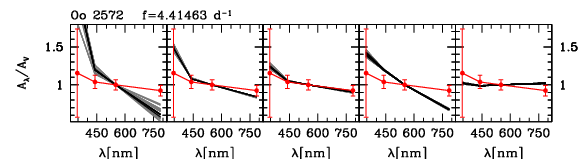}
\includegraphics[width=0.2\columnwidth]{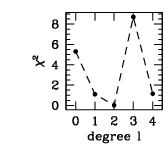}\\
\includegraphics[width=0.7\columnwidth]{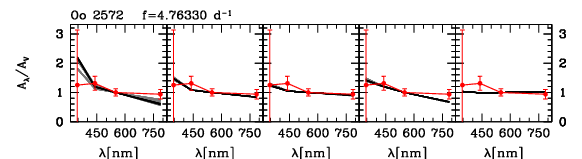}
\includegraphics[width=0.2\columnwidth]{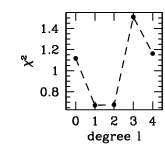}\\
\includegraphics[width=0.7\columnwidth]{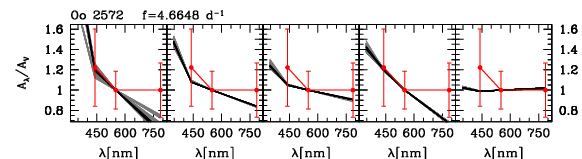}
\includegraphics[width=0.2\columnwidth]{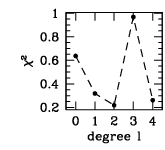}\\
\includegraphics[width=0.7\columnwidth]{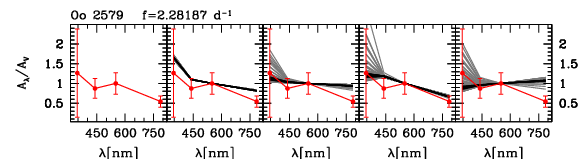}
\includegraphics[width=0.2\columnwidth]{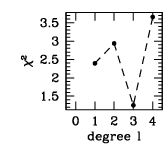}\\
\caption{Continued.}\end{figure*}\addtocounter{figure}{-1}\begin{figure*}
\centering
\includegraphics[width=0.7\columnwidth]{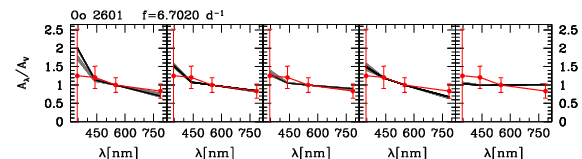}
\includegraphics[width=0.2\columnwidth]{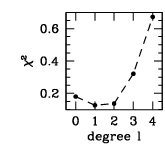}\\
\includegraphics[width=0.7\columnwidth]{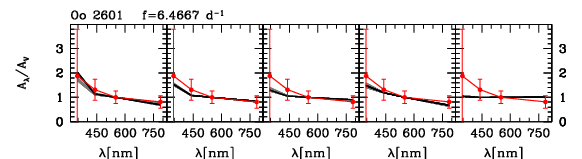}
\includegraphics[width=0.2\columnwidth]{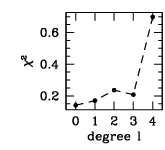}\\
\includegraphics[width=0.7\columnwidth]{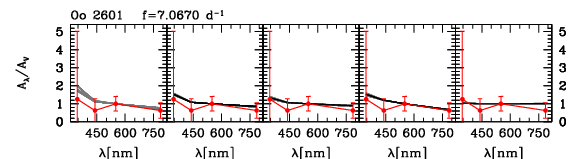}
\includegraphics[width=0.2\columnwidth]{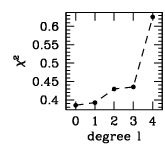}\\
\includegraphics[width=0.7\columnwidth]{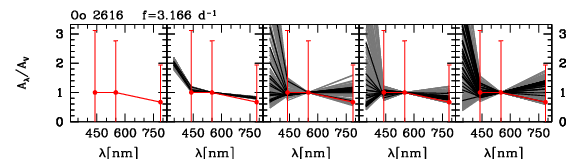}
\includegraphics[width=0.2\columnwidth]{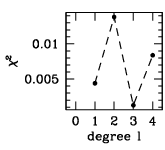}\\
\includegraphics[width=0.7\columnwidth]{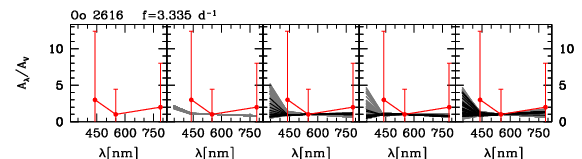}
\includegraphics[width=0.2\columnwidth]{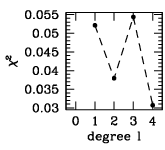}\\
\includegraphics[width=0.7\columnwidth]{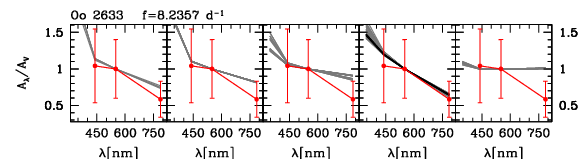}
\includegraphics[width=0.2\columnwidth]{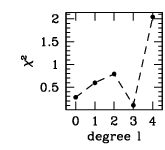}\\
\caption{Continued.}\end{figure*}\FloatBarrier
\addtocounter{figure}{-1}\begin{figure*}
\centering
\includegraphics[width=0.7\columnwidth]{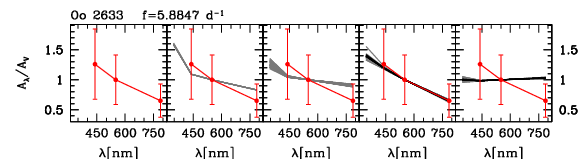}
\includegraphics[width=0.2\columnwidth]{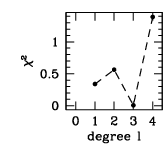}\\
\includegraphics[width=0.7\columnwidth]{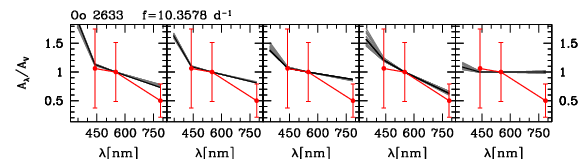}
\includegraphics[width=0.2\columnwidth]{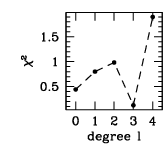}\\
\includegraphics[width=0.7\columnwidth]{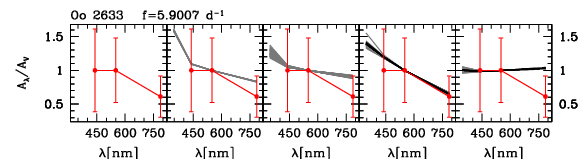}
\includegraphics[width=0.2\columnwidth]{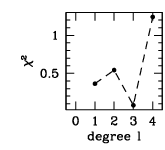}\\
\includegraphics[width=0.7\columnwidth]{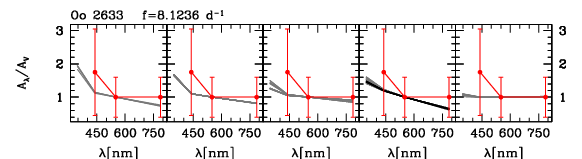}
\includegraphics[width=0.2\columnwidth]{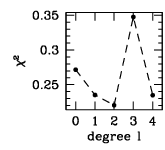}\\
\includegraphics[width=0.7\columnwidth]{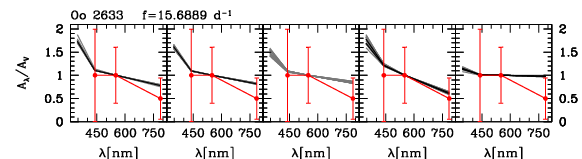}
\includegraphics[width=0.2\columnwidth]{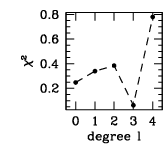}\\
\includegraphics[width=0.7\columnwidth]{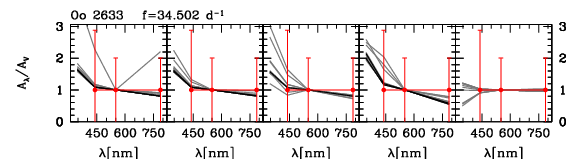}
\includegraphics[width=0.2\columnwidth]{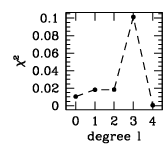}\\
\caption{Continued.}\end{figure*}\addtocounter{figure}{-1}\begin{figure*}
\centering
\includegraphics[width=0.7\columnwidth]{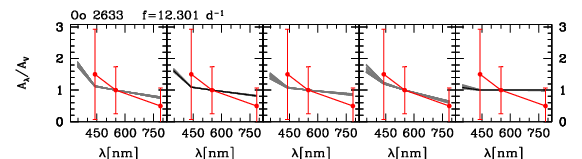}
\includegraphics[width=0.2\columnwidth]{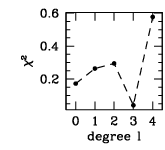}\\
\includegraphics[width=0.7\columnwidth]{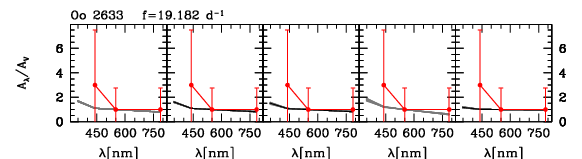}
\includegraphics[width=0.2\columnwidth]{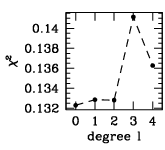}\\
\includegraphics[width=0.7\columnwidth]{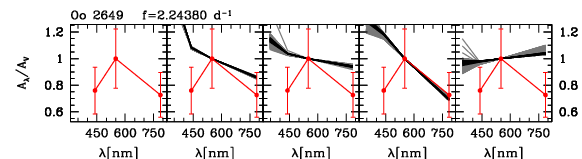}
\includegraphics[width=0.2\columnwidth]{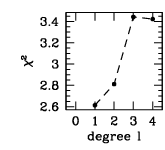}\\
\includegraphics[width=0.7\columnwidth]{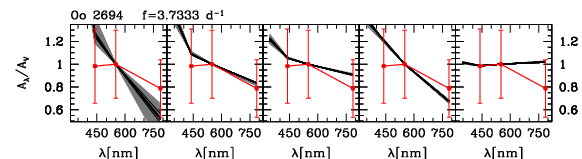}
\includegraphics[width=0.2\columnwidth]{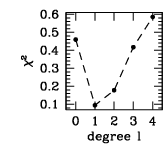}\\
\caption{Continued.}\end{figure*}

\end{document}